\documentclass[12pt]{article}
\usepackage{amsmath,amsthm,amssymb,euscript,epsf,epsfig,color}
\usepackage{array}
\usepackage{fancybox}
\usepackage{url}
\usepackage{hyperref}

\newcommand{\be}{\begin{equation}}
\newcommand{\ee}{\end{equation}}
\newcommand{\bea}{\begin{eqnarray}\displaystyle}
\newcommand{\eea}{\end{eqnarray}}

\newcommand{\nnm}{\nonumber}

\makeatletter
\@addtoreset{equation}{section}
\makeatother

\def\one{{\hbox{ 1\kern-.8mm l}}}
\def\zero{{\hbox{ 0\kern-1.5mm 0}}}

\def\bp{ { \bf p } }

  \def\cF{{\cal F}}
\def\cG{{\cal G}} \def\cH{{\cal H}} \def\cI{{\cal I}}
  
\def\cM{{\cal M}} \def\cN{{\cal N}} \def\cO{{\cal O}}
\def\cP{{\cal P}}  
\def\cS{{\cal S}} \def\cT{{\cal T}} 
  \def\cX{{\cal X}}
 
\def\mbX{ { \mathbb{X}}  } 
 
\def\s{ \sigma } 
\def\L{ \Lambda }

\def\mP{ \mathbb{P} }

\def\mbA{ \mathbb{A} } 
\def\mbC{ \mathbb{C}}
 
\def\mbZ{ \mathbb{Z} } 
\def\mbI{ \mathbb{I} }

\def\tr{ {\rm tr } } 
\def\Dim{ {\rm Dim} }
\def\Sym{ {\rm Sym} }
\def\tr{ {\rm tr } } 
\def\Str{ {\rm Str } } 
\def\Im{ {\rm Im } }
\def\Ker{ {\rm Ker } }
\def\Span{ {\rm Span } }

\def\BPS{ { \rm BPS} } 
\def\rmS{ {\rm S} }

\def\des{ { \rm D}  } 
\def\planar{ {\rm planar} }

\begin{document}
{}~
{}~
\hbox{QMUL-PH-10-06}
\break

\vskip .6cm

\centerline{{\LARGE \bf  From counting to construction of    }} 
\centerline{{\LARGE \bf   BPS states   in $\cN=4$ SYM }}

\medskip

\vspace*{4.0ex}

\centerline{ {\large \bf Jurgis Pasukonis}\footnote{j.pasukonis@qmul.ac.uk}
{ \bf  and }  {\large \bf Sanjaye Ramgoolam}\footnote{s.ramgoolam@qmul.ac.uk}  } 
\vspace*{4.0ex}
\begin{center}
{\large Department of Physics\\
Queen Mary, University of London\\
Mile End Road\\
London E1 4NS UK\\
}
\end{center}

\vspace*{5.0ex}

\centerline{\bf Abstract} \bigskip

We describe a universal  element  $ \mP $  in the group algebra 
of symmetric groups, whose characters provides the counting 
of quarter and eighth BPS states at weak coupling in $ \cN=4$ 
SYM, refined according to representations of the global symmetry group. 
 A related projector  $ \cP $ acting on the Hilbert space of the 
free theory is used to construct  the matrix of two-point functions 
of the states annihilated by 
 the one-loop dilatation operator,  at finite $N$ or in the large $N$ limit. 
The matrix is given simply in terms of Clebsch-Gordan coefficients 
of symmetric groups and dimensions of $U(N)$ representations. 
It is expected, by non-renormalization theorems, to 
contain  observables at strong coupling. Using the 
stringy exclusion principle, we  interpret a class of  its
eigenvalues and eigenvectors  in terms of  giant gravitons. 
We also give a formula for the action of the one-loop 
dilatation operator on the orthogonal basis of the free 
theory, which is manifestly covariant under the global symmetry.

\thispagestyle{empty}
\vfill

\eject


\tableofcontents

\section{Introduction}

The half-BPS sector of $ \cN=4$  $U(N)$ SYM has been a very fruitful 
area of study in the context of AdS/CFT \cite{malda,gkp,witten} from a number of different 
perspectives. It has shown how different classes of 
half-BPS local operators in the large $N$ theory can be 
mapped to different types of physical objects in space-time : 
Kaluza-Klein gravitons, 
strings, giant gravitons and classical space-time geometries
\cite{witten,minseib,freedmat,bbns,cjr,cr,bmn,llm}. 
The study of the two-point function has been central. It defines 
an inner product on the space of states corresponding 
to the local operators. The diagonalization of the inner product in 
the space of half-BPS operators in terms of Young diagrams
\cite{cjr} has provided the tools for identifying the  
physical objects in different semi-classical limits. 
For example we expect a state describing two distinct 
giant gravitons to be orthogonal to a state of  a single giant 
\cite{bbns,cjr}. 

Recent years have seen progress on the quarter and one-eighth 
BPS sectors. From the space-time point of view, there is a 
better understanding of the geometry of giant gravitons
and their connections to simple harmonic oscillator systems
\cite{mikhailov,beasley,kinney,biswas,mandsurya06}. 
From the point of view of gauge-invariant operators, 
we have a diagonalization at finite $N$ for  free Yang Mills i.e  
with the $g_{YM}^2 = 0 $  \cite{BHR1,BHR2,bhatt,kimram}.  
It is known that there is a jump in the spectrum of BPS 
operators from zero to weak coupling. It is conjectured
 that there is no further jump as the coupling is tuned to the strong coupling 
region \cite{kinney}.  This makes 
it extremely important to understand the weak coupling sector 
in complete  detail. The lessons we learn from these sectors 
will be useful for an understanding of the sixteenth BPS 
sector where we can access black holes with finite horizon areas.

Henceforth when we refer to BPS operators 
we have in mind quarter BPS operators or 
bosonic eighth BPS operators (more precisely eighth-BPS 
operators with bosonic lowest weight states)
 at weak coupling. 
The quarter BPS problem involves two matrices $ X_1 , X_2 $ 
transforming in the adjoint of $U(N) $ 
and a $U(2)$ global R-symmetry which rotates these two matrices.
The eighth BPS problem involves $ X_1 , X_2 , X_3 $ and a $U(3)$ global 
R-symmetry which rotates these. In our considerations below, we can keep 
the global symmetry general in the form $U(M)$. The general 
$U(M)$ case does not have a direct application in $ \cN =4$ 
SYM, but may have applications elsewhere in theories with 
$M$ copies of adjoints.

The  explicit construction of BPS operators was studied in
 \cite{hokryz1,HHHR03}.  A class of quarter BPS operators involving a small 
number of $X_1,X_2$ were  constructed in the $SU(N)$ theory at weak coupling.  
In \cite{BHR1,BHR2}, the diagonalization of the two-point function of
BPS operators in the free (zero-coupling) $U(N)$ theory at general finite $N$ 
was accomplished. The solution for the case of a general number $n$ 
of $ X_1 , X_2 $ was given in terms of the group theory of $S_n$. 
A Fourier basis constructed using Clebsch-Gordan coefficients
and matrix elements of $S_n$ was found to accomplish the diagonalization. 
The  one-loop BPS operators can be characterized as being 
orthogonal, in the free field theory metric,  to the descendants \cite{HHHR03}. 
Based on \cite{HH01} it was conjectured that these BPS operators have 
non-renormalized two- and three- point functions. 
The idea of characterizing BPS operators using orthogonality was
 developed in a  finite $N$ set-up, using the  zero coupling 
inner product to give expressions for the BPS operators \cite{BHR1}. 
The expressions were based on the notion of dual bases of operators 
\cite{copto,tomsu,robgwyn} and arrived at expressions involving 
inverse dimensions of $U(N)$ representations, 
 which will play a role later in this paper. 
A missing ingredient was an explicit description of the 
descendant operators. Exploiting some developments in BMN operators
\cite{vamver} a characterization of  the BPS operators 
in the $1/N$ expansion  was given in \cite{tom-recent} using  the
inner product in the planar limit. For related recent developments
on BMN operators,  see \cite{mxhuang}.

In this paper we describe the symmetrization operation 
on traces as a linear operator, which we call $\cP$, 
acting on the orthogonal Fourier basis 
for the free theory. This allows 
us to characterize the descendants as the kernel of $\cP$.
We use the symmetrization 
matrix to derive a formula for the two-point function 
of the BPS states to all orders in the $1/N$ expansion.
A manifestly finite $N$ construction is also given. 
It relies on an infinite dimensional Hilbert space which 
contains the Hilbert spaces for any $N$. The geometry 
of  intersections of subspaces in this Hilbert space 
plays a crucial role.

The paper is organised as follows. We will review the relevant 
background in Section~\ref{sec:reviews}. It will become clear that 
an explicit description of the  linear operator $\cP$ implementing 
the symmetrization will be important.  
The operation of symmetrizing traces appears
 in the simpler problem of counting BPS operators. 
 In Section~\ref{sec:counting} we review the standard counting 
results and  consider a refined counting according to representations
$ \Lambda $ of  $U(M)$. This is done with a new approach based 
on a universal element in 
$ \mbC ( S_n)$ related to trace symmetrization, which we denote  as $\mP$,  
and  whose characters provide this counting for any $U(M)$.
  We derive a generating function which gives the structure of 
$\mP$. This turns out to have a rich combinatoric structure, which 
we describe in Appendix~\ref{sec:integerseq}. We show in Section~\ref{sec:countconstr}  that analysing the {\it fine structure}  of   $ \mP$  leads 
naturally to the symmetrization operator $ \cP$. This allows the 
construction of the BPS operators in Section~\ref{sec:thecPop} 
and leads to the matrix of two-point functions in Section~\ref{sec:2pt}.  
 Section~\ref{sec:finiteN} gives a description of the finite $N$ 
construction. This discussion relies on a careful description of 
an infinite dimensional Hilbert space $ \cH $ spanned by
states $ | \L , M_{ \L } , R , \tau   \rangle  $ containing 
the labels of the free orthogonal basis.   An inner product 
which we call the $ S_{\infty} $ inner product plays a key role. 
The finite $N$ Hilbert space can be  realized as a quotient
of $\cH$ by the image of a finite $N$ projector. The $S_{\infty}$ 
inner product induces a simple  inner product on the
 finite $N$ Hilbert space, which is consistent with the 
finie $N$ cutoffs. Clarifying the  interplay between the finite 
$N$ projector and $\cP$, viewed as an 
operator in $ \cH$ is the key to the finite $N$ construction.   
In Section~\ref{sec:eigs2point}, we analyze the properties of the 
matrix of BPS two-point functions. Using the stringy exclusion 
principle \cite{malstrom,mst}, we interpret some of its 
eigenvectors in terms of giant gravitons and their spacetime excitations. 

We collect miscellaneous technical material 
in  the Appendices. Appendix~\ref{sec:onedil} also gives a new 
$U(3)$ covariant formula for the mixing of descendant 
operators under the action of the one-loop dilatation operator, 
which has a similar structure in terms of symmetric group 
Clebsch-Gordans as the symmetrization operator $ \cP$.

\section{Review  and Remarks on BPS operators } 
\label{sec:reviews}

We review here recent work on the construction of 
quarter and eighth (bosonic)  BPS operators.
It is known that the 
spectrum of BPS operators at zero coupling 
$ \cN =4 $ SYM (the free theory) is different from 
the spectrum at weak coupling. At zero coupling, 
any gauge invariant operator constructed from 
holomorphic combinations of the three adjoint complex scalar fields $ X_1 , X_2 , X_3 $ is BPS. 
At weak coupling only those operators in the 
kernel of one-loop dilatation operator 
$ \cH_2$  are  BPS \cite{Beisertcomplete}. 
In \cite{BHR1,BHR2}, the diagonalization of the 
 BPS operators in the free theory at finite $N$ was accomplished. 
We will begin with a  short summary of this work, 
where a Fourier basis was found to be central to  the free field 
diagonalization.  We will follow it with a summary of 
observations from \cite{HHHR03,BHR1,tom-recent} on the construction 
of BPS operators at one loop.   
This will motivate the construction of an operator $ \cP$ 
on the Fourier basis, which implements symmetrization of traces. 
In Section~\ref{sec:counting} we will obtain  new results 
on the counting of BPS operators, which provide 
an independent avenue and concrete steps towards the operator
 $ \cP $.

\subsection{The orthogonal basis in the free theory}

In diagonalizing the free two-point function \cite{BHR1,BHR2}, 
it was extremely useful to exploit the notion of 
Fourier transformation as applied in the context of 
the symmetric group. To set the stage, let us recall 
that the Fourier transform relation between a 
circle $S^1$ and the set of momenta $ \mbZ $ 
can be understood in group theoretic terms as a 
transformation relating  the group $U(1)$  to the 
space of inequivalent representations $  \mbZ  $.
The transformation from permutations of $n$ elements 
to representations is a generalization of this Fourier transformation to 
the symmetric group $S_n$. 

Gauge invariant multi-matrix operators can be written 
in the form 
\begin{equation} 
\cO_{ \vec a , \alpha } = \frac{  \tr_n ( \mbX_{ \vec{a} }~  \alpha ) }{ N^{ n/2} }  
 = \frac{ 1 }{ N^{ n/2} } 
 ( X_{a_1} )^{i_1 }_{ i_{\alpha(1) } } \cdots   ( X_{a_n} )^{i_n }_{ i_{\alpha(n) } } 
\end{equation} 
In the first expression, the operator  $ \mbX_{ \vec{a} } \equiv X_{a_1} \otimes X_{a_2} \cdots \otimes X_{a_n } $
where the matrix  $X_{a_k}$ can be viewed as  a linear map from the $N$-dimensional vector space $V$ to 
itself, with matrix elements $ ( X_{a_k} )^{i_k }_{ j_k  } $. The index $a$
 transforms in the fundamental $M$-dimensional 
 representation of $U(M)$.  By $ \tr_n $ 
we mean a trace of operators in $ V^{\otimes n } $, i.e 
 $\tr_n(\mbA) = \mbA^{i_1 \cdots i_n }_{i_1 \cdots i_n } $.  The permutation $ \alpha $ 
takes integers $ 1 \cdots n $ to $ \alpha(1) \cdots \alpha(n) $. 
In the context of representation theory, the $ \alpha$
label for operators  naturally Fourier-transforms using 
matrix elements $ D^R_{ij} ( \alpha ) $ to a set of labels $ R , i  , j $, 
where $ R $ runs over 
irreducible representations (irreps.) 
 of $S_n$ and $i,j$ run over states in the 
irrep.  The state indices $\vec a $ of the $U(M)$ symmetry 
can be transformed using Schur-Weyl duality into 
labels $  \L , M_{ \L } , m $ where $ \L $ labels 
simultaneously irreps of $U(M)$ and $S_n$,   $M_{ \L } $ 
labels a state in the irrep $ \L $ of $U(M)$, and $m$ labels 
a state in the irrep $ \L $ of $S_n$. The transformation is 
achieved by the Clebsch-Gordan coefficients 
$C^{ \vec a }_{ \L , M_{ \L } , k } $ for the decomposition 
\begin{equation} 
V^{ \otimes n } = \bigoplus_{ \L } V_{ \L }^{U(M) } \otimes V_{ \L }^{ S_n } 
\end{equation}  
This leads us to consider 
\begin{equation} 
\cO_{ \L  , M_{ \L }  , m , R , i , j } 
= \sum_{ \alpha, \vec{a} } D^{ R}_{ ij} ( \alpha ) 
C^{ \vec a }_{ \L , M_{ \L} , k }  \cO_{ \vec a , \alpha }
\end{equation} 
However the labels $ \vec a , \alpha $ are a redundant description
of the gauge invariant operators. Imposing the correct invariances
leads to operators 
\begin{equation}
\label{eq:ob_definition} 
\cO_{ \Lambda , M_{\Lambda} ,  R , \tau  } =  
\frac{ \sqrt {d_R } }{ n! }
\sum_{\alpha, \vec{a}}
S^{ ~ R~  R ~ \L ,~  \tau }_{ ~~  i ~ j ~ m } D^{R}_{ij} ( \alpha )
C^{ \vec a }_{ \Lambda , M_{\Lambda } ,
 m }  \cO_{ \vec a , \alpha }
\end{equation} 
where $ S^{ ~ R~  R ~ \L ,~  \tau }_{ ~~  i ~ j ~ m } $ are Clebsch-Gordan
coefficients for coupling $ R \otimes R \otimes \Lambda $ to 
the one-dimensional irrep. of $ S_n$. 
These operators also have the virtue of diagonalizing the free two-point function. We will call this the Fourier basis of the free theory. 
The inverse map giving the trace basis in terms of the Fourier basis is 
\begin{equation}\label{trans2}  
\cO_{ \vec a , \alpha }
=  \sum_{ \L , M_{ \L } ,  R , \tau  } \sqrt { d_R }   C_{ \vec a }^{ \L , M_{ \L } , m  } 
   D^R_{ij} ( \alpha  ) S^{ ~ R ~ R ~ \L ; \tau }_{ ~ ~i ~ j ~ m }
  \cO_{ \L , M_{ \L } ,  R , \tau  } 
\end{equation} 
 We defined 
$  C_{ \vec a }^{ \L ,  M_{ \L } ,  m  } =  (C^{ \vec a }_{ \Lambda , M_{\Lambda } ,  m })^*$ 
which satisfy standard orthogonality relations for Clebsch-Gordan coefficients 
\cite{BHR2}. 

Defining 
\begin{equation}\begin{split} 
 \cS_{ \vec a , \alpha }^{ \L , M_{ \L } , R , \tau } 
& =  \sqrt{ d_R }  C_{ \vec a }^{ \L , M_{ \L } , m } D^R_{ij} ( \alpha )  
  S^{ ~ R ~ R ~ \L ; \tau }_{ ~~ i ~ j ~ m } \\ 
\cT_{ \L , M_{ \L } , R , \tau }^{ \vec a , \alpha } 
& = \frac{ \sqrt { d_R } }{ n! }  C^{ \vec a }_{ \L , M_{ \L } , m }
 D^R_{ij} ( \alpha ) 
  S^{ ~ R ~ R ~ \L ; \tau }_{ ~~  i ~ j ~ m } 
\end{split}\end{equation} 
we may write a more compact version of the transformations 
(\ref{eq:ob_definition}) (\ref{trans2}) 
between trace and Fourier basis 
\begin{equation}\begin{split}\label{comptransf} 
& \cO_{ \vec a , \alpha } = 
 \cS_{ \vec a  , \alpha }^{ \L , M_{ \L } , R , \tau  }  
 \cO_{ \L , R , M_{ \L } , \tau } \\ 
&  \cO_{ \L , R , M_{ \L } , \tau } = 
 \cT_{ \L , M_{ \L } , R , \tau }^{ \vec a , \alpha }  \cO_{ \vec a , \alpha }
\end{split}\end{equation} 
We have 
\begin{equation} 
\sum_{\vec{a},\alpha} \cT_{ \L_1 , M_{\L_1} , R_1 , \tau_1 }^{ \vec a , \alpha } 
\cS_{ \vec a , \alpha }^{ \L_2 , R_2 , \tau_2  , M'_{\L_2} } 
= \delta_{ \L_1, \L_2 } \delta_{ M_{ \L_1} , M'_{ \L_2 } }
 \delta_{ R_1 , R_2 } \delta_{ \tau_1 , \tau_2 } 
\end{equation} 
and 
\begin{equation}\label{finiteNI} 
\sum_{\L , M_{ \L } ,  R , \tau } \cS_{ \vec a , \alpha }^{ \L , M_{ \L } , R , \tau  }  
\cT_{ \L , M_{ \L} , R , \tau }^{ \vec b , \beta } 
= \sum_{ \gamma  } \frac{ 1 }{ n! } \delta_{  \gamma ( \vec a )  ,  \vec b  } 
\delta_N ( \beta^{-1} \gamma^{-1} \alpha \gamma) 
\end{equation} 
The $ \delta_{ N } $ is defined as 
\begin{equation} 
\delta_N ( \alpha ) = \sum_{ R : c_1 (R) \le N } 
\frac{ d_R \chi_R ( \alpha ) }{ n! } 
\end{equation} 
When $ n < N $, this is the sum over all irreps 
of $S_n$, which is equal to
\begin{equation}\begin{split}  
 \delta ( \alpha )   & = 1 ~~ \hbox { for }  ~~ \hbox { $\alpha  ~ =$  identity element of}  ~ S_n \\ 
& = 0 ~~ \hbox{ otherwise } 
\end{split}\end{equation} 
 The right hand side of (\ref{finiteNI}) is the identity operator 
in the finite $N$ Hilbert space, which we will denote as 
$ \cI^{(N)} $ and return to in Section~\ref{sec:finiteN}.

The two-point function defines an inner product on the gauge-invariant operators
\begin{equation}
\langle \cO_1 | \cO_2  \rangle
= \lim_{ x_1 \rightarrow \infty  }  x_1^{2 n   }  
  \langle ( \cO_1 )^{\dagger} ( x_1 ) 
         ( \cO_2 ) ( 0  ) \rangle 
\end{equation}
Here and throughout the paper we use the \emph{zero-coupling} but \emph{finite $N$} two-point function and the associated inner product, unless specified otherwise. The inner product on the trace basis evaluated by Wick contractions
 is:
\begin{equation}\begin{split}
\label{eq:tr_2pt}
	\langle \cO_{ \vec b , \beta } | \cO_{ \vec a , \alpha } \rangle &= 
\sum_{ \gamma, \sigma \in S_n  } \delta_{ \gamma ( \vec a )  , \vec b  } 
N^{C(\sigma) - n} \delta( \beta^{-1} \gamma^{-1} \alpha \gamma \sigma)
\\
	&= \sum_{ \gamma \in S_n  } \delta_{ \gamma ( \vec a )  , \vec b  } 
\delta(\beta^{-1} \gamma^{-1} \alpha \gamma \Omega).
\end{split}\end{equation}
On the Fourier basis it is  diagonal  \cite{BHR2} 
\begin{equation}\label{diag2pt}\begin{split}  
\langle \cO_{ \Lambda_1 , M_{ \L_1} , R_1 , \tau_1 } | \cO_{ \L_2 , M'_{ \L_2} , R_2 , \tau_2 }  \rangle 
& = \frac{ n! \Dim R_1 }{ N^n d_{R_1}  } \delta_{ R_1 , R_2 } \delta_{ \tau_1 , \tau_2 }
 \delta_{ \L_1 , \L_2  } \delta_{ M_{ \L_1 } , M'_{ \L_2 } } \\ 
& = \frac{ \chi_{R_1} ( \Omega ) }{ d_{R_1} }
 \delta_{ R_1 , R_2 } \delta_{ \tau_1 , \tau_2 }
 \delta_{ \L_1,  \L_2  } \delta_{ M_{ \L_1 } , M'_{ \L_2 } }
\end{split}\end{equation} 
There is no $d_{ \L } $ as in eq. (110)  \cite{BHR2}  
because we defined our Clebsch-Gordans 
to be the coupling from $ R \otimes R \otimes \L $ 
to identity rather than $  R \otimes R \rightarrow  \L$. 
In addition because of the $ \sqrt { d_R } $ in the 
definition of the operator of $ d_R$ in the denominator 
rather than $d_R^2$. We explain the derivation of 
 the result (\ref{diag2pt}) in Appendix~\ref{sec:free2pt}.

The $ \Omega $ factor 
\begin{equation}\label{omegadef} 
\Omega = \sum_{ \sigma } N^{ C_{ \sigma } - n } \sigma 
\end{equation} 
is a { \it central  element}  in the group algebra $ \mbC ( S_n) $,  
i.e it commutes with any element of $  S_n  $. 
Defining the operator 
$ \cF $ on the Fourier basis as 
\begin{equation} 
\begin{split} \label{theFmatrix} 
(\cF)^{\L_2 , M'_{ \L_2 } , R_2 , \tau_2 }_{\L_1 , M_{ \L_1 } , R_1 , \tau_1}
& = \delta_{\L_1 ,  \L_2 } \delta_{ M_{ \L_1 } ,  M'_{\L_2 } } 
 \delta_{\tau_1  , \tau_2 } \delta_{R_1 ,  R_2 } \frac{ n! \Dim R_1 }{ N^n  d_{R_1}  }\\
& = \delta_{\L_1 ,  \L_2 } \delta_{ M_{ \L_1 } ,  M'_{\L_2 } } 
 \delta_{\tau_1  , \tau_2 } \delta_{R_1 ,  R_2 } \frac{ \chi_{R_1} ( \Omega )}{  d_{R_1} } 
\end{split} 
\end{equation} 
 We see that  
\begin{equation}\label{2ptF}  
\langle \cO_{ \Lambda_1 , M_{ \L_1} , R_1 , \tau_1 } | \cO_{ \L_2 , M'_{ \L_2} , R_2 , \tau_2 }  \rangle 
= 
(\cF)_{\L_2 , M_{ \L_2 } , R_2 , \tau_2 }^{\L_1 , M_{ \L_1 } , R_1 , \tau_1}
\end{equation} 
The operator $ \cG $ is defined with inverse matrix elements 
\begin{equation} \label{defG} 
\begin{split}
(\cG)_{\L_2 , M'_{ \L_2 } , R_2 , \tau_2 }^{\L_1 , M_{ \L_1 } , R_1 , \tau_1}
& = \delta_{\L_1 ,  \L_2 } \delta_{ M_{ \L_1 } ,  M'_{\L_2 } } 
 \delta_{\tau_1  , \tau_2 } \delta_{R_1 ,  R_2 } 
 \frac{ N^n d_{R_1 } }{ n! \Dim R_1   }  \\
& = \delta_{\L_1 ,  \L_2 } \delta_{ M_{ \L_1 } ,  M_{\L_2 } } 
 \delta_{\tau_1  , \tau_2 } \delta_{R_1 ,  R_2 } \frac{ \chi_{R_1} ( \Omega^{-1}  )}{  d_{R_1} } 
\end{split}
\end{equation} 
The second line is only meaningful in the region 
 $ n \le   N $, where $ \Omega $ can be inverted.  
The equation for the inverse dimension in terms of 
characters  $\frac{ \chi_{R_1} ( \Omega^{-1}  ) }{ d_{R_1} }$
is useful in two-dimensional Yang Mills theory (see 
e.g.  equation  (6.6)  in \cite{cmrii}). 
In the leading large $N$ (planar) limit, where $ n \ll N $, 
the two-point function approaches the identity 
\begin{equation} 
\begin{split} 
\label{eq:2ptplanar_def}
\langle \cO_{ \L_2 , M'_{ \L_2} ,  R_2 , \tau_2 } |  
\cO_{ \L_1 ,M_{\L_1} ,   R_1 , \tau_1 }  \rangle 
\rightarrow  
\langle \cO_{ \L_2 , M'_{ \L_2} ,  R_2 , \tau_2 } |
\cO_{ \L_1 ,M_{\L_1} ,   R_1 , \tau_1 }  \rangle_{\rm planar} 
=   \delta_{\L_1 ,  \L_2 } \delta_{ M_{ \L_1 } ,  M_{\L_2 } } 
 \delta_{\tau_1  , \tau_2 } \delta_{R_1 ,  R_2 }
\end{split} 
\end{equation} 
and we may write 
\begin{equation}\label{2pttoplanar} 
\langle \cO_{ \L_2 , M'_{ \L_2} ,  R_2 , \tau_2 } |  
\cO_{ \L_1 ,M_{\L_1} ,   R_1 , \tau_1 }  \rangle 
=
\langle \cO_{ \L_2 , M'_{ \L_2} ,  R_2 , \tau_2 } |  
\cF \cO_{ \L_1 ,M_{\L_1} ,   R_1 , \tau_1 }  \rangle_{\rm planar}
\end{equation}
Since the large $N$ limit of $ \Omega $ is $1$  the 
2-point function (\ref{eq:tr_2pt}) in the trace basis approaches 
\bea\label{tracelargeN} 
\langle \cO_{ \vec b , \beta } | \cO_{ \vec a , \alpha } \rangle_{ \planar } 
	= \sum_{ \gamma \in S_n  } \delta_{ \gamma ( \vec a )  , \vec b  } 
\delta(\beta^{-1} \gamma^{-1} \alpha \gamma ).
\eea 
This shows that the 2-point function in this limit  is zero unless the 
two operators have the same fields in the same trace structures. 
This is large $N$ factorization. 

\subsection{Towards  the construction of BPS  operators } 
\label{sec:review}

In this section we review earlier work on quarter and 
eighth-BPS operators at one-loop  \cite{HH01,HHHR03,BHR1,tom-recent}. 

 These  operators, which have vanishing one-loop anomalous dimensions,
 are annihilated by the one-loop dilatation 
operator $\cH_2$ \cite{BKPSS,Beisertcomplete}:
\begin{equation}
	\cH_2 \cO^{\rm BPS} = 0.
\end{equation}
where $\cH_2$ in the $U(3)$ sector is given by 
\bea 
\cH_2 = - { 1 \over 2 } \tr [ X_i , X_j ] [ \check X_i , \check X_j ] 
\eea 
Equivalently, the protected operators are precisely those 
which are  orthogonal,  in the inner product defined by
the zero-coupling two-point function, 
to  the SUSY-descendant operators $\cO^{\rm D}$:
\begin{equation}
\label{eq:od_obs_orth}
	\langle \cO^{{ \rm D } } | \cO^{\rm BPS} \rangle = 0.
\end{equation}
Indeed, from hermiticity of $\cH_2$, we have 
\bea 
 0 = \langle \cO^{\rm { any } } | \cH_2 \cO^{ \BPS} \rangle  =
    \langle \cH_2 \cO^{\rm { any } } | \cO^{ \BPS} \rangle
   =   \langle \cO^{{ \rm D } }  | \cO^{ \BPS} \rangle
\eea
It is expected that the 2- and 3-point functions of 
these operators receive no further corrections at higher loops.

The descendant operators are easily found: they are those which, when written as a gauge invariant product of single traces, contain a commutator of fields \cite{HHHR03}:
\begin{equation}
	\cO^{{ \rm D } } = \tr\left( [ X_{a_1}, X_{a_2} ] X_{a_3} X_{a_4} \ldots \right)
		\tr( \ldots ) \ldots
\end{equation}
This is true even when the total number of $X$'s denoted by $n$ is
 in the region 
 $n>N$, in which case, since there is no unique way to write an operator as a product of traces, an operator would be a descendant if there \emph{exists a way} to write it with a commutator. Since we have an explicit set of $\cO^{\rm D}$, the task of finding $\cO^{\rm BPS}$ is just finding the orthogonal subspace according to (\ref{eq:od_obs_orth}). This has already been pointed out and used in \cite{HHHR03}. The procedure, however, turns out to be complicated because of the non-planar inner product and does not provide a convenient way to express $\cO^{\rm BPS}$ when $n$ grows large.

In order to make progress  \cite{tom-recent,vamver} it is useful to use 
(\ref{2pttoplanar}) in order to  express the non-planar two-point function as
\begin{equation}
\label{eq:od_obps_planar}
	\langle \cO^{{ \rm D } } | \cO^{\rm BPS} \rangle =
	\langle \cO^{{ \rm D} } | \cF \cO^{\rm BPS} \rangle_{\rm planar} ,
\end{equation}
 Now pick the space of operators $\cO^{ \rm S } $ which are
 orthogonal to $\cO^{ \rm D} $ in the planar inner product   
\begin{equation}\label{orthoplanDS} 
	\langle \cO^{ \rm D }  | \cO^{ \rm S}  \rangle_{\rm planar} = 0 . 
\end{equation}
The space of protected operators can be written as
\begin{equation}
	\cO^{\rm BPS} = \cG  \cO^{ \rm S} .
\end{equation}
where $ \cG $ is defined in  (\ref{defG}) as the inverse of $\cF$. 
It follows simply from (\ref{eq:od_obps_planar}) that such operators will be orthogonal to $\cO^{\rm D}$ in the non-planar two-point function
\begin{equation}
	\langle \cO^{\rm D} |  \cG  \cO^{ \rm S}   \rangle
	= \langle \cO^{\rm D} | \cF \cG  \cO^{ \rm S}  \rangle_{\rm planar} 
	= \langle \cO^{\rm D} | \cO^{ \rm S}  \rangle_{\rm planar} 
	= 0
\end{equation}
and thus they are protected.
 The action of $ \cF  $, which acts diagonally 
 by multiplication with ${ n! \Dim R \over N^n d_R } $ in the Fourier basis
 is the operation 
\bea 
\cF ~ \tr (  \mbX_{ \vec a }~ \alpha ) 
= \tr  ( \mbX_{ \vec a } ~ \Omega ~  \alpha ) 
\eea 
where $ \Omega $ is the central element in the 
 algebra of $ \mbC ( S_n ) $ defined in (\ref{omegadef}). 
Indeed, using (\ref{trans2}),  
\begin{align}  
\cF ~   \cO_{ \vec a , \alpha } 
& = \cF  \sqrt { d_R }   C_{ \vec a }^{ \L , M_{ \L } , m  } 
  D^R_{ij} ( \alpha  ) S^{ ~ R ~ R ~ \L ; \tau }_{ ~ ~i ~ j ~ m }
  \cO_{ \L , M_{ \L } ,  R , \tau  } \nnm \\
&= { \chi_R ( \Omega ) \over d_R } \sqrt { d_R } 
  C_{ \vec a }^{ \L , M_{ \L } , m  } 
  D^R_{ij} ( \alpha  ) S^{ ~ R ~ R ~ \L ; \tau }_{ ~ ~i ~ j ~ m }
  \cO_{ \L , M_{ \L } ,  R , \tau  } \nnm \\ 
& =  \sqrt { d_R }   C_{ \vec a }^{ \L , M_{ \L } , m  } 
  D^R_{ij} ( \Omega \alpha  ) S^{ ~ R ~ R ~ \L ; \tau }_{ ~ ~i ~ j ~ m }
  \cO_{ \L , M_{ \L } ,  R , \tau  } \nnm \\
& =   \cO_{ \vec a , \Omega  \alpha  }   
\end{align} 
To go from the second  line to the third we use the fact that 
$ \Omega $ is a central element, hence proportional to the identity 
in any irrep.  $R$  by Schur's Lemma. Similar steps can be done 
with the $ \cG $ matrix. In the large $N$ limit, where $ n \le  N $, 
we can define $ \Omega^{-1} $ as an element in $ \mbC ( S_n ) $ and 
\begin{equation}
\label{eq:g_on_tr}
\cG ~  \tr (  \mbX_{ \vec a } ~ \alpha ) 
= \tr  ( \mbX_{ \vec a } ~  \Omega^{-1}  ~   \alpha )  
\end{equation}

We now want to make a final point, which was not fully expressed or   
exploited  in the earlier literature. The space of operators $ \cO^{\rm S}$  which 
are orthogonal to $\cO^{\rm D}$  in the planar inner product is  the space
 spanned by products of symmetrized traces. The \emph{symmetrization}
 operation  replaces each single trace
 with the sum over all permutation of
 elements inside the trace
\begin{equation}\label{symmetrize} 
	{\rm symm} \left[\, \tr(X_{a_1} X_{a_2} \ldots X_{a_n} ) \,\right] \equiv
	\frac{1}{n!} \sum_{\sigma\in S_n} 	
	\tr(X_{a_{\sigma(1)}} X_{a_{\sigma(2)}} \ldots X_{a_{\sigma(n)}})  \equiv 
    \Str ( X_{a_1 }  X_{a_{2}}  \ldots X_{a_n }  )
\end{equation}
The operators $ \cO^{\rm S}$ are linear combinations of 
\begin{equation}
\label{eq:os_from_str}
	 \Str\left( X_{a_1} X_{a_2} \ldots X_{a_k} \right) \Str\left(X_{b_1} X_{b_2} \ldots X_{b_l} \right) \ldots .
\end{equation}
The orthogonality to descendants  can be seen, 
using the property of large $N$ factorization, as follows. 
An operator $\cO^{\rm S}$, orthogonal to descendants,     is 
a sum of individual multitrace operators.   
Take one multitrace term of it
\begin{equation}
	\tr\left( X_a X_b A \right) B,
\end{equation}
where $A$ is a product of matrices and $B$ is some remaining multitrace
 operator. Now the whole $\cO^{\rm S}$ has to be orthogonal to a descendant
\begin{equation}
	\tr\left( [X_a, X_b] A \right) B.
\end{equation}
But in order for that to happen the $\cO^{\rm S}$ must also contain a term
\begin{equation}
	\tr\left( X_b X_a A \right) B,
\end{equation}
so that
\begin{equation}
	\langle	
		\tr\left( (X_a X_b - X_b X_a)  A \right) B \,|\, 
		\tr\left( (X_a X_b + X_b X_a) A \right) B
	\rangle_{\rm planar}  = 0 .
\end{equation}
Of course, it can be that $\tr\left( X_b X_a A \right)$ is already the same as $\tr\left( X_b X_a A \right)$ either because of $a=b$ or because of cyclicity of the trace, but that does not change the fact if one is contained in $\cO^{\rm S}$ then the other must be as well with the same coefficient. Therefore, since every two adjacent matrices in $\cO^{\rm S}$ have to be symmetrized, we  conclude that the whole $\cO^{\rm S}$ must be built from  fully symmetrized traces as in (\ref{eq:os_from_str}) This allows us to conclude that the space of one-loop BPS operators is spanned by
\begin{equation}
\label{eq:obps_from_str_sec2}
	\cO^{\rm BPS} = \cG  \left[ \Str\left( X_{a_1} X_{a_2} \ldots X_{a_k} \right) \Str\left(X_{b_1} X_{b_2} \ldots X_{b_l} \right) \ldots \right] .
\end{equation}

In this paper we will develop  (\ref{eq:obps_from_str_sec2})
and arrive at a description of the BPS operators 
in the $1/N $ expansion in terms of an  explicitly defined matrix 
$ \cG \cP $ in the free Hilbert space of SYM. 
The  matrix $ \cP $,  related to 
symmetrization,  will be described in Section~\ref{sec:thecPop}.
Incidentally, the properties of $ \cP$ will lead to a one-line 
 proof of   (\ref{orthoplanDS}) in Section~\ref{sec:propertiescP_sec4}. We
 will derive the two-point function on the BPS operators
as a matrix $ ( \cP \cG \cP ) $. In Section~\ref{sec:finiteN} 
we will extend these results to finite $N$, where we will identify a 
projector $ \cP_{I}$, which approaches $ \cP$ in the large $N$ limit. 
In the next section, we will derive some new results 
on the counting of BPS operators, which will lead naturally to 
the operator $ \cP $.

\section{Counting and the universal element in $\mbC(S_{ n }) $  }
\label{sec:counting} 

\subsection{Chiral ring counting }

Here we briefly review some  known results about counting of BPS operators.

The expected number of bosonic gauge invariant eighth-BPS
 operators with given R-charges $(n_1, n_2, n_3)$ can be
 calculated using either the Chiral ring counting arguments \cite{kinney} or alternatively using the description of the BPS sector as a Hilbert space of $N$ 3D harmonic oscillators \cite{Berenstein:2005aa}. In the chiral ring
gauge invariant operators  made from  $X_1 , X_2 , X_3$ are counted 
modulo  equivalences following from  
\begin{equation}\label{vaceqs} 
	[X_1, X_2] = [X_2, X_3] = [X_3, X_1] = 0,
\end{equation}
It is argued that this  leads to a 
counting of traces of diagonal matrices \cite{kinney} : 
\begin{equation}
\label{eq:symm_traces}
	\tr \left((X_1)^{a_1}(X_2)^{b_1}(X_3)^{c_1}\right) 
	\tr \left((X_1)^{a_2}(X_2)^{b_2}(X_3)^{c_2}\right) \ldots  
	\tr \left((X_1)^{a_k}(X_2)^{b_k}(X_3)^{c_k}\right)
\end{equation}
or as a symmetric polynomial of the $N$ eigenvalues of the three matrices.
This counting of states also arises from the collective coordinate 
quantization of the coordinates of the BPS moduli space 
determined by (\ref{vaceqs}) upon reduction to $ 0 +1 $ 
dimensions on $ S^3$  \cite{Berenstein:2005aa}.  
In the limit of large $N$ the
 counting is given  by the partition function $Z(x_1,x_2,x_3)$:
\begin{equation}
\label{eq:gxy_definition}
	Z(x_1,x_2,x_3) = \prod_{i+j+k>0 }^{\infty} { \frac{1}{1 - x_1^i x_2^j x_3^k}  } = \sum_{n_1,n_2,n_3=0}^{\infty} Z_{n_1n_2n_3}\, x_1^{n_1} x_2^{n_2} x_3^{n_3},
\end{equation}
with numbers $Z_{n_1n_2n_3}$ providing the number of operators with charges $(n_1, n_2, n_3)$. This will be valid as long as $n_1+n_2+n_3 \le N$.

For a  $U(N)$ gauge group at finite $N$,  the counting is given by the generating function $Z_N(x_1,x_2,x_3)$ which can  be expressed as:
\begin{align} 
	Z(\nu;x_1,x_2,x_3) & = \prod_{i,j,k=0}^{\infty} { \frac{1}{1 - \nu x_1^i x_2^j x_3^k} } = 1 + \sum_{N=1}^{\infty} Z_N(x_1,x_2,x_3) \nu^N  \\ 
 Z_N ( x_1 , x_2 , x_3 ) & = \sum_{n_1,n_2,n_3=0}^{\infty} Z_{N ; n_1n_2n_3}\, x_1^{n_1} x_2^{n_2} x_3^{n_3}
\end{align} 

In terms of counting this is equivalent to limiting the number of traces $k$ appearing in the operators such as (\ref{eq:symm_traces}) to at most $N$ (this is 
certainly not true in the actual construction of BPS states at finite $N$ 
as can be easily checked using the 
 examples of Appendix~\ref{sec:app_example}). 
 This limit arises from the fact that there are only $N$ distinct eigenvalues to build the symmetric polynomials, or, in the 3D harmonic oscillator picture, from the fact that the collection of labels $(a_i,b_i,c_i)$ indicate the excitation numbers of $k$ particles, and we are limited to at most $N$ excited particles.

For the purposes of this paper we would like one extra refinement, that is, to distinguish operators in different representations $\Lambda$ of the global symmetry group $U(3)$, which rotates the three matrices $X_a$. We will represent $\Lambda=[\lambda_1,\lambda_2,\lambda_3]$ by a Young diagram with at most three rows of lengths $\lambda_1 \ge \lambda_2 \ge \lambda_3$ of $U(3)$. 
Operators transforming in the representation $ \Lambda $ contain 
 a total number of matrices equal to the number of boxes in the
 Young diagram. 
\begin{equation}
	\lambda_1 + \lambda_2 + \lambda_3 = n_1 + n_2 + n_3 \equiv n
\end{equation}
 We denote by $\cM_\Lambda$ the  multiplicity 
of   $\Lambda$. 
 Each representation $\Lambda$ contributes to the full partition function the \emph{Schur polynomial} in variables $x_i$, 
therefore the partition function can be expanded as \cite{dolan07,tom-recent}: 
\begin{equation}\label{schurexpand} 
	Z(x_1, x_2, x_3) = \sum_{\Lambda} \cM_\Lambda \, \chi_\Lambda(x_1, x_2, x_3).
\end{equation}
This can be inverted to calculate $\cM_\Lambda$ (see for example (A.23) in \cite{fulton}):
\begin{equation}
\label{eq:m_from_zx}
	\cM_{[\lambda_1,\lambda_2,\lambda_3]} = 
	\left[
		Z ( x_1 , x_2 , x_3)
		\prod_{i<j}(x_i - x_j)
	\right]_{ x_1^{\lambda_1+2} x_2^{\lambda_2+1}x_3^{\lambda_3} }
\end{equation}
where the square brackets with the subscript is an instruction  to take the coefficient of the indicated monomial from the function inside.
Equivalently we can define 
\begin{equation}
	\cM(x_1,x_2,x_3) 	
	\equiv  Z ( x_1 , x_2 , x_3 )
		\left(1 - \frac{x_2}{x_1}\right)
		\left(1 - \frac{x_3}{x_1}\right)
		\left(1 - \frac{x_3}{x_2}\right)
\end{equation} 
and write 
\begin{equation} 
\label{eq:mx_from_zx}
 \cM_{[\lambda_1,\lambda_2,\lambda_3]} = 
 \bigg[
 	\cM(x_1,x_2,x_3)  
\bigg]_{  x_1^{\lambda_1} x_2^{\lambda_2} x_3^{\lambda_3}  }	.	
\end{equation}
Organizing the  operators into representations of $U(3)$ works 
equally well for  finite $N$. We can  define $\cM_{N ; \Lambda}$ as   the multiplicity of the 
 $\Lambda$ representation in the space of operators at finite $N$.
We can again define
\begin{equation}
	\cM_N(x_1,x_2,x_3) 	
	= Z_N ( x_1 , x_2 , x_3 )
		\left(1 - \frac{x_2}{x_1}\right)
		\left(1 - \frac{x_3}{x_1}\right)
		\left(1 - \frac{x_3}{x_2}\right) 
\end{equation}
and write for finite $N$ multiplicities $\cM_{N ; [\lambda_1,\lambda_2,\lambda_3]}$
\begin{equation}
\cM_{N ; [\lambda_1,\lambda_2,\lambda_3]} = 
 \bigg[
 	\cM_N (x_1,x_2,x_3)  
\bigg]_{  x_1^{\lambda_1} x_2^{\lambda_2} x_3^{\lambda_3}  }
\end{equation} 
We refer the reader to (\ref{tab:multiplicities}) in Appendix~\ref{sec:app_multiplicities} for some examples of $\cM_\L$.

We can also refine the counting for specified $U(1)$ charges $ n_1, n_2, n_3$ 
into those belonging to specified $U(3)$ representations $ \Lambda $.  
Let  $Z_{  n_1 n_2 n_3}^\Lambda$,   $Z_{N ;  n_1 n_2 n_3}^\Lambda$
be these refined multiplicities at large and finite $N$ respectively. 
They are given by 
\begin{equation}\begin{split}
Z_{ n_1 n_2 n_3}^\Lambda &=  \cM_{ \Lambda } ~ g ( \Lambda , [n_1] , [n_2 ] , [n_3] )
\\
Z_{N ; n_1 n_2 n_3}^\Lambda & = \cM_{ N ;  \Lambda } ~ g ( \Lambda , [n_1] , [n_2 ] , [n_3] )
\end{split}\end{equation}
 where $ g (  [n_1] , [n_2 ] , [n_3] ; \Lambda )$ standard group theoretic 
 numbers, namely the Littlewood-Richardson 
coefficient for combining three single-row Young diagrams of 
lengths $ n_1,n_2,n_3$ to give the Young diagram $ \Lambda $
(see for example \cite{fulton}). 

The representation multiplicities  $\cM_\Lambda$, $\cM_{ N  ;  \Lambda } $ 
thus  provide a refinement of the counting $ Z_{ n_1 n_2 n_3} $,
$Z_{N ; n_1 n_2 n_3}$ : 
\begin{equation}\begin{split}
Z_{n_1, n_2 , n_3 } &= \sum_{ \Lambda } Z_{  n_1 n_2 n_3}^\Lambda
\\
Z_{ N; n_1 n_2 n_3 } &=  \sum_{ \Lambda } Z_{N ; n_1 n_2 n_3}^\Lambda
\end{split}\end{equation}
In the rest of the paper we will focus  on  these 
representation multiplicities $\cM_\Lambda , \cM_{ N ; \Lambda } $.

\subsubsection{Universality from  counting formulae }

The above arguments can be run for the case of quarter BPS 
operators with $U(2)$ symmetry. We have the partition functions 
\begin{equation}\begin{split}
\label{eq:gxy_definition-u2}
	Z(x_1,x_2) = \prod_{i+j>0 }^{\infty} { \frac{1}{1 - x_1^i x_2^j }  }
&  = \sum_{n_1,n_2=0}^{\infty} Z_{n_1n_2}\, x_1^{n_1} x_2^{n_2}
\\ 
Z_{N} ( x_1 , x_2 ) = \bigg[  \prod_{i,j =0 }^{\infty}
 { \frac{1}{1 - \nu x_1^i x_2^j } }\bigg]_{ \nu^N }  & =  
\sum_{n_1,n_2=0}^{\infty} Z_{N ; n_1n_2}\, x_1^{n_1} x_2^{n_2}
\end{split}\end{equation} 
and representation multiplicities 
\begin{equation}\begin{split}
\cM_{[\lambda_1,\lambda_2]} & = 
	\bigg[
		Z ( x_1 , x_2 ) ( x_1 - x_2 ) 
	 	\bigg]_{ x_1^{ \lambda_1 + 1 } x_2^{\lambda_2} } 
\\	 	 
\cM_{N ; [\lambda_1,\lambda_2]} &= 
   	\bigg[
		Z_N  ( x_1 , x_2 )
	( x_1 - x_2 ) 
	\bigg]_{ x_1^{ \lambda_1 + 1 } x_2^{\lambda_2} }
\end{split}\end{equation} 
These can be usefully expressed in the form  
\begin{equation}\begin{split} 
\cM_{[\lambda_1,\lambda_2]} & = Z_{ \lambda_1 , \lambda_2 } - 
Z_{ \lambda_1 +1  , \lambda_2 -1 }
\\ 
\cM_{ N ; [\lambda_1,\lambda_2]} & = Z_{ N  ; \lambda_1 , \lambda_2 } - 
Z_{N ;  \lambda_1 +1  , \lambda_2 -1 }
\end{split}\end{equation}
An important point is that these expressions in fact follow 
from the ones for $U(3)$ simply by setting $ \lambda_3 = 0$. 
If we consider generating functions with $M$ variables, 
as appropriate for $U(M)$ symmetry,   for 
sufficiently large $M$, they contain all the information 
about the multiplicities $ \cM_{ \Lambda }$ for any Young 
diagrams $ \Lambda $ having $n$ boxes, with $ n < M$. 
Of course when $M < n $, we are only interested in Young diagrams 
with less than or equal to $n$ rows.   All Young diagrams $\Lambda $
with $n$ boxes correspond to the complete set of 
representations of $S_n$, so we 
may expect structures entirely in $S_n$ which contain information 
about the multiplicities $ \cM_{ \Lambda } , \cM_{ N ; \Lambda } $ 
for all $ \Lambda $ with $n$ boxes which are irreps. of 
 $ S_n$. In the next section, we will show that 
 $\cM_{ \Lambda } $ is related to a universal element $ \mP$ 
 and $  \cM_{ N ; \Lambda }$ to a finite $N$ version thereof $ \mP^{(N)} $.

\subsection{Universal element }
\label{sec:univelem} 

We will  now present an alternative method of counting the BPS
 operators and multiplicities $\cM_\Lambda$ using the 
symmetric group techniques. We will develop a notion
 of \emph{universal element} $\mP$ in the symmetric group algebra
 whose characters give the multiplicities. The element is easy to define 
but the precise coefficients  of different permutations in $ \mP$ (which 
only depend on the conjugacy class of these permutations) are 
not easy to infer directly from the definition. 
We will find that the  generating function of BPS states $\cM(x_1,x_2,x_3)$ 
in (\ref{eq:mx_from_zx}) can, after a change of variables, be used to 
derive a generating function for these coefficients. Analysing the structure of 
$\mP$ will lead to the construction of BPS
 operators in Section~\ref{sec:countconstr}.

First consider the space of abstract  strings of $n$ operators $X_1, X_2, X_3$ without putting any traces:
\begin{equation}
\label{eq:xastring}
	X_{a_1} \otimes X_{a_2} \otimes \ldots \otimes X_{a_n},
\end{equation}
where $a_i$ take values 1 to 3. Most of what we will discuss is true for any global symmetry group $U(M)$ with indices running over 1 to $M$, in the case 
at hand we will specialize to $ M =2$ for the quarter BPS sector
and $M=3$ for the eighth-BPS sector. 
 Each ``letter" $X_a$ can be thought of as a state in the vector space $V_M$, which is the fundamental of $U(M)$. The whole string is then a state in the tensor product representation $V_M^{\otimes n}$. 

There is the natural action of the symmetric group $S_n$ on these states where the permutation $\sigma$ shuffles the indices:
\begin{equation}
	X_{a_{\sigma(1)}} \otimes X_{a_{\sigma(2)}} \otimes \ldots \otimes X_{a_{\sigma(n)}}.
\end{equation}
By  Schur-Weyl duality the tensor product representation can be decomposed into irreducible representations of both $U(3)$ and $S_n$:
\begin{equation}
\label{eq:shcurweyl}
	V_M^{\otimes n} = \oplus_\Lambda V^{U(M)}_\Lambda \otimes V^{S_n}_\Lambda.
\end{equation}
This immediately tells us that if our operators were defined by ordered strings of $X_a$ as  in (\ref{eq:xastring}) then the multiplicity of the $U(M)$ representation $\Lambda$ would be just $d_\Lambda$ -- the dimension of $V^{S_n}_\Lambda$.

In our case of interest the operators are instead products of traces like (\ref{eq:symm_traces}), where the order of operators inside the trace does not matter. Let us still view them as states in $V_M^{\otimes n}$, but with extra structure, for example:
\begin{equation}
\label{eq:xatraces}
	\tr(X_{a_1} X_{a_2}) \tr(X_{a_3}X_{a_4}) \tr(X_{a_5}).
\end{equation}
The groups $U(3)$ and $S_n$ act on the indices just like before. The only difference now between counting states like (\ref{eq:xatraces}) and those in (\ref{eq:xastring}) is that if we change the ordering of the operators within a trace or if we swap all operators between two traces of the same length it is still considered to be the same state. This restriction can be formulated in terms of the $S_n$ action: in our example we want to consider the states which are invariant under permutations $\sigma=(12)$, $\sigma=(34)$ and $\sigma=(13)(24)$.

Consider a general trace structure which we represent by a partition $p=[1^{p_1},2^{p_2},\ldots]$. This means there are $p_i$ traces of length $i$, and we have $\sum_i i p_i = n$. We will first count how many states there are with the given $p$ and later sum over it. The set of permutations which leave the state invariant given the trace structure $p$ will form a subgroup of $S_n$ which we denote $G(p)$. It consists of all permutations within traces:
\begin{equation}
	G_1(p) = \times_i (S_i)^{p_i},
\end{equation}
with each $S_i \subset S_n$ understood to be embedded in $S_n$ in such a way to permute its corresponding trace. Regarding $ S_n$ as the permutations 
of integers $1 \cdots  n $, the $1$-cycles correspond to $S_1^{p_1}$ 
acting on $ 1 \cdots p_1 $, the $2$-cycles correspond to $ S_2^{p_2}$
acting on $ p_1+1  \cdots p_1 + 2p_2 $. $ G(p) $ also contains  
permutations which exchange traces:
\begin{equation}
	G_2(p) = \times_i S_{p_i},
\end{equation}
where each $S_{p_i}$ permutes all the elements in the $p_i$ traces of the same length $i$. The full group which leaves the state invariant is the one generated by $G_1(p)$ and $G_2(p)$, which can be seen as a semi-direct product\footnote{
It is a semi-direct product in the sense that $G_2$ provides an automorphism of $G_1$ by conjugation in the $S_n$: $g_2 G_1 g_2^{-1} = G_1$ for $g_2 \in G_2$. Then group multiplication of two elements in $G_1 \ltimes G_2$ can be defined as $(g_1, g_2)\cdot(h_1,h_2) = (g_1 (g_2 h_1 g_2^{-1}), g_2 h_2)$ where $g_1, h_1 \in G_1$ and $g_2, h_2 \in G_2$.
}:
\begin{equation}
\label{eq:gp_definition}
	G(p) = G_1(p) \ltimes G_2(p).
\end{equation}
For completeness note that:
\begin{equation}
	|G_1(p)| = \prod_i (i!)^{p_i} \,, \quad
	|G_2(p)| = \prod_i (p_i)! \,, \quad
	|G(p)| = |G_1(p)|\,|G_2(p)| \,.
\end{equation}
Note that we have chosen a specific embedding of $ G(p)$. 
Different permutations $ \alpha $ in the conjugacy class 
$p$ give different embeddings $ G ( \alpha ) $ of the same group in $ S_n$. 
We can now define a projector which will project onto the subspace of $V_M^{\otimes n}$ invariant under $G(p)$:
\begin{equation}
\label{eq:pp_definition}
\begin{split}
\mP_p 
	&= \frac{1}{|G(p)|} \sum_{\sigma \in G(p)} \sigma \\
	&= { 1 \over |G_1 ( p )|  |G_2 ( p )|  } \sum_{ \sigma \in G_1 (p) } \sum_{ \tau \in G_2 (p) } \sigma \tau.
\end{split}	
\end{equation}
Choosing different $\alpha $ also leads to different $ \mP_{ \alpha } $, 
a special case of which is $\mP_p$ : we will return to this freedom 
in the next section. The element $\mP_p$ 
 is exactly what we need to account for the multiplicity of 
 distinct operators with the trace structure $p$: the dimension of the space projected to is just the trace of the projection operator
\begin{equation}
	\cM_p = \tr_{V_M^{\otimes n}} (\mP_p).
\end{equation}
What we in fact want to know is how many times a particular $U(M)$ representation $\Lambda$ appears in this space. For that we also project onto the $V^{U(M)}_\Lambda$ and divide by its size:
\begin{equation}
\label{eq:Mlambdap}
	\cM_{\Lambda, p} = \frac{1}{\Dim \Lambda} \tr_{V_M^{\otimes n}}( \mP_p \mP_\Lambda ),
\end{equation}
where $\Dim \Lambda$ is the dimension of $U(M)$ representation $\Lambda$.
Because of the Schur-Weyl duality (\ref{eq:shcurweyl}) we know that $\mP_\Lambda$ just projects onto the $V^{U(M)}_\Lambda \otimes V^{S_n}_\Lambda$ factor, and so:
\begin{equation}\begin{split}
	\cM_{\Lambda, p} 
		&= \frac{1}{\Dim \Lambda} \tr_{V^{U(M )}_\Lambda \otimes V^{S_n}_\Lambda}( \mP_p ) 
\\
		&= \frac{1}{\Dim \Lambda} \tr_{V^{U( M )}_\Lambda}( \mbI ) \tr_{V^{S_n}_\Lambda}( \mP_p ) 
\\		
		&=  \chi_\Lambda( \mP_p ).
\end{split}\end{equation}
Here $\chi_\Lambda$ is the symmetric group character and this is an expression entirely in terms of $S_n$ quantities! The second line follows from the fact that $\mP_p$ consists of permutations, therefore, it acts only in the $V_\Lambda^{S_n}$ factor. Then the trace factors into $U(M)$ and $S_n$ parts, with the trace over identity in the $U(M)$ canceling $\Dim \Lambda$. This result
 relates nicely to the fact we mentioned  earlier,
 that for states  living in $V_3^{\otimes n}$,  the multiplicity of representation $\Lambda$ would be just $d_\Lambda = \chi_\Lambda(\mbI)$. Instead here because of the invariance  of products of traces under some permutations captured by $\mP_p$ we lose some states, which is reflected in counting by taking trace of the projector $\mP_p$ instead of identity.

We are just a short step away now from the formula for the total multiplicity $\cM_\Lambda$. All we need to do is to sum over all partitions $p$ describing different trace structures. Also note that the precise form of $\mP_p$ depends on how the $G(p)$ group is embedded in $S_n$, i.e. how the traces are ordered. However, the counting does not depend on that, and so we want to make things more uniform by summing over all embeddings. That gives us the final form for what we will call the \emph{universal element} $\mP$ in the algebra of the symmetric group $\mbC(S_n)$:
\begin{equation}
\label{eq:p_definition}
\begin{split}
	\mP &= \sum_{p \vdash n}\left( \frac{1}{n!} \sum_{\gamma \in S_n} \gamma \mP_p \gamma^{-1} \right)  \\
		&= \sum_{p \vdash n}\left( \frac{1}{n!\,|G(p)|} \sum_{\gamma \in S_n} \sum_{\sigma \in G(p)} \gamma \sigma \gamma^{-1} \right).
\end{split}\end{equation}
Note that $\mP$ depends on $n$ but we will leave that 
dependence implicit. 

Now the formula for counting multiplicities is just
\begin{equation}\label{eq:M_lambda_from_P} 
\fbox{   
$ \displaystyle{ 
	\cM_{\Lambda} = \chi_\Lambda (\mP).
}$} 
\end{equation}

Just to emphasize, this is exactly the same $\cM_\Lambda$ as in (\ref{eq:m_from_zx}), but now we have expressed it as a character of an element $\mP$ of $\mbC(S_n)$. The first nice thing about this result is that the same $\mP$ will give multiplicity for any $U(M)$ representation $\Lambda$. When the global symmetry 
is $U(2)$, we are only interested in $ \Lambda = [ \lambda_1 , \lambda_2 ]  $, i.e in Young diagrams with at most two rows;  for $U(3)$ we apply the formula 
for $  \Lambda = [ \lambda_1 , \lambda_2 , \lambda_3 ]$.  
 Second, the $\mP$ itself is an interesting object to study. It is universal in the sense that it is defined uniquely for every $S_n$ and there is no additional data that goes into it. Since $\mP$ contains averaging over conjugation by $\gamma$, its terms will be  sums over conjugacy classes, 
and we will find that the coefficients weighting these conjugacy classes
 are integers (after multiplication by $n!$) with interesting properties.

Before we go on let us provide a short example of $\mP$ and $\cM_\Lambda$'s to get a better feeling for them. Take $n=4$, states made from four matrices. Using (\ref{eq:p_definition}) we can explicitly calculate:
\begin{equation}
	\mP(S_4) = \frac{1}{4!} \left(
		3 \Sigma_{[4]} + 3 \Sigma_{[3,1]} + 7 \Sigma_{[2,2]}
		+ 7 \Sigma_{[2,1,1]} + 15 \Sigma_{[1,1,1,1]}
	\right),
\end{equation}
where $\Sigma_p$ is the sum over all elements in the conjugacy class $p$:
\begin{equation}
	\Sigma_p \equiv \sum_{\sigma \in p} \sigma .
\end{equation}
Now if we go and compute the characters of $\mP$ we get:
\begin{equation}\begin{split}
	\cM_{[4]} &= \chi_{[4]} (\mP) = 5 \\
	\cM_{[3,1]} &= \chi_{[3,1]} (\mP) = 2 \\
	\cM_{[2,2]} &= \chi_{[2,2]} (\mP) = 2.
\end{split}\end{equation}
These are indeed the right multiplicities. Converting to the  more familiar numbers of operators for different $(n_1,n_2,n_3)$ according to 
(\ref{schurexpand}) gives $Z_{400}=5,\, Z_{310}=7,\, Z_{220}=9$.

 Note that $ \chi_{ \Lambda } ( \mP_p ) $ has some vanishing  
 properties which are easy to understand by thinking about 
 symmetries and trace structures. When $p$ is a single cycle, 
then the character vanishes for any $ \Lambda $ with more than 
 one row. When it has two cycles, the character is zero for 
any $ \Lambda $ with more than two rows. This reflects the well-known fact that 
 the simplest quarter (eighth) BPS operator has two (three)  traces at large $N$. 

We refer the reader to Appendix~\ref{sec:app_multiplicities} for explicit examples of $\mP$ at various $n$.

\subsection{Generating function for $\mP$ } 

As already mentioned, $\mP$ will be an important ingredient in the construction of operators in this paper, so we would like to know more about it than the formula (\ref{eq:p_definition}), from which it is hard to see what the sums evaluate to. First, we find that it can always be written as
\begin{equation}\label{mPexpandtp}
	\mP(S_n) = \frac{1}{n!} \sum_{p \vdash n} t_p \Sigma_p,
\end{equation}
where the coefficients $t_p$ are \emph{integers}. These are numbers assigned to each possible partition $p$, and, therefore, have various  combinatorial properties, which might be of interest in themselves. For certain simple sequences 
of partitions they belong to integer sequences tabulated in \cite{OEIS}.   We discuss  them further in  Appendices \ref{sec:app_multiplicities},  \ref{sec:integerseq}  and provide more explicit examples there.

Consider an  \emph{exponential generating function} $t(\vec{y})$ for $t_p$ in the sense that:
\begin{equation}
	t(y_1,y_2,y_3,\ldots) = \sum_p
		t_{p=[1^{p_1}2^{p_2}3^{p_3}\ldots]}
		{ y_1^{p_1} \over p_1! }
		{ y_2^{p_2} \over p_2! }
		{ y_3^{p_3} \over p_3! }
		\ldots
\end{equation}
so that power of $y_i$ indicates the number of cycles of length $i$ in the partition. We claim that this generating function is in closed form:
\begin{equation}
\label{eq:TPcompact}
	t(\vec{y}) = \exp \left(
		\sum_{d=1}^{\infty} \frac{1}{d} \left(
			e^{d\sum_i y_{di}} - 1
		\right)
	\right) = \prod_{d=1}^{\infty}   \exp \left(
		 \frac{1}{d} \left(
			e^{d\sum_i y_{di}} - 1
		\right) \right) 
\end{equation}
or more explicitly
\begin{equation}
	t(\vec{y}) = \exp \left( 
		\left( e^{y_1 + y_2 + y_3 + \ldots} - 1 \right)
	+ \frac{1}{2} \left( e^{2y_2 + 2y_4 + 2y_6 + \ldots} - 1 \right)
	+ \frac{1}{3} \left( e^{3y_3 + 3y_6 + 3y_9 + \ldots} - 1 \right)
	+ \ldots
\right).
\end{equation}
This is as close as we can get to an explicit formula  for $\mP$.

We will prove that (\ref{eq:TPcompact}) generates the coefficients of $\mP$ correctly by verifying the counting formula (\ref{eq:M_lambda_from_P}) with the assumed $t(\vec{y})$ against the known result (\ref{eq:mx_from_zx}). The expression $\chi_\Lambda(\mP)$ can be evaluated directly from the generating function $t(\vec{y})$ with the help of Frobenius character formula. It states that a symmetric group character can be written as \cite{fulton}:
\begin{equation}
	\chi_\Lambda( p ) = \left[
		\prod_{i<j}(x_i - x_j)
		\prod_k \left( x_1^k + x_2^k + \ldots + x_n^k \right)^{p_k}		
	\right]_{x_1^{n-1+\lambda_1}x_2^{n-2+\lambda_2}\ldots},
\end{equation}
where $p$ is a partition. Using this we can express $\chi_\Lambda({\mP})$ as:
\begin{equation}\begin{split}
	\chi_\Lambda({\mP}) 
		&= \frac{1}{n!} \sum_{p \vdash n} t_p \chi_\Lambda(\Sigma_p)
 \\		
		&= \sum_{p \vdash n} \frac{t_p}{|\Sym(p)|} \chi_\Lambda(p)
 \\
		&= \left[
		\prod_{i<j}(x_i - x_j)
		\sum_{p \vdash n} \frac{t_p}{|\Sym(p)|} 
		\prod_k \left( x_1^k + x_2^k + \ldots + x_n^k \right)^{p_k}		
	\right]_{x_1^{n-1+\lambda_1}x_2^{n-2+\lambda_2}\ldots}.
\end{split}\end{equation}
In the second line we used the fact that $\Sigma_p$ contains $n!/|\Sym(p)|$ elements all in conjugacy class $p$, where $\Sym(p)$ is the automorphism group of permutations in $p$ and
\begin{equation}
	|\Sym(p)| = \prod_i (p_i !) i^{p_i} .
\end{equation}
Let us now perform a change of variables
\begin{equation}
\label{eq:xk_to_yk_substitution}
	y_k = \frac{1}{k} \left( x_1^k + x_2^k + \ldots + x_n^k \right).
\end{equation}
We get
\begin{equation}
\label{eq:multfromF}\begin{split}
	\chi_\Lambda({\mP}) 	
	&= \left[
		\prod_{i<j}(x_i - x_j)
		\sum_{p \vdash n} \frac{t_p}{|\Sym(p)|} 
		\prod_k k^{p_k} (y_k)^{p_k}		
	\right]_{x_1^{n-1+\lambda_1}x_2^{n-2+\lambda_2}\ldots}
 \\
	&= \left[
		\prod_{i<j}(x_i - x_j)
		\sum_{p \vdash n} t_p
		\prod_k \frac{(y_k)^{p_k}}{p_k !}
	\right]_{x_1^{n-1+\lambda_1}x_2^{n-2+\lambda_2}\ldots}	
 \\	
	&= \left[
		\prod_{i<j}(x_i - x_j)
		\times 
		t(\vec{y})
	\right]_{x_1^{n-1+\lambda_1}x_2^{n-2+\lambda_2}\ldots}	
\end{split}\end{equation}
with exactly the generating function $t(\vec{y})$, with the understanding 
  that  the $y_k$ variables have to be substituted according to (\ref{eq:xk_to_yk_substitution}).

We can evaluate  the function $t(\vec{y})$  in terms of $x_k$ variables: 
\begin{equation}
	t(\vec{x}) 
	\equiv t\left(y_k = \frac{1}{k}\sum_{i=1}^{n} (x_i)^k \right)
	= \exp \left\{
		\sum_{d=1}^{\infty} \frac{1}{d} \left[
			\exp \left(d\sum_k \frac{1}{dk} \sum_i (x_i)^{dk}  \right)
			- 1
		\right]
	\right\}.
\end{equation}
Let us simplify this using $\sum_{k=1} (x^k)/k = \log(1/(1-x))$ and some other infinite series relationships:
\begin{equation}
\label{eq:sy_derivation}\begin{split}
	t(\vec{x}) &= \exp \left\{
		\sum_{d=1}^{\infty} \frac{1}{d} \left[
			\exp \left(\sum_i \log \frac{1}{1 - (x_i)^{d}}  \right)
			- 1
		\right]
	\right\}
\\
	&= \exp \left\{
		\sum_{d=1}^{\infty} \frac{1}{d} \left[
			\prod_{i=1} \frac{1}{1 - (x_i)^{d}}
			- 1
		\right]
	\right\}
\\
	&= \exp \left\{
		\sum_{d=1}^{\infty} \frac{1}{d} \left[
			\sum_{i+j+k+\ldots>0} (x_1^i x_2^j x_3^k \ldots)^d
		\right]
	\right\}
\\
	&= \exp \left\{		
			\sum_{i+j+k+\ldots>0} 
			\log \left( \frac{1}{1 - (x_1^i x_2^j x_3^k \ldots)} \right)
	\right\}
\\
	&= \prod_{i+j+k+\ldots>0} 
			\frac{1}{1 - (x_1^i x_2^j x_3^k \ldots)}.
\end{split}\end{equation}
We find, in fact, that $t(\vec{x})$ in terms of $x_i$ variables is precisely the partition function $Z(x_1,x_2,x_3)$ from (\ref{eq:gxy_definition}) only with arbitrary number of variables  
\begin{equation}
	t(x_1,x_2,x_3;x_{i>3}=0) = Z(x_1,x_2,x_3).
\end{equation}
Evaluating (\ref{eq:multfromF}) for the specific case of counting $U(3)$ representations $\Lambda=[\lambda_1,\lambda_2,\lambda_3]$ we then immediately find
\begin{equation}\begin{split}
	\chi_{[\lambda_1,\lambda_2,\lambda_3]}(\mP ) &= 
	\left[
		Z ( x_1 , x_2 , x_3)
		\prod_{i<j}(x_i - x_j)
	\right]_{ x_1^{\lambda_1+2} x_2^{\lambda_2+1}x_3^{\lambda_3} }
.
\end{split}\end{equation}
This is precisely $\cM_{[\lambda_1,\lambda_2,\lambda_3]}$ according to (\ref{eq:m_from_zx})! That means we have explicitly shown $\chi_\Lambda(\mP ) = \cM_\Lambda$ and, therefore, proved (\ref{eq:TPcompact}).

\subsection{Finite N  counting } 

We can also introduce a finite-$N$ version of  the universal element, denoted 
as  $\mP^{(N) } $, and defined as  
\begin{equation}
\label{eq:mPN_defined}
\mP^{(N)}  = \sum_{p \vdash n   }^{(N)} \left( \frac{1}{n!} \sum_{\gamma \in S_n} \gamma \mP_p \gamma^{-1} \right) 
\end{equation}
 
The superscript on the summation symbol indicates that the 
summation over partitions $p = [ 1^{p_1} 2^{p_2} \cdots ] $ is restricted by 
$ \sum p_i \le N$. This constraint follows from the description 
of the chiral ring counting in terms of $N$ particles in 
a simple harmonic oscillator, which can be described 
as the counting of products of 
symmetrized traces with no more than $N$ factors. 
By repeating the arguments explained in deriving 
(\ref{eq:M_lambda_from_P}), we obtain 
\begin{equation}
	\cM_{N;\Lambda} = \chi_\Lambda(\mP^{(N)} ).
\end{equation}
As in (\ref{mPexpandtp}) there is an  expansion in conjugacy classes
\begin{equation} 
\mP^{(N)} = { 1 \over n! } \sum_{p \vdash n } t_p \Sigma_p  
\end{equation} 
Let us introduce conjugate variable $\nu$ to $N$ again
 and define a deformed generating function $t(\nu;\vec{y})$:
\begin{equation}
\label{eq:tnux_definition}
	t(\nu; \vec{y}) ={ 1 \over 1 - \nu }  \exp \left(
		\sum_{d=1}^{\infty} \frac{\nu^d }{d} \left(
			e^{d\sum_i y_{di}} - 1
		\right)
	\right).
\end{equation}
Then the picking off the coefficient of $\nu^N$ in $t(\nu;\vec{y})$ will provide a generating function $t_N(\vec{y})$ for the coefficients $t_{p}$ in 
$\mP^{(N)}$:
\begin{equation}
	t(\nu; \vec{y}) =  \sum_N \nu^N t_N(\vec{y}).
\end{equation}
This can be proved by noting first that the derivation (\ref{eq:sy_derivation}) with $t(\nu;\vec{y})$ instead of $t(\vec{y})$ gives $\cM(\nu;x_1,x_2,x_3)$. Then we see that
\begin{equation}\begin{split}
	\chi_\Lambda(\mP^{(N)} ) 
		& = \left[ \chi_\Lambda(\mP(\nu)) \right]_{\nu^N}  \\
		& = \left [  t ( \nu; \vec{y})
 ( x_1 - x_2 )( x_1 - x_3 ) ( x_2 - x_3 ) \right ]_{\nu^N x_1^{\lambda_1+2 } x_2^{\lambda_2 + 1 } x_3^{\lambda_3}}  \\ 
 &= \cM_{N;\Lambda},
\end{split}\end{equation}
where by $\mP(\nu)$ we just mean the object generated by $t(\nu;\vec{y})$. This is again in agreement with Chiral ring counting results.

\section{From counting to operators on the Hilbert space  } 
\label{sec:countconstr} 

The counting of BPS operators in Section~\ref{sec:counting}  used a universal element $ \mP \in \mbC ( S_n ) $ defined in (\ref{eq:p_definition}). 
It is a sum of projectors $\mP_p$ over conjugacy classes $p$.  Equivalently it can be viewed as a sum over the symmetric group of projectors 
$\mP_{\alpha }$, where $\alpha\in S_n$ is a permutation specifying a concrete embedding of the cycle structure $p$. We examine this in
 Section~\ref{sec:p_map} and also derive a related operator $\bp$.
The operator $\bp$  is then used in Section~\ref{sec:thecPop} to construct an operator $ \cP $ which acts on the free basis $ \cO_{ \L , M_{ \L } , R , \tau  } $ 
and produces  linear combinations of symmetrized traces. With the help of $\cP$ we then construct the BPS basis valid at $N \ge n$.
We then pause to  discuss the properties of $\cP$ in Section~\ref{sec:propertiescP_sec4} and conclude with the calculation of the two-point function on the BPS states in Section~\ref{sec:2pt}.
  
In Section~\ref{sec:finiteN} we will also see how the operator $\cP$ allows the construction of BPS operators at finite $N$.

\subsection{A $\bp$- map from $ \mbC ( S_n ) \rightarrow \mbC ( S_n ) $ } 
\label{sec:p_map}

 As we remarked in Section~\ref{sec:counting}  the element 
 $\mP_p$ in $ \mbC( S_n) $, which counts the number of 
 $U(M)$ representations $ \L $ 
among trace structures specified by the cycle structure $p$, 
depends on a choice of embedding of the group $ G ( p) $ 
in $S_n$. Different permutations $ \alpha $ 
in the conjugacy class $p$ are associated with different 
groups $ G ( \alpha ) $. We can define an element 
\begin{equation} 
\label{eq:mpa_defined}
\mP_{ \alpha}  = { 1 \over |G( \alpha )|  }
 \sum_{ \sigma \in G ( \alpha ) } \sigma 
\end{equation}
There is a related  linear  map $ \bp : \mbC ( S_n ) \rightarrow \mbC ( S_n ) $ 
\begin{equation}
\label{eq:bp_defined}
\bp  ( \alpha ) = \mP_{ \alpha } = \sum_{ \beta } \bp_{ \beta , \alpha } \beta 
\end{equation}
A  trace $ \tr_n ( \mbX_{ \vec a }  \alpha ) $ is symmetrized by 
the action of $ \mP_ { \alpha } $ 
\begin{equation}
\label{pforsymm}  
{ \rm { symm } } [ \tr_n  ( \mbX_{ \vec a } \alpha ) ] = 
\sum_{ \beta } \bp_{ \beta, \alpha } \tr_n   ( \mbX_{ \beta ( \vec a )  } \alpha )
\end{equation}
We extend the definition to sums of traces by linearity, i.e 
the symmetrization of a sum of multi-traces is the sum of terms obtained 
by symmetrization each multi-trace.

The universal element in the previous section (\ref{eq:p_definition}) is a linear combination 
\begin{equation}\begin{split}
\mP &= \frac{1}{n!}\sum_{p \vdash n}\sum_{\gamma\in S_n} \gamma \mP_p \gamma^{-1}
\\
&= \sum_{ \alpha \in S_n } { |\Sym( \alpha )|  \over n! }   \mP_{ \alpha } 
\\
& = \sum_{ \alpha ,\beta } { |\Sym( \alpha )|  \over n! } \bp_{ \beta , \alpha  } \beta 
\end{split}\end{equation}

We can define the coefficients in the Fourier transformed basis 
for $  \mbC ( S_n ) $. 
\begin{equation}\label{fourierp}  
\bp^{ R i j }_{ S k l } \equiv  \sum_{ \alpha , \beta } 
 D^S_{kl}  ( \beta )  \bp_{ \beta , \alpha  } D^{R}_{ ij} ( \alpha ) 
\end{equation} 
Then 
\begin{equation} 
\bp_{\beta , \alpha } = \sum_{ R , S , i,j, k, l } { d_R\over n! }
 { d_S \over n! } ~ D^S_{kl} ( \beta ) ~   \bp^{ R ~ i ~ j }_{ S  ~ k ~ l }   
~ D^R_{ij} ( \alpha ) 
\end{equation} 
\begin{equation} 
\bp ( \alpha  )  = \sum_{ R , S , i,j, k, l } { d_R\over n! }
 { d_S \over n! } 
\bp^{ R i j }_{ S k l } ~ D^R_{ij} ( \alpha ) ~  D^S_{kl} ( \beta ) ~  \beta  
\end{equation} 

Figure \ref{fig:ppft} gives a diagrammatic expression of the equation 
(\ref{fourierp}). On the right hand side we use the 
standard diagrammatic form for operators. 
On the left we we have  short-hand which we will use subsequently.    
\begin{figure}
\begin{center}
\input{pdiag.pstex_t}
\caption{ Diagram for $\bp$  } \label{fig:pdiag}
\end{center}
\end{figure}

\subsection{The $ \cP$ operator and the BPS basis }
\label{sec:thecPop} 

In this section we proceed with the explicit construction of the BPS operators, annihilated by the one loop dilatation operator $\cH_2$. 
As explained in Section~\ref{sec:review}, in order to do that, we have to consider operators of the form (\ref{eq:obps_from_str_sec2}). The construction involves two tasks: symmetrization and the application of $\cG$.
A priori it is not clear how to accomplish both in a way to arrive at a closed form expression. The reason is that if we start off with a naive basis of multitrace operators, the symmetrization is easy to implement, but the action of $\cG $ (\ref{eq:g_on_tr}) is complicated, involving the inverse $\Omega^{-1}$ which is not well defined for $n>N$. 
On the other hand if we start with the diagonal basis of the free theory $\cO_{\L,M_\L,R,\tau}$, the $\cG$ action is straightforward (\ref{defG}), but it is not clear how to impose the symmetrization. Here we will show how the $\bp_{\beta,\alpha}$ defined in the previous section allows us to do just that. The procedure described in this section is illustrated in the Appendix~\ref{sec:app_example} with concrete examples of $\Lambda=[2,2]$ and $\L=[3,2]$.

We define the following basis for the BPS operators
\begin{equation}
\label{eq:obps_definition}
	\cO^{\rm{BPS }}_{ \Lambda , M_{\Lambda} ,  R , \tau  } =
	\cG  ~  {\rm symm} \left[
		\cO_{ \Lambda , M_{\Lambda} ,  R , \tau  }
	\right].
\end{equation}
In order to evaluate this expression, the key step is to calculate how the symmetrization acts on 
$\cO_{ \Lambda , M_{\Lambda} ,  R , \tau}$. We use (\ref{pforsymm}) and the fact that it acts linearly:
\begin{equation}\label{eq:cp_derriv1} \begin{split}
	{\rm symm} & \left[
		\cO_{ \Lambda , M_{\Lambda} ,  R , \tau  }
	\right] 
=
	{ \sqrt {d_R } \over n! } \sum_{\alpha, \vec{a}}
		S^{ R~  R ~ \L ,\;  \tau }_{ ~ i ~ j ~ m } 
		D^{R}_{ij} ( \alpha ) 
		C^{ \vec a }_{ \Lambda , M_{\Lambda } , m } 
		{\rm symm} \left[ \tr ( \mbX_{ \vec a } ~ \alpha ) \right]
\\ &=		
	{ \sqrt {d_R } \over n! } \sum_{\alpha, \beta, \vec{a}}
		S^{ R~  R ~ \L ,\;  \tau }_{ ~ i ~ j ~ m } 
		D^{R}_{ij} ( \alpha ) 
		C^{ \vec a }_{ \Lambda , M_{\Lambda } , m } 
		\bp_{\beta, \alpha} \tr ( \mbX_{ \beta(\vec a) } ~ \alpha )
\\ &=		
	{ \sqrt {d_R } \over n! } \sum_{\alpha, \beta, \vec{a}}
		S^{ R~  R ~ \L ,\;  \tau }_{ ~ i ~ j ~ m } 
		D^{R}_{ij} ( \alpha ) D^\Lambda_{m m'}(\beta )
		C^{ \vec a }_{ \Lambda , M_{\Lambda } , m' } 
		\bp_{\beta, \alpha} \tr ( \mbX_{ \vec a } ~ \alpha )		
\end{split} 
\end{equation} 
Now re-express the trace in terms of the Fourier basis, 
use orthogonality of the $U(M)\times S_n $  Clebsch, 
the fusion property of the $ S_n $ Clebsch :   
\begin{equation} \label{eq:cp_derriv2}
\begin{split}
	{\rm symm}  & \left[
		\cO_{ \Lambda , M_{\Lambda} ,  R , \tau  }
                	\right] 
  = 	{ \sqrt {d_R } \over n! } \sum_{\alpha, \beta, \vec{a}} 
		S^{ R~  R ~ \L ,\;  \tau }_{ ~ i ~ j ~ m } 
		D^{R}_{ij} ( \alpha ) D^\Lambda_{m m'}(\beta )
		C^{ \vec a }_{ \Lambda , M_{\Lambda } , m' } 
		\bp_{\beta, \alpha} 
\\ & \hskip2.5cm 
	\times \sum_{\L_1  , \tilde M_{ \L } ,  R_1 , \tau_1}
	 \sqrt { d_{R_1}  }   C_{ \vec a }^{ \L_1 , \tilde M_{ \L_1 } , m_1  } 
  D^{R_1}_{a b  } ( \alpha  ) S^{ ~ R_1 ~ R_1 ~ \L_1 ; \tau_1 }_{ ~~a   ~~ b   ~~ m_1 }
  \cO_{ \L_1  , \tilde M_{ \L } ,  R_1 , \tau_1  } 
 \\ 
& = \sum_{R_1,\tau_1}	{ \sqrt {d_R d_{R_1} } \over n! } \sum_{\alpha, \beta, S , \tau_S } 
		S^{ R~  R ~ \L ,\;  \tau }_{ ~ i ~ j ~ m } 
		D^{R}_{ij} ( \alpha ) D^\Lambda_{m m_1 }(\beta )
 D^{R_1 }_{ a b   } ( \alpha  )	\bp_{\beta, \alpha}  S^{ ~ R_1 ~ R_1 ~ \L ; \tau_1 }_{ ~~a   ~~ b  
 ~~ m_1 }  \cO_{ \L , M_{ \L } ,  R_1 , \tau_1  }\\ 
&= \sum_{R_1,\tau_1} { \sqrt {d_R d_{R_1} } \over n! }\sum_{\alpha, \beta , S , \tau_S } 
	d_S 
	S^{ R~  R ~ \L ,\;  \tau }_{ ~ i ~ j ~ m } 
	S^{ R~ S ~ R_1 ,\, \tau_S }_{ ~ i ~ k ~ a } 
		S^{ R~ S ~ R_1 ,\, \tau_S }_{ ~ j ~ l ~ b } D^{S}_{kl} ( \alpha ) 
                 D^\Lambda_{m m_1 }(\beta ) \bp_{\beta, \alpha}	
\\ & \hskip2.5cm                  
      \times  S^{ ~ R_1 ~ R_1 ~ \L ; \tau_1 }_{ ~~a   ~~ b  
 ~~ m_1 } \cO_{ \L  ,
  M_{ \L } ,  R_1 , \tau_1  }
 \\
& = 	\sum_{S, \tau_S, R_1, \tau_1}
		\frac{\sqrt {d_R d_{R_1} } d_S }{n!}
		\bp^{S ~ k ~\, l}_{\Lambda \, m m_1 }
		S^{ R~  R ~ \L ,\;  \tau }_{ ~ i ~ j ~ m } 
		S^{ R~ S ~ R_1 ,\, \tau_S }_{ ~ i ~ k ~ a } 
		S^{ R~ S ~ R_1 ,\, \tau_S }_{ ~ j ~ l ~ b } 		
		S^{ R_1 ~ R_1 ~ \Lambda ,\, \tau_1 }_{ ~ a ~ b ~ m_1 } 	
		\cO_{ \Lambda , M_{\Lambda} ,  R_1 , \tau_1  }	.
\end{split} 
\end{equation}
In the final line we recognized the sum over $ \alpha , \beta $ as a 
Fourier transform of the $\bp_{ \beta , \alpha} $

This allows us to define a matrix $\cP$:
\begin{equation}\begin{split}
\label{eq:cp_definition}
&(\cP)^{\Lambda_2 , M_{\L_2}, R_2 , \tau_2}_{\Lambda_1 , M_{\L_1}, R_1 , \tau_1 }
\\
&= \delta_{\L_1,\L_2}\delta_{M_{\L_1},M_{\L_2}}  \sum_{ S , \tau} \sum_{ m_{ \Lambda} , m'_{\Lambda } } 
 { d_S \sqrt{ d_{ R_1} d_{R_2} } \over n!  } 
S^{R_1 ~ S ~ R_2 ; ~ \tau }_{ i_1 ~ k_1 ~ j_1 } 
  \bp^{ S ~ k_1 ~ k_2 }_{ \Lambda ~ m_{ \Lambda } ~ m'_{ \Lambda } } 
 S^{ R_1 ~ S ~ R_2 ; ~ \tau }_{~  i_2 ~ k_2 ~ j_2 } 
 S^{ R_1 ~ R_1 ~ \Lambda ; ~ \tau_1 }_{ ~ i_1 ~ i_2 ~ m_{ \Lambda } } 
 S^{ R_2 ~ R_2 ~ \Lambda ; ~ \tau_2 }_{ j_1 ~ j_2 ~ m_{\Lambda}' }
\end{split}
\end{equation}
and write the symmetrization as
\begin{equation}
\label{eq:osymm_definition}
 \cO^{ \rm S }_{ \L , M_{ \L } , R , \tau } 
 \equiv 
 {\rm symm}  \left[
		\cO_{ \Lambda , M_{\Lambda} ,  R , \tau  }
	\right] 
 =  \sum_{R_1,\tau_1} (\cP)_{\Lambda, M_\L, R, \tau }^{\Lambda ,M_\L, R_1  , \tau_1 }
 \cO_{ \L , M_{ \L } , R_1 , \tau_1 } .
\end{equation}
We have thus been able to express the symmetrization of individual traces as the transformation of the free orthogonal basis by a matrix $\cP$. 

Let us make a remark about the notation. Since $\cP$ is diagonal in the $M_\L$ label and also independent of it, we will often use a shorthand matrix $(\cP)^{\L_2,R_2,\tau_2}_{\L_1,R_1,\tau_1}$ understood as $\cP$ without the $\delta_{M_{\L_1},M_{\L_2}}$ factor:
\begin{equation}
	(\cP)^{\Lambda_2 , M_{\L_2}, R_2 , \tau_2}_{\Lambda_1 , M_{\L_1}, R_1 , \tau_1 } \equiv \delta_{M_{\L_1},M_{\L_2}} (\cP)^{\L_2,R_2,\tau_2}_{\L_1,R_1,\tau_1} .
\end{equation}
On the other hand, even though $\cP$ is diagonal $\L$, it does depend on it, so we will keep the $\L$ label. The same applies to the matrices $\cF$ and $\cG$. Note also that whenever a repeated $\L$ or $M_\L$ index appears, the summation convention \emph{does not} apply.

Evaluation of (\ref{eq:obps_definition}) is now straightforward. Using (\ref{eq:osymm_definition}) the BPS operators can be written as
\begin{equation} 	
\label{eq:obps_result}
	\cO^{\rm BPS }_{ \Lambda , M_{\Lambda} ,  R , \tau  }   = 
\sum_{R_1,\tau_1} ( \cG \cP )_{\L , R , \tau}^{\L ,  R_1 , \tau_1} \cO_{ \L , M_{ \L } , R_1 , \tau_1 }
\end{equation}
or more explicitly 
\begin{equation}
\label{eq:obps_result_expl}
	\cO^{\rm BPS }_{ \Lambda , M_{\Lambda} ,  R , \tau  } 
=  	\sum_{R_1,\tau_1}
	(\cP)^{\L , R_1 , \tau_1}_{\L , R , \tau} 
	\frac{N^n d_{R_1}}{n! \Dim R_1}
 	\cO_{ \L , M_{ \L } , R_1 , \tau_1 }
\end{equation}

The equation (\ref{eq:obps_result}) is a  key equation
for BPS operators  and we will devote the rest of
 the paper to analyzing these BPS operators in detail.
We have written an expression for the  space of  operators annihilated by the 
one-loop dilatation operator
\begin{equation}
	\cH_2 \cO^{\rm BPS}_{ \Lambda , M_{\Lambda} ,  R , \tau  } = 0
\end{equation}
in terms of known group theory quantities.
 They are nicely packaged in $\cP$ (\ref{eq:cp_definition})
 involving Clebsch-Gordan coefficients and $\bp_{\beta,\alpha}$, 
and in the Fourier basis of operators (\ref{eq:ob_definition})
for the free theory.
To be more precise, the above space of operators  provides 
an  overcomplete basis.     The labels of
 $ \{ \Lambda , M_{\Lambda} ,  R , \tau\}$ 
uniquely label the free basis, but we know that the kernel
 of $\cH_2$ has lower dimensionality. A linearly independent basis 
can be found by analysing the inner product  on the the BPS 
operators. We will give an expression for the inner product in 
Section~\ref{sec:2pt} and discuss its eigenvalues and eigenvectors 
in the context of the dual space-time physics in Section~\ref{sec:eigs2point}.   

The expressions (\ref{eq:obps_result}), (\ref{eq:obps_result_expl}) should be understood as valid for $ n \le N $. They already contains hints 
on  the problems we might encounter at  $n > N$. 
The $\Dim R$ factor in the denominator will be 0 whenever $c_1(R)>N$ making the basis ill-defined.
This is related to the fact that $\Omega^{-1}$ appearing in (\ref{eq:g_on_tr}) is not well defined as $n > N$.
We will solve the finite $N$ problem in  Section~\ref{sec:finiteN}. 


\subsection{Properties of $\cP$ }
\label{sec:propertiescP_sec4}

The matrix $\cP$ will be a crucial object throughout the rest of the paper and so we pause here to note a few of its properties. In addition to this section we provide further analysis of $\cP$ in Appendix~\ref{sec:propertiescP}.
\noindent

\begin{itemize} 

\item 
 The operator $ \cP$  can be seen essentially as the operator $\bp_{\beta, \alpha}$ transformed appropriately into the basis of free operators labeled by $|  \L, M_\L, R , \tau \rangle$. As is apparent from the definition (\ref{eq:cp_definition}), the transformation is done by applying four Clebsch-Gordan coefficients. We can also visualize this transformation by expressing it diagrammatically (see Figure~\ref{fig:Popdef}).
\begin{figure}
\begin{center}
\input{DefP.pstex_t}
\caption{Definition of $ \cP$  } \label{fig:Popdef}
\end{center}
\end{figure}
We can also write  the matrix $\cP$ as a sum over conjugacy classes 
$ [ \alpha ] $ in $ \mbC ( S_n ) $, which replaces the sum over $R$ 
in  (\ref{eq:osymm_definition}). See Appendix~\ref{sec:propertiescP}. \\

\item 
 The matrix $ (\cP)^{\L_1 , M_{\L_1} , R_1 , \tau_1}_{\L_2 , M'_{\L_2} , R_2 , \tau_2}$ is 
proportional to $ \delta_{ \L_1, \L_2} \delta_{ M_{ \L_1} , M'_{ \L_2} } $ 
and is symmetric under exchange of $ R_1 , \tau_1 $ 
with $ R_2 , \tau_2 $.
\begin{equation} 
(\cP)^{\L , R_1 , \tau_1}_{\L , R_2 , \tau_2}
=
(\cP)^{\L , R_2 , \tau_2}_{\L , R_1 , \tau_1}
\end{equation} 
This means that it has the hermiticity 
property 
$$ \cP^{\dagger} = \cP $$
under any  inner product where the $ ( \L, M_\L, R , \tau )   $
label an orthonormal basis. 
The free field inner product does not have this property, 
except in the planar limit  ($ n \ll N$) where it simplifies to 
(\ref{eq:2ptplanar_def}).  
In  the finite $N$ construction of 
 Section~\ref{sec:finiteN} we will define an inner product 
which has this property for any $n$.

\item 
Since $ \cP $ is constructed as the operator on the free field Hilbert space 
which projects to symmetrized traces, we expect it satisfies the projector 
equation 
$$ \cP^2 = \cP $$ 
We prove this equation directly  in the Appendix~\ref{sec:propertiescP}.

\item
The hermiticity of $\cP$ in the planar inner product
 gives an elegant way to see that the symmetrised traces are orthogonal to 
the descendants in the planar limit - a property we first proved in 
Section~\ref{sec:review}.  States in $ \Ker ( \cP ) $ 
are descendants. States in $ \Im ( \cP ) $ are symmetrised traces. 
The hermiticity of the projector $ \cP$ 
ensures that  $ \Im (\cP) $  is orthogonal to $ \Ker (\cP) $. Explicitly
\bea 
0 = \langle \cP  \cO^{ \rm D } | \cO^{ \rm S }  \rangle_{\planar}  =
  \langle  \cO^{ \rm D }  | \cP \cO^{ \rm S }  \rangle_{\planar}  =
  \langle  \cO^{ \rm D }  |  \cO^{ \rm S }  \rangle_{ \planar } 
\eea

\item
For each $ \Lambda , M_{ \Lambda } $, the states $ \cO_{ \Lambda , M_{ \Lambda } , R , \tau } $ span a space of dimension 
\begin{equation} 
\sum_{R} C ( R , R , \Lambda ) 
\end{equation} 
where $ C ( R , R , \Lambda )  $ is the multiplicity of 
the one-dimensional irrep. in the $S_n$ tensor product 
$ R \times R \times \Lambda $. Note the symmetry under 
transposition of the Young diagram 
\begin{equation}\begin{split} 
C ( R , R , \Lambda ) &  = 
{ 1 \over n ! } \sum_{ \sigma } \chi_R ( \sigma )
 \chi_R ( \sigma ) \chi_{ \L } ( \sigma ) \\
& = { 1 \over n ! } \sum_{ \sigma }  (-1)^{ \sigma } \chi_{R^T}  ( \sigma )
(-1)^{ \sigma }  \chi_{R^T}  ( \sigma ) \chi_{ \L } ( \sigma ) \\
& = C ( R^T  , R^T  , \Lambda ) 
\end{split}\end{equation} 
We expect that, given a choice of orthogonal basis 
labelled by $ \tau $ in $ {\rm Inv} ( R \otimes R \otimes \L ) $,
 there is a corresponding choice 
 $ \tau^T $ in $ {\rm Inv} ( R^T \otimes R^T \otimes \L ) $, such that 
\begin{equation}\begin{split} 
(\cP)^{\L ,R_1 , \tau_1}_{\L , R_2 , \tau_2} =
\pm (\cP)^{\L , R_1^T , \tau_1^T}_{\L , R_2^T , \tau_2^T}
\end{split}\end{equation} 
This will be demonstrated in examples by the explicit construction 
in Appendix~\ref{sec:app_example}, using bases
in $ \tau$ space  constructed as eigenstates of 
generalized Casimirs defined in \cite{kimram3}.

\end{itemize} 

\subsubsection{Counting Check } 

The counting of states in a given irrep. $ \Lambda $ 
of $U(M)$ is given, using (\ref{eq:p_definition}) by 
\begin{equation} 
\cM_{ \Lambda } = \sum_{ p \vdash n } 
 \chi_{ \Lambda } ( \mP_{ p } ) 
= \sum_{ \alpha \in S_n } 
 \chi_{ \Lambda } ( \mP_{ \alpha } ) { |\Sym(\alpha)| \over n!  } 
\end{equation}   
It is useful to write this as 
\begin{equation} 
\label{eq:ml_ab_sum}
 \cM_{ \Lambda } = \sum_{ \alpha, \beta \in S_n } 
 \chi_{ \Lambda } ( \mP_{ \alpha } ) { 1  \over n!  } 
 \delta  (  \beta \alpha \beta^{-1} \alpha^{-1} )   
\end{equation}   

On the other hand $(\cP)^{\L , M_{\L} , R_2 , \tau_2}_{\L , M_{\L} , R_1, \tau_1}$ at fixed $\L,M_\L$ is a projector matrix to the space of symmetrized traces. The dimension of this space should be the large-$N$ number of BPS operators $\cM_\L$. In general, the dimension of the projection must be given by the trace of the projector matrix. We can thus perform a consistency check by calculating the trace of $\cP$:
\begin{equation}\begin{split} 
\sum_{ R , \tau } 
(\cP)^{\L , M_{\L} , R , \tau}_{\L , M_{\L} , R , \tau}  
& = \sum_{ \tau , \tau_1 , S } \bp_{ \beta , \alpha } 
 D^{ \Lambda }_{ i_3 j_3 } ( \beta ) D^{S}_{ k_1 k_2 } ( \alpha ) 
 { d_R d_S \over n! } S^{ R ~ R ~ S , \tau }_{~  i_2 ~ j_2 ~ k_2 } 
 S^{ R ~ R ~ S , \tau  }_{~  i_1 ~ j_1 ~ k_1 }  
 S^{ R ~ R ~ \Lambda , \tau_1 }_{ i_1 ~ i_2 ~ i_3 } 
 S^{ R ~ R  ~ \Lambda , \tau_1 }_{ j_1 ~ j_2 ~ j_3 } \\
& =   \sum_{ \tau , \tau_1 , S } \sum_{\sigma_1,\sigma_2} \bp_{ \beta , \alpha } 
 D^{ \Lambda }_{ i_3 j_3 } ( \beta ) D^{S}_{ k_1 k_2 } ( \alpha ) 
 { d_R d_S \over n! } { 1 \over n! } D^R_{i_2 i_1} ( \s_1 ) 
 D^R_{j_2 j_1} ( \s_1 ) D^S_{k_2 k_1} ( \s_1 )\\ 
& \qquad \qquad \qquad  { 1 \over n! }  D^R_{i_1 j_1} ( \s_2 ) 
 D^R_{i_2 j_2} ( \s_2 ) D^{ \L } _{i_3 j_3 } ( \s_2 )\\ 
& = \sum_{\sigma_1,\sigma_2\in S_n} { 1 \over n! } \chi_{ \L } ( \mP_{ \s_1} \s_2 ) 
 \delta ( \s_1 \s_2 \s_1^{-1} \s_2^{-1} ) 
\end{split}\end{equation} 
Now given the form of $  \mP_{ \s_1} $ we know that if $ \s_2 $ commutes with 
$\s_1 $ then $ \tr (\mP_{ \s_1} \s_2) = \tr (\mP_{\s_1}) $. 
This is then equal to $\cM_\L$ according to (\ref{eq:ml_ab_sum}) as required. 

\subsection{The two-point function for BPS operators } 
\label{sec:2pt} 

Given the expression of BPS operators as (\ref{eq:obps_result}) we may easily evaluate the two-point function matrix. 
\begin{equation} 
\begin{split} 
 \langle \cO^{\BPS}_{ \L_1 , M_{\L_1 } ,  R_1 , \tau_1 } |
  \cO^{\BPS}_{ \L_2  , M^{\prime}_{\L_2} , R_2  ,  \tau_2 } \rangle &=
 \langle \cG \cP \cO_{ \L_1 , M_{\L_1 } ,  R_1 , \tau_1 } |
  \cG \cP \cO_{ \L_2  , M^{\prime}_{\L_2} , R_2  ,  \tau_2 } \rangle \\
& =   \langle \cO_{ \L_1 , M_{\L_1 } ,  R_1 , \tau_1 } |
 \cP \cG \cF \cG \cP  \cO_{ \L_2  , M^{\prime}_{\L_2} , R_2  ,  \tau_2 }
 \rangle_{\rm planar}       \\
& =   \langle \cO_{ \L_1 , M_{\L_1 } ,  R_1 , \tau_1 } |
 \cP \cG  \cP  \cO_{ \L_2  , M^{\prime}_{\L_2} , R_2  ,  \tau_2 }
 \rangle_{\rm planar}  \\
 & = \delta_{\L_1,\L_2} \delta_{M_{ \L_1},M'_{ \L_2}}
 ( \cP \cG \cP )^{\L_1, R_1 , \tau_1 }_{ \L_1, R_2 , \tau_2 } 
\end{split} 
\end{equation} 
We have used the fact that the two-point function of $ \cO_{ \Lambda , M_{ \Lambda } , R ,  \tau } $ is diagonal in the labels  and given by the matrix 
$ \cF $ in equation (\ref{2ptF}), which is an inverse of $\cG$.
This leads to  the formula
\begin{equation} 	
\label{eq:twopointresult}
\fbox{$ \displaystyle{
\langle \cO^{\rm BPS}_{ \L_1 , M_{ \L_1} ,  R_1 , \tau_1 } | 
 \cO^{\rm BPS}_{ \L_2 , M_{\L_2 } ,  R_2 , \tau_2 } \rangle 
= \delta_{\L_1,\L_2} \delta_{M_{ \L_1},M'_{ \L_2}}
 ( \cP \cG \cP )^{\L_1,  R_1 , \tau_1 }_{ \L_1, R_2 , \tau_2 } 
}$}
\end{equation}
for the two-point function on 
of BPS operators, valid at all orders in the  $1/N$ expansion. 
This is the main result of this paper.

Since the Clebsch-Gordan coefficients of the symmetric group 
can be chosen to be real \cite{hammermesh}, the matrix $ \cP \cG \cP $
 is a real symmetric matrix. We will develop a  manifestly 
finite $N$ construction in the next section to arrive at
 $ \cP_{I} \cG_N \cP_{I}$. In Section~\ref{sec:eigs2point} we
 will use analyticity arguments 
to show that the matrix $ \cP \cG \cP$ can be used to extract finite 
$N$ properties.

\section{Finite N and geometry in Hilbert spaces } 
\label{sec:finiteN}

%
%

The derivation of  the two-point function (\ref{eq:twopointresult})
is valid to all orders in the $1/N$
expansion but requires  $N\ge n$. Here we will describe 
a framework which allows the derivation of a formula 
for the 2-point function  at finite $N$,  
with no restriction on $n$. This treatment contains some general 
insights, notably the geometry of the interplay between different 
projectors in appropriate Hilbert spaces,  
 which should   be useful in thinking more generally 
about finite $N$ issues. e.g in the sixteenth BPS case 
in $ \cN =4 $ or in more general theories.

\subsection{Finite N Hilbert space as a quotient and BPS operators from intersections  } 
\label{sec:inters} 

Start with an infinite dimensional Hilbert space $ \cH$, graded by a
natural number $n$.   For each $n$, 
there are finitely many states.   
The states  are labeled by $ | \L , M_{ \L } , R , \tau \rangle  $ for 
each natural number $n$.  
 $ \L , R $  are Young diagrams with $n$ boxes,  $M_{\L} $ a state 
 in the $U(M)$ representation  corresponding to $ \Lambda $, 
 $ \tau $ runs over the multiplicity $ C ( R , R , \Lambda ) $ 
 which is the number of times the one-dimensional  representation of $S_n$ 
appears in the tensor product $  R \otimes R \otimes \L $. 
We can regard the states as made of gauge invariant polynomials in variables 
$ (\cX_1)^i_j , (\cX_2)^i_j  , (\cX_3)^i_j  $ where $ i , j $ extend from $1$ to 
$\infty$. These are traces e.g $ \tr ( \cX_1 \cX_2 \cX_3 ) $.  
The relation between the traces and the group 
theoretic labels is given by the equation  
(\ref{comptransf}) involving the matrices $ \cS , \cT $, 
involved in the free-field diagonalization problems solved in \cite{BHR1,BHR2}, 
constructed from Clebsch-Gordan coefficients of $ U ( M) \times S_n $, 
matrix elements and Clebsch-Gordan coefficients of  $S_n$. 
They make no reference to the   $U(N)$ gauge group 
hence they make sense in the $N = \infty $ set-up we are describing here.  
We can define a simple inner product on $\cH$
\begin{equation} 
 \langle  \L^{\prime} , M'_{ \L'} , R' , \tau'  | \L , M_{ \L } , R , \tau 
\rangle_{S_\infty} 
= \delta_{ \L' , \L } \delta_{ M_{ \L}  , M'_{\L'} } 
 \delta_{ R , R' } \delta_{ \tau , \tau' },
\end{equation} 
which we will call the $ S_{\infty} $ inner product. Note that its form matches the planar limit inner product (\ref{eq:2ptplanar_def}), but there is a subtle difference in that $\langle.|.\rangle_{S_\infty}$ is defined on a manifestly $N = \infty$ Hilbert space $\cH$, while the planar product is defined as the 
limit of the finite large $N$ inner product for traces  
involving a total of $n$ matrices chosen from $ X_1, X_2, X_3$
when $ n \ll N$.

Finite  $N$ physics can be obtained by replacing the
formal variables $ (\cX_a)^i_j $ 
with matrix elements  $ (X_a)^i_j $ of $N \times N $ 
matrices $ X_a $.  
It was shown in \cite{dolan07,BHR1}   that the finite $N$ counting of 
gauge-invariant operators can be achieved simply by 
cutting off $ R $ with the condition that its first 
column does not exceed $N$ in length, i.e $c_1 (R) \le N$. 
It is useful therefore to introduce a projection operator 
\begin{equation}
\label{eq:IN_defined}
 \cI^{(N)}  | \L , M_{ \L } , R , \tau \rangle = 
 \begin{cases}
 	| \L , M_{ \L } , R , \tau \rangle  & \rm {if } ~~ c_1 ( R ) \le  N  \\ 
  	0 & \rm  {if }~~  c_1 ( R ) >     N
 \end{cases}
\end{equation}
The finite $N$ Hilbert space is 
isomorphic to $ \cH / \Ker ( \cI^{(N)} )  $. We find that 
a very useful way to think about finite $N$ physics is to 
consider projectors in $\cH$. Alongside $ \cI^{(N)} $ we will 
also be interested in $ \cP$, as defined by the equation 
(\ref{eq:cp_definition}). In the finite $N$ Hilbert space, we do not think 
of $ \cP $ as a linear operator defined by its action on traces, because 
the traces do not form a good basis, hence the emphasis on 
(\ref{eq:cp_definition}). 
 Understanding the interplay 
between $ \cP $ and  $\cI^{(N)} $ 
will be the key to understanding finite $N$ physics
of BPS operators. Since  $ \cP$ is real and symmetric in the $ \L , R , \tau $ 
basis,  it is a Hermitian operator with respect to 
the $ S_{\infty} $ metric. The operator $ \cI^{(N)} $ is also 
Hermitian with respect to the same metric. In the following 
 we will  describe the construction of 
a new  orthogonal projector $\cP_I$, where orthogonality is meant  
with respect to the $ S_{\infty } $ metric. 

The $ S_{\infty} $ metric induces a metric on the quotient 
$ \cH / \Ker ( \cI^{(N)} )   $, which is isomorphic to the finite $N$ 
Hilbert space $ \cH^{(N)} $  of finite $N$ matrices. On this 
finite $N$ quotient it is 
\begin{equation} 
 \langle \L_1 , M_{\L_1 } , R_1 , \tau_1 | \cI^{(N) } |  
     \L_2 , M_{\L_2 } , R_2 , \tau_2 \rangle_{ S_{\infty} } 
=  \delta_{ \L_1 \L_2 } \delta_{ M_{\L_1}  , M_{\L_2}  } 
   \delta_{ R_1 R_2 } \delta_{ \tau_1 \tau_2 } ( \cI^{(N) } )_{R_1 R_2}
\end{equation}
This metric can be evaluated on the trace basis using (\ref{finiteNI}) to be
\begin{equation}\begin{split} 
\langle \cO_{\vec b , \beta } | \cI^{(N)} | \cO_{ \vec a , \alpha } \rangle_{ S_{\infty} }  
& = \sum_{ \gamma  } { 1 \over n! } \delta_{ \gamma ( \vec a )  , \vec b  } 
\delta_N ( \beta^{-1} \gamma^{-1} \alpha \gamma )
\\ 
& = \sum_{ \gamma }  \delta_{ \gamma ( \vec a )  , \vec b  } 
   \sum_{ R : c_1 ( R ) \le N } d_{ R } \chi_R (  \beta^{-1} \gamma^{-1} \alpha \gamma)
\end{split}\end{equation} 
For operators with $n \ll  N$, this is just the 
delta function expressing large $N$ factorization. 
Operators only overlap if they are a product of like 
traces with like field content.

Whether at large $N$ or finite $N$, the descendants are in $ \Ker ( \cP  ) $. 
Whenever a single trace contains commutators, it is a descendant 
and it is annihilated by the symmetrization operation $ \cP$. 
 For example in the case of $ \L = [ 2,2] $ 
in the Appendix~\ref{sec:app_example}, we have the expression 
$ { 1\over 2 } \cO_{[3,1]} - { 1\over \sqrt{2} } \cO_{ [ 2,2]} +
 { 1\over 2} \cO_{[2,1^2]} $ for the descendant in the Fourier 
basis. For finite  $N$ matrices,
 the last term is identically zero, so we may equally 
write  $ { 1\over 2 } \cO_{[3,1]} - { 1\over \sqrt{2} } \cO_{ [ 2,2]}$. 
Equivalently, without ever talking about finite $N$ matrices, 
but only about projectors in $ \cH$, we obtain the 
same expression by considering the image of  $ \cI^{(N)}  ( 1-  \cP )  $
in $ \cH $ which contains the linear combination 
 $ { 1\over 2 } \cO_{[3,1]} - { 1\over \sqrt{2} } \cO_{ [ 2,2]}$.
We look for a hermitian (under the $S_{\infty} $ metric) 
 projection  operator $ \cP_{ I } $ which will annihilate the 
descendants, so we have $ \cP_{ I } \cI^{(N)}  ( 1-  \cP ) = 0 $. 
We also would like this projector to map  states in $ \cH^{(N)}  $
to $ \cH^{(N)}$, so that $ \cI^{ (N)  } \cP_I  = \cP_I  \cI^{(N)} = \cP_I   $. 
Combining these we have  $ \cP \cP_{ I } = \cP_I \cP = \cP_I $.
 Together, we see that 
these equations characterise a projector for the intersection 
\bea 
 \Im ( \cP ) \cap \Im ( \cI^{(N)} ) = \Im ( P_I )
\eea

An equivalent way to arrive at this is to consider  the orthogonal complement 
in $ \cH $ to $ \Ker (I^{(N)}) \cup \Ker (\cP)  $ in the 
$ S_{\infty} $ metric.  The dimension of this  
complement is equal to the number of independent 
BPS operators, since imposing the  finite $N$  condition 
in $\cH $ sets to zero states in $  \Ker (I^{(N)}) $ 
and setting descendants to zero imposes the vanishing 
of states in  $\Ker (\cP)  $. 
This orthogonal complement is 
$ \Im(I^{(N)}) \cap \Im(\cP) $ so we have the orthogonal 
decomposition 
\begin{equation} 
\cH =  \left( \Ker (I^{(N)}) \cup \Ker (\cP) \right) \oplus \left(  \Im(\cI^{(N)}) \cap \Im(\cP) \right).
\end{equation} 

A natural question arises how, given two projectors $\cI^{(N)}$ and $\cP$, one
explicitly  constructs a hermitian projector for their intersection space
\begin{equation}
\label{eq:cpi_definition}
	\cP_I \equiv \cP_{ \Im(\cI^{(N)}) \cap \Im(\cP) }.
\end{equation}
Let us consider the null space of $\cI^{(N)}(1-\cP)\cI^{(N)}$. It will be composed of two orthogonal subspaces: the null space of $\cI^{(N)}$ itself and the null space of $(1-\cP)$ inside of $\Im(\cI^{(N)})$:
\begin{equation}\begin{split}
\label{eq:kerx_decomposition}
	\Ker(\cI^{(N)}(1-\cP)\cI^{(N)}) &= \Ker(\cI^{(N)}) \oplus \left( \Im(\cI^{(N)}) \cap \Ker(1-\cP)  \right)
\\
	&= \Ker(\cI^{(N)}) \oplus \left( \Im(\cI^{(N)}) \cap \Im(\cP)  \right).
\end{split}\end{equation}
This means if we had a projector for $\Ker(\cI^{(N)}(1-\cP)\cI^{(N)})$ 
then we would just multiply by $\cI^{(N)}$ on each side to extract the second component
\begin{equation}
\label{eq:pi_from_ipkeri}
	\cP_I = \cI^{(N)} \cP_{\Ker(\cI^{(N)}(1-\cP)\cI^{(N)})} \cI^{(N)}.
\end{equation}
The question of constructing $\cP_I$ is then reduced to the question of, given an arbitrary matrix $X$, constructing the projector for its null space $\cP_{\Ker(X)}$. It can be written as follows
\begin{equation}
	\cP_{\Ker(X)} = \frac{1}{\prod_{i=1}^{k} (-\lambda_i) } \prod_{i=1}^{k} (X - \lambda_i),
\end{equation}
where $i$ runs over the \emph{non-zero eigenvalues} of $X$ and $k={\rm Rank}(X)$. This, in turn, can be rewritten in terms of Schur polynomials
\begin{equation}
	\cP_{\Ker(X)} = \frac{1}{\chi_{[1^k]}(X)} \sum_{j=0}^{k} (-1)^{k-j} X^{k-j} \chi_{[1^j]}(X).
\end{equation}
The characters $ \chi_{[1^j]}(X) $ are the Schur Polynomials for 
Young diagrams with $j$ rows of length $1$ or  one column of length $j$.  
This manipulation then allows us to write the intersection projector
 $\cP_I$ fairly  explicitly in terms of matrix multiplications as
\begin{equation}\begin{split}
	\cP_I &= \frac{1}{\chi_{[1^k]}(X)} \sum_{j=0}^{k} (-1)^{k-j} \left(\cI^{(N)} X^{k-j} \cI^{(N)} \right) \chi_{[1^j]}(X) ,
\\
	X &\equiv \cI^{(N)}(1-\cP)\cI^{(N)} ,
\\
	k &= {\rm Rank}(X) .
\end{split}\end{equation}
Also note that using the decomposition (\ref{eq:kerx_decomposition}) we can express the number of operators $\cM_{N;\L}$ as
\begin{equation}\begin{split}
	\cM_{N;\L} &= | \Im(\cI^{(N)}) \cap \Im(\cP) |
\\
	&= 	| \Ker(\cI^{(N)}(1-\cP)\cI^{(N)}) | - | \Ker(\cI^{(N)}) |
\\	
	&= {\rm Rank}(\cI^{(N)}) - {\rm Rank}(\cI^{(N)}(1-\cP)\cI^{(N)}).
\end{split}\end{equation}

\subsection{Finite $N$  construction of BPS operators } 

We will prove that the BPS operators at finite $N$ 
are given by the image of $  \cG P_{ I }$ as an operator in $ \cH$.

We first restate the construction of the BPS 
operators at large $N$ in language that will generalize 
at finite $N$. The large $N$ construction starts with two key observations. 
The relation between the planar inner product $\langle . | . \rangle_{S_\infty} $  and the 
$1/N$ corrected inner product $\langle . | . \rangle$  is given by (see (\ref{2pttoplanar})):
\begin{equation} 
  \langle \cO_{1} | \cO_2 \rangle = \langle \cO_{1} | \cF  \cO_{2} \rangle_{S_\infty}
\end{equation} 
The second observation is that descendant operators 
contain commutators $ [ X_a , X_b ] $ 
inside a trace. These operators are annihilated by
the operator  $ \cP $ we constructed. From these facts and the Hermiticity of $\cP$ with respect to $S_\infty$ we easily see that the operators in the image of $\cG\cP$ are orthogonal to descendants, and thus BPS:
\begin{equation}\begin{split} 
&
\langle  \cO^{ \des } |   \cG \cP   \cO \rangle =
\langle \cO^{\des } | \cF \cG  \cP  \cO  \rangle_{S_\infty} 
\\ 
&     =   \langle \cO^{\des } |  \cP \cO  \rangle_{S_\infty}  = 0  
    =  \langle \cP \cO^{\des } |  \cO  \rangle_{S_\infty}  = 0
\end{split}\end{equation}

To develop a finite version of this argument, we need to 
carefully consider the  definition of  $\cG $
at finite $N$. 
Define $ \cG_N \equiv \cI^{ (N) } \cG \cI^{ (N) }$ 
which is manifestly zero for $ c_1 (R) > N $ and well defined 
at finite $N$.  Acting on a Hilbert space $ \cH $ 
with states  $ | { \L , M_{ \L } , R , \tau }  \rangle $ 
it is  non-zero only on the operators obeying 
$ c_1 ( R ) \le N $, i.e where the Young diagram $R$ 
has first column of length  smaller than $N$. The second ingredient is the projector $\cP_I$ defined in (\ref{2pttoplanar}).

Note the following useful properties of $ \cG , \cF , \cI^{(N)} , \cP, \cP_I$ 
\begin{equation}\begin{split} 
& \cG_N \cF = \cF \cG_N = \cI^{(N)}
\\ 
& \cP_I \cI^{(N) } = \cI^{(N) } \cP_I  =  \cP_{I}
\\
& \cP_{I} \cP = \cP \cP_{I} = \cP_{I}
\\ 
& \cF \cG \cP_{I} = \cF \cG \cI_{N} \cP_{I} = \cP_{I} 
\end{split}\end{equation} 
Consider operators $  \cG_N \cP_{I}  \cO $. These operators 
are manifestly well-defined at finite $N$ and have the following 
2-point function with descendants
\begin{equation}\begin{split} 
\langle \cO^{\des} |  \cG_N  \cP_{I}    \cO \rangle
& = \langle \cO^{\des} | \cF   \cG_N  \cP_{I}    \cO \rangle_{S_{\infty} }
\\ 
& = \langle \cO^{\des} |  \cI^{(N)}  \cP_{I}    \cO \rangle_{ S_{\infty} }
\\ 
& = \langle \cO^{\des} |  \cP_{I } \cO \rangle_{S_{\infty} }
\\ 
& =  \langle  \cO^{\des} | \cP \cP_{ I } \cO  \rangle_{S_{\infty}}  
\\
&= 
  \langle  \cP \cO^{\des} | \cP_{I} \cO  \rangle_{S_{\infty}}  
\\ &= 0 
\end{split}\end{equation}  
In the first line we have used the fact that the 
metric of the free theory is related to the 
$S_{\infty} $ metric by the matrix $ \cF$. 
In the fourth line, we have used the 
fact that $ \cP $ is Hermitian with respect to 
the $ S_{\infty} $ metric, and annihilates the 
descendants. 
This shows that the $ \cG_N \cP_{I} \cO $ construct BPS operators. 
\begin{equation}\label{finiteNBPSconst} 
\cO^{\rm BPS} =   \cP_{I} \cG_N  \cO 
\end{equation}
The number of these operators is $ | \Span ( \cP_I ) | $, 
which as explained in section \ref{sec:inters},  correctly matches the finite $N$ counting. 

The two point function on BPS operators is the 
matrix $ \cP_I \cG_N \cP_I $:
\begin{equation} \label{finiteN2pt} 
\begin{split} 
& \langle \cG_N  \cP_{I}    \cO_{ \L_2 , M'_{ \L_2 } , R_2 , \tau_2  } |   \cG_N  \cP_{I}    \cO_{ \L_1 , M_{ \L_1 } , R_1  , \tau_1 } \rangle \\  
& = \langle \cO_{ \L_2 , M'_{ \L_2 } , R_2 , \tau_2  } | \cP_{ I } \cG_N \cF \cG_N \cP_{I} ~ \cO_{ \L_1 , M_{ \L_1 } , R_1  , \tau_1 } \rangle_{S_{\infty} } \\
& = ( \cP_I  \cG_N I^{(N)} \cP_{I})_{ \L_1 , M_{ \L_1 } , R_1  , \tau_1 }^{ \L_2 , M_{ \L_2 } , R_2 , \tau_2}  \\ 
& =  ( \cP_I \cG_N  \cP_{I})_{ \L_1 , M_{ \L_1 } , R_1  , \tau_1 }^{ \L_2 , M_{ \L_2 } , R_2 , \tau_2 } \\
& = \delta_{ \L_1 , \L_2 } \delta_{ M_{ \L_1 } , M'_{\L_2} } 
 ( \cP_I \cG_N  \cP_{I})^{ \L_1 , R_2  , \tau_2 }_{ \L_1 , R_1 , \tau_1 } 
\end{split} 
\end{equation} 

It is interesting to note that the geometry of intersections 
 $ \Im ( \cI^{(N)} ) \cap \Im ( \cP )  $ has played a key role. 
The projector $ \cP $ was constructed from the universal element 
$ \mP $ which gave the large $N$ counting. Although it is easy to give 
 indirect arguments, based on points already discussed, 
 that the above finite $N$ construction with $ \cP_I $ 
agrees with the finite $N$ counting from $ \mP^{(N)} $, there was 
no direct finite $N $ construction following directly from  $ \mP^{(N)} $, 
 the way the large $N$ construction based on $ \cP $  followed from $ \mP$.  
This was not apriori obvious and is intriguing.

\section{The matrix $ \cP_I  \cG \cP_I  $ and space-time physics } 
\label{sec:eigs2point} 

According to non-renormalization theorems \cite{HH01,HHHR03,heslopchar}, we expect that 
the two-point functions of BPS states,  constructed 
in this paper as the matrix $ \cP_I \cG \cP_I  $ 
acting in the free Hilbert space,  will be unchanged as the coupling 
$g_{YM}^2$ is increased from weak to strong coupling.
We will attempt to extract aspects of the 2-point functions 
at one-loop which are most likely to have a strong coupling 
interpretation.

The set of eigenvalues, and their multiplicities, 
being independent of  a choice of basis are of particular interest. 
In the regime $ n < N $, the matrix  $ \cP_I \cG \cP_I  $
simplifies to $ \cP \cG \cP $. We will also argue that, by 
analytic continuation, that information about  $ \cP_I \cG \cP_I  $
can be extracted from $ \cP \cG \cP $. In particular, 
there is a simple way to get the characteristic equation
of the former from the latter. The behaviour of 
the spectrum for $ n $ near $N$ is specially interesting
in view of the stringy exclusion principle (SEP)  \cite{malstrom,jevram} 
and the related 
property of the growth in size of giant gravitons with angular momentum
\cite{mst}. The matching of these spacetime phenomena with 
gauge theory operators is well-understood in the half-BPS 
case using a Young diagram classification of 
the operators \cite{cjr}. We will use the physics of the SEP 
to make a start towards the identification of gauge 
theory operators corresponding to BPS operators beyond the half-BPS case. 
We will then turn to a discussion of how to find a 
labelling analogous to the Young diagrams of the half-BPS case
in the case of quarter and eighth-BPS operators at hand.

\subsection{Eigenvalues of $ \cP \cG \cP $ and $ \cP_I \cG  \cP_I $ }
\label{PIandP} 

 The matrix $\cP\cG\cP$ is derived as the matrix of two-point 
functions of BPS operators
 calculated in the regime where $N \ge n$. 
 As described in Section~\ref{sec:finiteN}, when $N<n$ the two-point function of the well-defined BPS operators is instead given by a modified matrix $\cP_I \cG_N \cP_I$. However, the eigenvalues of $\cP\cG\cP$,
 even though calculated in the large $N$ regime,
 are explicit functions of $N$. Therefore, one may ask, what happens to the eigenvalues as we continuously decrease $N$ below $n$. For an example of the functions see (\ref{eq:l22_pgp_evalues}) in Appendix~\ref{sec:app_example}.

First, let us review the half-BPS case $\Lambda=[n]$. 
Since $\cP=\mbI$, the BPS operators are just rescaled free operators
\begin{equation}
	\cO^{\rm BPS}_{\Lambda=[n];R} = \frac{N^n d_R}{n! \Dim R}\cO_{\Lambda=[n];R}
\end{equation}
and the two-point function 
\begin{equation}
	(\cP\cG\cP)^{\L=[n];R_1}_{R_2}  = \frac{ N^n d_{R_1}  \delta_{R_1,R_2}}{n! \Dim R_1}.
\end{equation}
The eigenvectors coincide then with the  orthogonal half-BPS basis
labelled by  Young diagram $R$ \cite{cjr}. The eigenvalues  
$ \lambda_R \equiv \frac{N^n d_R}{n! \Dim R}$
 are equal to $N^n$ divided by a polynomial 
$ \prod_{i,j } ( N- i + j ) $, where the product runs over boxes in the 
Young diagram $R$, with $i$ labelling the row 
 and $j$ labelling the column. The eigenvalue $ \lambda_R $ 
diverges when the first column has length $c_1 ( R ) $ 
in the region  $ c_1 ( R ) > N $. The eigenvalue of a BPS 
state at large $N$, when analytically continued to a regime where that BPS 
state is not part of the finite $N$ Hilbert space, \emph{diverges}. 
In the  original half-BPS construction or in the free orthogonal
 basis the states were normalized  so that the eigenvalues 
continued  to \emph{zero}. In our construction of 
the general  quarter and bosonic eighth-BPS 
states,  the divergent normalization is more natural.

Interestingly, we find the same behavior to be general for any $\Lambda$.
That is, if we calculate eigenvalues  at large $N$ and 
then take the limit for any finite $N$, 
the divergent eigenvalues will correspond to the states 
that drop out from the Hilbert space. 
On the other hand, each finite (but non-zero) eigenvalue 
corresponds to a state that remains BPS. Equivalently, 
we can  read off the finite $N$ characteristic 
equation by factoring out a common denominator and dropping terms 
in the numerator which vanish at finite $N$ (see the discussion around 
(\ref{eq:l32_pgp_poly})). In fact, the 
surviving finite eigenvalues of $\cP\cG\cP$ 
with the corresponding eigenvectors precisely 
match those of $\cP_I \cG_N \cP_I$, calculated 
in the manifestly finite $N$ framework. In the discussion here we are not interested in zero eigenvalues of $\cP\cG\cP$, because they just span $\Ker(\cP)$ and do not depend on $N$. Also, since $\Ker(\cP)$ is contained in $\Ker(\cP_I)$, the zero eigenvalues of $\cP\cG\cP$ will always correspond to zero eigenvalues of $\cP_I\cG_N\cP_I$.

This correspondence between eigenstates
 of $\cP \cG \cP$ and $\cP_I \cG_N \cP_I$ can be explained as follows. 
Consider the matrix $\cP\cG\cP$ at a value $N_\epsilon=N+\epsilon$ approaching some finite $N$ in the regime $ N < n $. 
The matrix $\cG$ will contain  finite and divergent parts:
\begin{equation}
	\cG = \cG_N + O(\epsilon^{-1}).
\end{equation}
The finite directions are those in $\Im(\cI^{(N)})$ and the divergent
 ones are in $\Im(1-\cI^{(N)})$,where the projector 
 $\cI^{(N)}$ is defined in (\ref{eq:IN_defined}). 
Now, $\cP\cG\cP$ can be viewed as a metric induced by $\cG$ on the
 subspace $\Im(\cP)$. The divergent direction of $\cG$ will induce
 a divergent direction on $\Im(\cP)$ which is a projection of
 $\Im(1-\cI^{(N)})$ to $\Im(\cP)$ and can be expressed as
 $\Im(\cP(1-\cI^{(N)})\cP)$. Therefore, as $\epsilon\rightarrow 0$ 
there will  be a set of eigenvectors of $\cP\cG\cP$ with
 divergent eigenvalues which span $\Im(\cP(1-\cI^{(N)})\cP)$. The
 remaining non-zero eigenvalues, however, should be finite and
 correspond to non-divergent directions. Since eigenvectors with
 different eigenvalues have to be orthogonal (in the $S_{\infty}$ metric which
is used in the subsequent discussion), they will span a
 space in $\Im(\cP)$ which is orthogonal to $\Im(\cP(1-\cI^{(N)})\cP)$
 and that is precisely
\begin{equation}
	\Im(\cP) \cap \Im(\cI^{(N)}) = \Im(\cP_I).
\end{equation}
That means that the finite-eigenvalue eigenvectors of $\cP\cG\cP$ live in $\Im(\cP_I)$ and on that subspace $\cP\cG\cP = \cP_I\cG_N\cP_I$. Therefore, the finite eigenvalues  of the two operators match. It is also plausible,
and justified by examples we have studied, that the precise forms 
of the eigenvectors of $\cP_I\cG_N\cP_I$ can be recovered as 
limits of those of  $\cP\cG\cP$.

As an additional evidence that interesting strong-coupling spacetime physics
can be expected from  the eigenvalues of $ \cP \cG \cP $, note that the 
2-point function in the half-BPS case is
 $ { \chi_R(\Omega^{-1})  \over d_R  } $. The ${ 1\over N } $ expansion of 
$ \Omega^{-1} $ is related to conjugacy classes of $S_n$ 
and in turn to Casimirs \cite{cmrii,ori,kimram3}.   In the spacetime picture 
they are related to conserved charges related to multipole moments 
of LLM geometries \cite{integinfo}.

\subsection{Identifying giant graviton states } 
\label{idgiants} 

We now turn to a discussion of the physics of BPS states 
which disappear as $n$ is reduced to values below $N$. 
We expect on physical grounds \cite{mst,mikhailov} that there will be states 
in the region of $ n = N + k $ with $ k $ order 
$1$ as $ N \rightarrow \infty $, which can be interpreted 
as  { \it single }  quarter  or eighth-BPS giant gravitons along with 
excitations of the space-time background : single and multi-particle 
gravitons.

In the half-BPS sector,  with Young diagrams of type 
$ \Lambda = [ n  , 0 ] $,  there are  Young diagrams $R$ 
which have a long row of order $N$, and a few rows with order $1$ 
boxes. These correspond to a giant graviton large in the $AdS_5$ directions
along with background spacetime excitations. 
 Then there are states close to a sphere giant (large in the $S^5$ directions) 
 with one column of length order $N$, and a 
few columns with order $1$ boxes. There are also 
states consisting of multiple giants, each having a 
finite fraction of the total angular momentum, along with
background excitations.  These correspond to Young diagrams 
with a few long columns along with shorter columns. 
There are also generic states corresponding to
Young diagrams of shape far from the above  types. 

More generally, we can consider   a sequence of representations of 
the form $ \Lambda = [ n - \lambda , \lambda ] $, where $ \lambda $ is 
order $1$. We expect that a single giant graviton will have   well-defined 
$U(2)$ quantum numbers.  As $n$ increases for fixed $ \lambda$, the 
total energy of the giant will increase, and the energy available 
for excitations of the background spacetime will increase. 

The distinguishing characteristic of the states 
which consist of a giant expanding in the $S^5$, 
along with order $1$ spacetime excitations is that 
as $k = n - N $  increases from $0$,  their
 size will grow as the angular momentum increases
and  they will start disappearing from the 
spectrum. In fact all the states that disappear as $k$ 
increases from $0$, will be precisely the states consisting of 
a sphere giant along with space-time excitations.  
Consider then the difference
\begin{equation}  
\Delta \cM_{ N = n-k ; [ n - \lambda , \lambda ]}
 \equiv \cM_{ N = \infty ; [ n - \lambda , \lambda ] } - 
 \cM_{ N = n - k ; [ n - \lambda , \lambda ] }
\end{equation}
This will measure the number of states in the irrep.
 $ [ n - \lambda , \lambda ] $ which disappear due to finite $N$ 
effects. A first consequence of our discussion is that 
this number should become independent of $n$ as $n$ increases. 
 We have accumulated extensive numerical evidence for this behaviour. 
We will return to these properties of $\Delta \cM_{ N = n-k ; [ n - \lambda , \lambda ]}$ and related counting formulae in future work.

Along the lines of the discussion based on continuity 
 in  section \ref{PIandP}, we expect that  states that are close to 
the cutoff, will have finite but  large eigenvalues of $ \cP_I  \cG \cP_I $
(or $\cP \cG \cP $).  Similarly we 
would expect that the states with the smallest eigenvalues of 
$ \cP_I \cG \cP_I $ will be associated with single AdS giants and 
their excitations. 

Clearly an important open problem is to find 
a precise description of the eigenstates of $ \cP_I  \cG \cP_I $ 
which will allow a more precise identification of
the giant graviton states, notably to distinguish  which 
states just correspond to single or multiple quarter-BPS  giants and which 
correspond to giants  along with excitations of the background spacetime. 
 We turn to a discussion of avenues towards this 
goal.

\subsection{Orthogonal BPS basis  }

Eigenvectors of the 2-point function matrix 
are precisely the ones that allow an identification
to semi-classical states \cite{bbns,cjr,llm} in the half-BPS 
case. The same line of argument has been used above for states close to 
giant gravitons. We expect that the same can be applied to 
corresponding to deformations of the bulk geometry, i.e
quarter and eighth-BPS generalizations of LLM.

In Section~\ref{sec:2pt} we have derived 
the two-point function on the BPS states as  the matrix $\cP\cG\cP$:
\begin{equation}
	\langle \cO^{\rm BPS}_{ \L_1 , M_{ \L_1} ,  R_1 , \tau_1 } | ~
 \cO^{\rm BPS}_{ \L_2 , M'_{\L_2 } ,  R_2 , \tau_2 } \rangle = 
 \delta_{ \L_1 , \L_2 } \delta_{ M_{\L_1}  , M'_{\L_2}  }  
 ( \cP  \cG \cP )^{\L_1 ,  R_1 , \tau_1 }_{\L_1  R_2 , \tau_2 }.
\end{equation}
Consider an eigenvector of  $\cP\cG\cP$ with eigenvalue 
$ \lambda_{\mu} ( N )  $  and components  $B_{ \mu , \nu }^{\Lambda;R\tau}$. The label $ \nu $ is a multiplicity label for eigenstates with a 
fixed eigenvalue $ \lambda_{\mu} $.  The eigenvector equation is : 
\begin{equation}
	\sum_{R_2,\tau_2} ( \cP \cG \cP )^{\L , R_1 , \tau_1 }_{ \L ,  R_2 , \tau_2 } B_{\mu,\nu}^{\Lambda;R_2\tau_2} = \lambda_\mu B_{ \mu , \nu }^{\Lambda;R_1\tau_1} .
\end{equation}
We know there will be $\cM_{ \Lambda ; N } $ of eigenvectors for which $\lambda_\mu$ is non-zero and finite. 
The remaining eigenvectors having zero eigenvalues are 
just the ones spanning the kernel of $\cP$ 
and so they correspond to the descendant 
non-BPS operators, while the eigenvectors with divergent eigenvalues drop out of the finite $N$ Hilbert space. An orthogonal basis for the BPS operators is then
\begin{equation}
\label{eq:obps_orth}
	\cO_{\L,M_\L; \mu , \nu }^{\rm BPS} = \sum_{R,\tau} B_{ \mu , \nu } ^{\Lambda;R\tau} \cO^{\rm BPS}_{ \L , M_{ \L};  R , \tau }
\end{equation}
with the two-point function given by 
\begin{equation}
\langle \cO^{\rm BPS}_{ \L , M_{\L};  \mu_1 , \nu_1 }  | ~
 \cO^{\rm BPS}_{ \L , M_{\L};  \mu_2 , \nu_2 } \rangle 
 =
  B_{\mu_1 , \nu_1 }^{\Lambda;R_1\tau_1} 
  ( \cP \cG \cP )^{\L ,  R_1 , \tau_1 }_{\L ,  R_2 , \tau_2 }
  B_{\mu_2 , \nu_2 }^{\Lambda;R_2\tau_2} 
 =
 \lambda_{\mu_1  }
 B_{\mu_1 , \nu_1 }^{\Lambda;R_1\tau_1} B_{\mu_2 , \nu_2 }^{\Lambda;R_1\tau_1}.
\end{equation}

These states will be orthogonal as long as the eigenvalues 
differ $\lambda_{\mu_1} \neq \lambda_{\mu_2}$.
In the half-BPS case,  $\cP=\mbI$ and $\cP\cG\cP \sim (\Dim R)^{-1}$ while the 
states are labeled by $R$ and so all eigenvalues are  distinct.
 Also for all cases $\Lambda=[n,m]$ with $m \ge 2$ that we checked 
explicitly we did not find any degeneracy among the eigenvalues.
 The only exception seems to be the sector $\Lambda=[n,1]$. 
With only one $X$ operator in
 the traces symmetrization doesn't act and $\cP\cG\cP \sim (\Dim R)^{-1}$
 again. But now there is a possible multiplicity in $\tau$ and two 
distinct states with the same $R$ will be eigenstates with the 
\emph{same} eigenvalue $(\Dim R)^{-1}$. This happens first 
at $\Lambda=[5,1]$ where we find operators 
$\cO_{R=[3,2,1],\tau=1}, \cO_{R=[3,2,1],\tau=2}$. 
If it is true that this pattern is general, namely that 
the cases of degenerate eigenvalues for fixed $ \Lambda $ 
only arise with   $\Lambda$ of the form $[n,1]$, then 
for most $\Lambda $ the eigenvalues of $ \cP \cG \cP $ 
can provide a way to label the orthogonal eigenvectors of the 2-point function.

Even though we can write the orthogonal basis formally as (\ref{eq:obps_orth}), finding the eigenvectors $B_\mu^{\Lambda;R\tau}$ in practice is of course a very difficult task. We provide an example result in the Appendix~\ref{sec:app_l22_orth} for the simplest nontrivial case $\L=[2,2]$. In general they come out to be complicated expressions involving $N$. We will not have much more to say about the precise form of the eigenvectors.

\subsection{ Towards  a complete  set of labels for the BPS operators  } 
\label{sec:pg1p}

We will now describe a matrix related to $ \cP \cG \cP $ 
whose diagonalisation should be simpler 
and which may provide a more practical way of arriving 
at a labelling of the eigenstates of $ \cP \cG \cP $. 
Consider the large $N$ expansion of  $\cG$ : 
\begin{equation}
	\cG \equiv \mbI + \frac{1}{N}\cG_1 + O\left(\frac{1}{N^2}\right).
\end{equation}
\begin{equation}
	(\cG_1)^{R_1}_{R_2} = - \delta_{R_1,R_2} \frac{\chi_{R_1}(\Sigma_{[2]})}{d_R}
\end{equation}
with $\Sigma_{[2]}$ being the sum over all permutations with
 cycle structure $[2]$. Note that $\frac{\chi_{R_1}(\Sigma_{[2]})}{d_R}$
is just a number and not an $N$-dependent function. 
The leading terms of $\cP\cG\cP$ are  then
\begin{equation}
	\cP\cG\cP = \cP + \frac{1}{N}\cP\cG_1\cP + O\left(\frac{1}{N^2}\right).
\end{equation}
 $\cP$ and $\cP\cG_1\cP$ commute due to $\cP^2=\cP$ and they can be simultaneously diagonalized. This implies that the eigenstates of $\cP\cG\cP$ in the limit $N\rightarrow\infty$ will approach those of simultaneous eigenstates of $\cP$ and $\cP\cG_1\cP$. The BPS states will correspond to the subspace of $\cP$ eigenvalue 1 and their eigenvalues will be
\begin{equation}
	\lambda_\mu = 1 + \frac{1}{N}\widetilde{\lambda}_\mu + O\left(\frac{1}{N^2}\right)
\end{equation}
where the $\widetilde{\lambda}_\mu$ are the eigenvalues of $\cP\cG_1\cP$.

It is conceivable that, for fixed $ \Lambda $, the eigenvalues 
of $ \cP \cG \cP $ do not cross when plotted as a function of 
$N$. Indeed, this is the case in the examples we have studied
concretely. In that case, the eigenvalues of $ \cP \cG_1 \cP $
would be adequate as a way to label the states 
and identify them systematically with giant gravitons and 
 their excitations in the regime $ n \sim N$, 
and with LLM-like geometries in the regime $ n \sim N^2 $.
This deserves further investigation.

Progress on the dictionary  between space-time giant graviton and LLM-type
states will also require the computation of 
appropriately normalized three-point functions. For example 
we may expect the three point function of an operator 
corresponding to a giant graviton along with some spacetime excitations 
to be of order one when the other operators in the 3-point function 
are the giant in question and the small operator for the space-time excitation. 
The techniques of \cite{skentay} using correlators in connection 
with the asymptotics of SUGRA solutions should  be useful 
in relating correlators of eigenstates of $ \cP \cG \cP $  
to bulk spacetime.

\section{Summary and Outlook }
\label{sec:sumout} 

We have described an elegant method for a 
refined counting of quarter and bosonic eighth-BPS states, 
where states are organized according to representations of 
$ U(2)$ or $U(3)$. The method extends to general $U(M)$. 
It relies on the construction of an element $ \mP$
living in the group algebra of $S_n$. Exploiting
the Schur-Weyl duality relation between representations of 
$U(M)$ and symmetric groups, the characters of 
$ \mP $ in different representations of $S_n$ provide the 
multiplicities of different $U(M)$ representations  $ \Lambda $ 
among BPS states constructed from $n$ matrices, chosen 
from $ X_1 , X_2 , \cdots X_M$. This result is
given in (\ref{eq:M_lambda_from_P}).  The coefficients of 
different conjugacy classes in the expansion of $ \mP$ 
match in simple cases known integer sequences, and generalize 
them (see Appendix \ref{sec:integerseq}).  
Recent results on the generating functions for
BPS states \cite{kinney,Berenstein:2005aa} in terms of  simple harmonic 
oscillator states in $M$ dimensions, after a judicious 
change of variables (\ref{eq:xk_to_yk_substitution}), 
provide the generating functions for these coefficients 
of conjugacy classes in $ \mP$. 

The fine-structure of $ \mP$ encoded in a map
$\bp : \mbC ( S_n ) \rightarrow \mbC ( S_n ) $ with matrix 
elements $ \bp_{ \beta , \alpha   }$ lead  to the construction 
of a linear  operator $ \cP $ acting on the Hilbert space of 
the free $ \cN =4 $ SYM theory (i.e the theory at $g_{YM}^2 =0 $) 
which projects to the symmetrized traces  which are BPS 
states in the planar limit, annihilated by the one-loop dilatation operator
\cite{Beisertcomplete}.   The matrix elements 
of $ \cP $ are given in terms of a representation theoretic 
basis for the free theory, which diagonalizes the free two-point 
functions at finite $N$ \cite{BHR1,BHR2}. The free basis is a Fourier transform 
of the trace basis. The expression for $ \cP$ in this Fourier 
basis is given in terms of Clebsch-Gordan coefficients and matrix elements 
of the symmetric group  (\ref{eq:cp_definition}).

BPS states to all orders in the large $N $ expansion
are given in (\ref{eq:obps_result})  as the image of $ \cG \cP $, 
where $\cG$ has simple matrix elements in the Fourier 
basis, given in terms of inverse dimensions of representations
of $U(N)$.  The two-point functions of the BPS 
states are given by the matrix $ \cP \cG \cP $ (\ref{eq:twopointresult}). 
The matrix $ \cP $ also has the virtue that its kernel 
is spanned by the descendants, which are not annihilated by $ \cH_2$. 
As a result it allows a manifestly finite $N $ construction 
of BPS states (\ref{finiteNBPSconst}) and their
two-point functions (\ref{finiteN2pt}) $ \cP_I \cG \cP_I $. 
The eigenstates of $ \cP_I \cG \cP_I $ can be used to 
construct higher point correlators in a manifestly 
finite $N$ setting. 
Analyticity makes it possible to reconstruct eigenvalues and, conjecturally, 
 eigenvectors of 
the $ \cP_I \cG \cP_I$ from the matrix  $ \cP \cG \cP$. 
Non-renormalization theorems allow us to argue that the eigenvalues of 
the matrix $ \cP \cG \cP $ contain information about gauge theory 
operators for giant gravitons  \cite{mikhailov}  and their 
spacetime excitations and LLM-like geometries 
\cite{donosquarter,luninquarter,chenetal,lin2010}. The use of 
the stringy exclusion principle allows some progress 
in this direction.   An important 
open problem is to find a labelling of the eigenvectors of 
$ \cP \cG \cP $, which will allow the identification of 
different types of semiclassical objects : giant gravitons and their 
spacetime excitations, LLM geometries and their fluctuations. 
We discussed some avenues in this direction in section \ref{sec:eigs2point}. 
The importance of the matrix $\cP \cG \cP $ cannot be overstated\footnote{  
We will in fact confess that an earlier  title of this paper 
which did not make the final cut was  :``The Matrix : $\cP \cG \cP $.''}.

The route leading from $\mP$, which furnishes results 
in counting, to the matrix $ \cP $ and the two-point 
functions $ \cP \cG \cP $ is very interesting. It seems 
to capture a sort of categorification \cite{bado}, where numbers (multiplicities) 
are promoted to explicit constructions of quantum states 
in string theory, in a manner that can make semiclassical 
limits and space-time physics transparent. It will be interesting to 
find out if a similar route can be followed in more general 
systems in string theory, to lead from results in counting to 
explicit operators, states and semi-classical limits.

We end by mentioning some additional  avenues for further research. 

\begin{itemize} 

\item 
Generalization of $ \mP$,  $ \cP $ and $ \cP_I \cG \cP_I $ 
to the case of the most general eighth-BPS states 
with global symmetry  group $ U ( 3| 2)$. Other generalizations 
to consider are  for the $SU(N), O(N) , Sp (N) $ gauge groups.

\item 
Developing the identification of eighth-BPS giant graviton states 
further and finding operators corresponding 
to strings attached to giants, generalizing the 
use of the Young diagram description to characterize strings 
attached to half-BPS giants \cite{BBFH04,BCV06,KSS1,KSS2,KSS3} 

\item 
The matrix $ \cP \cG \cP $ can be described in 
other bases which diagonalize the free 2-point functions. 
The transformation between the restricted Schur basis
\cite{BBFH04,bck08,bck08II}  and the Fourier basis 
is described in \cite{collins08}. 
A similar transformation can be used to connect to the 
Brauer algebra basis \cite{kimram,krt}. 
This should shed light on the construction 
of \cite{kimuraquarter} which leads to an elegant description 
of a subset of quarter-BPS states in terms of projectors 
in Brauer algebras. This could provide another angle on the 
problem of finding a neat labelling of the eigenstates 
of $ \cP \cG \cP $ which connects to spacetime physics.

\item 
By exploiting a result of \cite{Bellucci:2004ru} 
on the one-loop dilatation operator for the $U(2)$ sector, 
the structure of Clebsch-Gordan coefficients 
and symmetric group elements which we used  to describe
$\cP$ can also be used to describe the action of 
the one-loop dilatation operator on the 
finite $N$ Fourier basis  in a manifestly $U(2)$ covariant
 way (see Appendix~\ref{sec:onedil}). 
The study of this  mixing in the Fourier basis 
was initiated in \cite{tom-loop}. The equation  (\ref{mixresult})  should shed light on the one-loop mixing problem 
of the descendants, which are 
 in $ \Ker ( \cP ) $. The  mixing problem at planar level 
and beyond has been a very active subject of research, notably 
 in connection with integrability. A few references, by no means complete, 
giving the flavour of the subject are 
\cite{vamver,conspost,GMR,BKPSS,jan02,minzar,GGR,lopez} 

\item 
A better understanding of the quarter and eighth-BPS 
sector is expected to have implications for black hole physics \cite{bbjs08,simon}. The  identification  of gauge theory operators,  
involving impurities added  to the eigenstates of 
$ \cP \cG \cP $,    corresponding to black hole solutions will 
be interesting to explore. In \cite{janik}, there was evidence for 
commuting variables in the counting of sixteenth BPS states, with analogies 
to the eighth BPS sector. This suggests something akin to the 
$\cP$ operator could be useful for the sixteenth BPS case.

\end{itemize}

{ \Large 
{ \centerline { \bf Acknowledgments }  } } 

\vskip.2in 
We thank  Tom Brown, Robert de Mello Koch, Paul Heslop,  Yusuke Kimura  
for discussions/correspondence.  We are also grateful to the string theory 
group at Wits University for organising the ``Second Joburg Workshop on String Theory: Correlation Functions and the AdS/CFT Correspondence'' 
where we benefited from numerous conversations on the subject of the 
one-loop dilatation operator.  SR is supported 
by an STFC grant ST/G000565/1. JP is supported
by a Queen Mary, University of London studentship. 

\vskip.5in

\begin{appendix}

\section{Further properties of $\cP$  and related operators  }
\label{sec:propertiescP} 

The operators $ \cP $ represents the action of the 
symmetrization operation in the basis $ \cO_{ \L , M_{ \L } , R , \tau } $ 
which diagonalizes the free two-point function. 
We describe the properties of $ \cP $ and of  some 
closely related projectors which are used in counting
of BPS operators. 

\subsection{Glossary of operators related to $\cP$  } 

~~~ $ \mP $ : Introduced in  (\ref{eq:p_definition}). A universal element in $\mbC(S_n)$.   Key counting result (\ref{eq:M_lambda_from_P}).\\

$ \mP_p $ : Introduced in  (\ref{eq:pp_definition}). It is a projector in $\mbC(S_n)$ which depends on the partition $p$. 
    Relation to $ \mP $ given in (\ref{eq:p_definition}).\\

$ \mP_{\alpha} $ : Introduced in (\ref{eq:mpa_defined}), also a projecctor in $\mbC(S_n)$. It is a slightly refined version of $\mP_p$ with a specified embedding of $p$.  \\

$\bp$ : Introduced in (\ref{eq:bp_defined}). It is  a linear map from $ \mbC ( S_n) $ to $ \mbC ( S_n)$ with
 $\bp_{\beta, \alpha } $ as the matrix elements of the map. 
$\bp^{R,i,j}_{S,k,l} $ is the  Fourier transform of  $\bp_{\beta, \alpha } $.  \\

$ \mP^{(N)} $ : Introduced in (\ref{eq:mPN_defined}). The finite $N$ version of the universal element $ \mP$
captures in symmetric group language the fact that the chiral ring 
counting is equivalent to counting states of $N$ particles 
in a simple harmonic oscillator. \\

$ \cP $ : Introduced in (\ref{eq:cp_definition}). It is a matrix built from $ \bp $, using Clebsch-Gordan coefficients of $S_n$. Key property: it is a projector implementing trace-symmetrization in 
  the Fourier-basis. \\

$ \cP_{[\alpha]} $ : Introduced in (\ref{eq:cpa_definition}). A summand that appears in the expression for $ \cP$. Also a projector in the Fourier basis which symmetrizes only the given trace structure $[\alpha]$.

\subsection{Poperties of $\bp : \mbC ( S_n ) \rightarrow \mbC ( S_n ) $ } 
The equation $ \mP_{ \alpha }^{2} = \mP_{ \alpha  } $
implies that 
\begin{equation}\label{projeq} 
\sum_{ \beta } \bp_{ \beta  , \alpha } \bp_{\beta^{-1} \tau , \alpha } 
= \bp_{ \tau , \alpha } 
\end{equation}

\begin{figure}
\begin{center}
\input{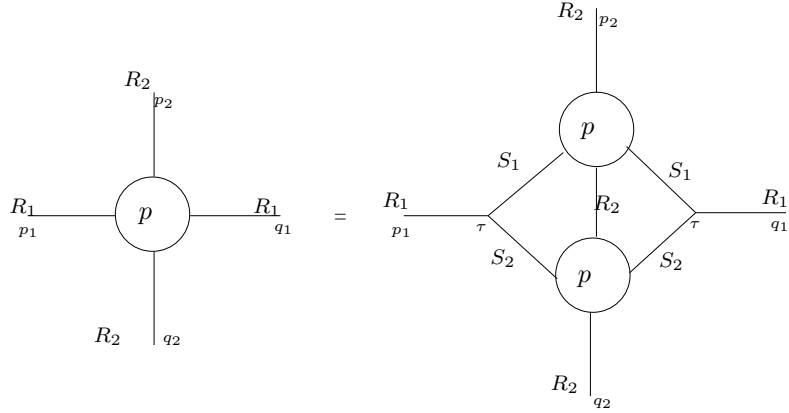}
\caption{projector relation } \label{fig:ppft}
\end{center}
\end{figure}

We can convert the equation (\ref{projeq}) into an equation in terms of 
the Fourier transformed quantity. 
\begin{equation}\label{peqppss}  
\bp^{ R_1 p_1 q_1}_{ R_2 p_1 q_2 } 
= \sum_{ S_1 , S_2 , i_1 , i_2 , j_1 , j_2 } 
  { d_{ S_1} d_{S_2} \over n! d_{R_1} } 
  \bp^{ S_1 ~ i_1 ~  j_1 }_{R_2~  p_2 ~  m }  
~ \bp^{ S_2 ~ i_2 ~ j_2 }_{R_2 ~   m  ~  q_2  } ~ 
  S^{ S_1  , S_2  ; R_1  , \tau }_{ ~j_1 ~ j_2  ~ q_1 } 
     S^{ S_1 , S_2  ; R_1  , \tau }_{~  i_1 ~ i_2  ~ p_1 }
\end{equation}  
Figure \ref{fig:ppft} gives a diagrammatic expression of the equation.

The coefficients $\bp_{ \beta , \alpha }$  have the symmetry 
\begin{equation} 
\bp_{ \beta , \alpha } = \bp_{ \gamma \beta  \gamma^{-1}  , \gamma \alpha \gamma^{-1} } 
\end{equation} 
Hence $ \sum_{ \alpha , \beta } \bp_{ \beta  , \alpha } ~ 
\beta  \otimes \alpha  $ defines 
an element of $ \mbC ( S_n ) \otimes \mbC ( S_n ) $ 
which commutes with the diagonal $  \mbC ( S_n ) $, 
and contains all the information about the counting 
$ \chi_{ \L } ( \mP_{ \alpha }  ) $  of irreps 
$ \L $ among the symmetrized traces of trace structure detrmined 
by the cycles of $ \alpha $. Such elements were considered in \cite{kimram3} 
in order to resolve the $ \tau $ multiplicity of symmetric group
Clebsch-Gordans.

\begin{equation} 
\bp_{ \beta^{-1}  , \alpha } = \bp_{ \beta , \alpha } 
\end{equation} 
Using $ D^S_{ ij} ( \beta^{-1}  ) =  D^S_{ ji } ( \beta ) $.  
This means that 
\begin{equation}  
\bp^{ R ~ i ~ j }_{ S ~ k ~ l } 
= \bp^{ R ~ i ~ j }_{ S ~ l ~ k }
\end{equation}

We also have 
\begin{equation}\begin{split}
& \bp_{ \beta , \alpha } =  \bp_{ \gamma \beta , \alpha } = 
 \bp_{  \beta \gamma  , \alpha } ~~ \hbox{ for } ~~
                                            \gamma \in G ( \alpha )
\end{split}\end{equation}  
where $ G ( \alpha ) $ is defined in section \ref{sec:p_map}.  

\subsection{Projectors $\cP_{[\alpha]}$}

The matrix $\cP$ in (\ref{eq:cp_definition}) can alternatively be written as:
\begin{equation}\label{eq:cP_alternative}\begin{split}
(\cP)^{\L  , R_1 , \tau_1}_{\L, R_2, \tau_2}
= \frac{\sqrt{d_{R_1} d_{R_2}}}{n!}
		\sum_{\alpha \in S_n} D^\Lambda_{ij}(\mP_\alpha)
		\left( D^{R_1}_{k_1 l_1}(\alpha) S^{ R_1 ~ R_1 ~ \L  , ~ \tau_1 }_{~  k_1 ~ l_1 ~ i } \right)
		\left( D^{R_2}_{k_2 l_2}(\alpha) S^{ R_2 ~ R_2 ~ \L  , ~ \tau_2 }_{~  k_2 ~ l_2 ~ j } \right).
\end{split}\end{equation}
This is seen by writing out $\bp^{ S ~ k_1 ~ k_2 }_{ \L ~ m_\L ~ m'_\L } = \sum_{\alpha} D^\L_{m_\L m'_\L}(\mP_\alpha) D^S_{k_1k_2}(\alpha)$ and then performing the sum over the representation $S$. Figure~\ref{fig:cPalphafig} expresses the RHS of (\ref{eq:cP_alternative}) in a diagramatic form.
\begin{figure}
\centering 
\input{cPalphaDiag.pstex_t} 
\caption{The operator $ \cP_{\alpha}$  } \label{fig:cPalphafig}
\end{figure}

The first thing to note is that the argument of the $\alpha$ sum in (\ref{eq:cP_alternative}) is invariant under conjugation $\alpha \rightarrow \gamma \alpha \gamma^{-1}$. That is because
\begin{equation}
	D^\Lambda_{ij}(\mP_{\gamma \alpha \gamma^{-1}}) 
	=
		D^\Lambda_{ij}(\gamma \mP_{\alpha} \gamma^{-1}) =
		D^\Lambda_{i'j'}(\mP_\alpha) D^\Lambda_{ii'}(\gamma) D^\Lambda_{jj'}(\gamma)
\end{equation}
and
\begin{equation}
	D^{R_1}_{k_1 l_1}(\gamma \alpha \gamma^{-1}) S^{ R_1 ~ R_1 ~ \L  , ~ \tau_1 }_{~  k_1 ~ l_1 ~ i }
	=
		 D^{R_1}_{k_1' l_1'}(\alpha) 
		 D^{R_1}_{k_1 k_1'}(\gamma) D^{R_1}_{l_1 l_1'}(\gamma) 
		 S^{ R_1 ~ R_1 ~ \L  , ~ \tau_1 }_{~  k_1 ~ l_1 ~ i }
	=
		 D^{R_1}_{k_1 l_1}(\alpha) 
		 S^{ R_1 ~ R_1 ~ \L  , ~ \tau_1 }_{~  k_1 ~ l_1 ~ i_1 }
		 D^{\Lambda}_{i_1 i}(\gamma^{-1})
\end{equation}
and so the resulting $D^\L(\gamma)$ matrices multiply to the identity.
This invariance means that the sum over $\alpha$ can be really seen as a sum over the conjugacy classes $[\alpha]$ each coming with a factor of $n!/|\Sym(\alpha)|$. It is then convenient to define a new matrix $\cP_{[\alpha]}$:
\begin{equation}	
\label{eq:cpa_definition}
(\cP_{[\alpha]})^{\L  , R_1 , \tau_1}_{\L, R_2, \tau_2}
	\equiv
	\frac{\sqrt{d_{R_1} d_{R_2}}}{|\Sym(\alpha)|}
		D^\Lambda_{ij}(\mP_\alpha)
		\left( D^{R_1}_{k_1 l_1}(\alpha) S^{ R_1 ~ R_1 ~ \L  , ~ \tau_1 }_{~  k_1 ~ l_1 ~ i } \right)
		\left( D^{R_2}_{k_2 l_2}(\alpha) S^{ R_2 ~ R_2 ~ \L  , ~ \tau_2 }_{~  k_2 ~ l_2 ~ j } \right),
\end{equation}
where the right hand side can be evaluated with any permutation $\alpha$ in the same conjugacy class, and thus $\cP_{[\alpha]}$ only depends on $[\alpha]$. The full matrix $\cP$ is then just a sum over conjugacy classes or partitions:
\begin{equation}
\label{PsumPa} 
	\cP
	= \sum_{[\alpha] \vdash n} \cP_{[\alpha]}
\end{equation}

The newly introduced matrix $\cP_{[\alpha]}$ can be seen as $\mP_\alpha$ or $\mP_p$ projector in $\mbC(S_n)$ appropriately transformed to the $|  \L  , R , \tau \rangle$ basis in the same way that $\cP$ is a transformed $\mP$. The most remarkable fact is that $\cP_{[\alpha]}$ itself is a projector with each $[\alpha]$ projecting to a \emph{distinct} subspace! That is
\begin{equation}\label{PaPbproj}
	\cP_{ [ \alpha] }  \cP_{ [ \beta ] }  = \cP_{[\alpha]} \delta_{[\alpha],\, [\beta]}.
\end{equation}
This can be shown by evaluating
\begin{equation}
	\sum_{R,\tau} d_R 
	\left( D^{R}_{k_1 l_1}(\alpha) S^{ R ~ R ~ \L  , ~ \tau }_{~  k_1 ~ l_1 ~ i } \right)
	\left( D^{R}_{k_2 l_2}(\beta) S^{ R ~ R ~ \L  , ~ \tau }_{~  k_2 ~ l_2 ~ j } \right) 
	=
	\sum_{\sigma \in S_n} D^\Lambda_{ij}(\sigma) \delta(\alpha \sigma \beta^{-1} \sigma^{-1})
\end{equation}
and then using this result to take the product:
\begin{equation}\begin{split}
	\sum_{R,\tau} &
	(\cP_{[\alpha]})^{\L  , R_1 , \tau_1}_{\L, R, \tau}
	(\cP_{[\beta]})^{\L  , R , \tau}_{\L, R_2, \tau_2}		
\\		
	&= \frac{\sqrt{d_{R_1} d_{R_2}}}{|\Sym(\alpha)|^2}
		\sum_{\sigma \in S_n}
		\delta(\alpha \sigma \beta^{-1} \sigma^{-1})
		\left( D^{R_1}_{k_1 l_1}(\alpha) S^{ R_1 ~ R_1 ~ \L  , ~ \tau_1 }_{~  k_1 ~ l_1 ~ i } \right)
		D^\Lambda_{ij}(\mP_\alpha \sigma \mP_\beta)
		\left( D^{R_2}_{k_2 l_2}(\beta) S^{ R_2 ~ R_2 ~ \L  , ~ \tau_2 }_{~  k_2 ~ l_2 ~ j } \right)
\\
	&= \frac{\sqrt{d_{R_1} d_{R_2}}}{|\Sym(\alpha)|^2}
		\sum_{\sigma \in S_n}
		\delta(\alpha \sigma \beta^{-1} \sigma^{-1})
		\left( D^{R_1}_{k_1 l_1}(\alpha) S^{ R_1 ~ R_1 ~ \L  , ~ \tau_1 }_{~  k_1 ~ l_1 ~ i } \right)
		D^\Lambda_{ij}(\mP_\alpha)
		\left( D^{R_2}_{k_2 l_2}(\alpha) S^{ R_2 ~ R_2 ~ \L  , ~ \tau_2 }_{~  k_2 ~ l_2 ~ j } \right)
\\
	&= (\cP_{[\alpha]})^{\L  , R_2 , \tau_2}_{\L, R_1, \tau_1}
		\sum_{\sigma \in S_n}
		\frac{ \delta(\alpha \sigma \beta^{-1} \sigma^{-1}) }{|\Sym(\alpha)|}	 
\\
	&= (\cP_{[\alpha]})^{\L  , R_2 , \tau_2}_{\L, R_1, \tau_1}
	\, \delta_{[\alpha],\, [\beta]}.
\end{split}\end{equation}
In the third line we used the fact that with the delta function present we could replace $\beta=\sigma^{-1}\alpha\sigma$ and that $\mP_\alpha^2 = \mP_\alpha$.

It follows from (\ref{PaPbproj}) and (\ref{PsumPa}) that 
$ \cP $ itself is a projector
\begin{equation} 
\cP^2 = \cP 
\end{equation} 
which is a nice consistency check.

It is interesting to ask what is the significance of the individual projectors $\cP_{[\alpha]}$, that is, what operation on the traces do they correspond to. More precisely, let us define a linear operation on the Hilbert space
\begin{equation}
\label{eq:symmsig_definition}
	{\rm symm}_{[\sigma]}\left[
		\cO_{ \Lambda , M_{\Lambda} ,  R , \tau  }
	\right] 
 =  \sum_{R_1,\tau_1} (\cP_{[\sigma]})^{\Lambda, M_\L, R_1 , \tau_1}_{\Lambda ,M_\L, R , \tau }
 \cO_{ \L , M_{ \L } , R_1 , \tau_1 }
\end{equation}
and ask what
\begin{equation}
	{ \rm { symm } }_{[\sigma]} [ \tr_n  ( X_{ \vec a } \alpha ) ]
\end{equation}
evaluates to. From the derivation (\ref{eq:cp_derriv1}) and (\ref{eq:cp_derriv2}) it can be seen that in order to get $\cP_{[\sigma]}$ on the RHS rather than $\cP$ the sum over $\alpha$ has to run only over the conjugacy class $[\sigma]$. That implies that the following action is needed on the trace basis instead of (\ref{pforsymm}):
\begin{equation}
	{ \rm { symm }_{[\sigma]} } [ \tr_n  ( X_{ \vec a } \alpha ) ] = \delta_{[\sigma],[\alpha]} 
\sum_{ \beta } \bp_{ \beta, \alpha } \tr_n   ( X_{ \beta ( \vec a )  } \alpha ).
\end{equation}
The interpretation is clear: ${\rm symm}_{[\sigma]}$ and its representation on the Fourier basis $\cP_{[\sigma]}$ acts by symmetrizing only the multitraces with the given trace structure $[\sigma]$. All traces with different structures are annihilated. This explains (\ref{PaPbproj}): if $[\alpha]$ and $[\beta]$ are different, there are no traces which survive. Also (\ref{PsumPa}) has a straightforward interpretation that full symmetrization is performed by symmetrizing all trace structures.

\section{One-loop dilatation operator in the Fourier basis}
\label{sec:onedil} 

In this section we derive an expression for the one-loop dilatation operator $\cH_2$ in the quarter-BPS sector acting on the Fourier basis. We use the same methods that led us to the $\cP$ matrix for symmetrization. The goal is to find the matrix $(\cH_2)^{\L_1,M_{\L_1},R_1,\tau_1}_{\L_2,M'_{\L_2},R_2,\tau_2}$ such that
\begin{equation}
	\cH_2 \cO_{\L,M_\L,R,\tau} = 
	\sum_{\L_1,M_{\L_1},R_1,\tau_1}
	(\cH_2)^{\L_1,\tilde M_{\L_1},R_1,\tau_1}_{\L,M_{\L},R,\tau}
	\cO_{\L_1,\tilde M_{\L_1},R_1,\tau_1}
\end{equation}
and to express it in terms of symmetric group objects. The matrix $(\cH_2)$ could allow us to go beyond the analysis of BPS states (which correspond to its zero-eigenvalue eigenvectors) and to calculate the anomalous scaling dimensions of the non-BPS operators.

Inspecting the derivation of $\cP$ in (\ref{eq:cp_derriv1}), (\ref{eq:cp_derriv2}) we find that the main ingredient there is the action of the linear operator of interest on the $\cO_{\vec{a},\alpha}$ basis. In the case of $\cP$ that was the symmetrization operator
\begin{equation}
	{ \rm { symm } } [ \tr_n  ( \mbX_{ \vec a } \alpha ) ] = 
\sum_{ \beta } \bp_{ \beta, \alpha } \tr_n   ( \mbX_{ \beta ( \vec a )  } \alpha ).
\end{equation}
The $\cP$ matrix is then a transformation of this action to the $|\L,M_\L,R,\tau\rangle$ basis.

The one-loop dilatation operator in the $U(3)$ sector 
 acting on multitrace operators can be written as 
\begin{equation}
	\cH_2 = - { 1 \over 2 } \tr([X_i,X_j][\check{X}_i,\check{X}_j ])
\end{equation}
with
\begin{equation}
	(\check{X}_a)^i_j \equiv \frac{\partial}{\partial (X_a)^j_i}.
\end{equation}
This can be translated to the $\cO_{\vec{a},\alpha}$ basis as \cite{Bellucci:2004ru}
\begin{equation}\label{soich} 
	\cH_2 \tr_n  ( \mbX_{ \vec a } \alpha ) = 
	 \sum_{i\neq j} 
	\tr_n  ( \mbX_{ P_{ij}(\vec a) }\, (i \alpha(j))' \alpha\, ).
\end{equation}
Here $P_{ij}$ is an element of the symmetric group algebra
\begin{equation}
	P_{ij} \equiv \mbI - (ij),
\end{equation}
with the action on $\mbX_{\vec{a}}$ defined as $\mbX_{P_{ij}(\vec{a})}=\mbX_{\vec{a}}-\mbX_{(ij)(\vec{a})}$. The term $(i \alpha(j))'$ is a permutation interchanging $i$ and $\alpha(j)$ which multiplies $\alpha$, \emph{except} for when $i=\alpha(j)$. In that case it should be understood as a factor of $N$:
\begin{equation}
	(ij)' \equiv 
	\begin{cases}
		(ij) & {\rm if} ~~ i\neq j, \\
		N & {\rm if} ~~ i=j.
	\end{cases}
\end{equation}
It is an element of $ \mbC ( S_n ) $ when $N$ is viewed as $N \in \mbC $. 
We can view (\ref{soich}) as expressing the fact that $\cH_2$ 
is determined by an element 
\bea 
\sum_{ \alpha} \sum_{ i \ne j  } ~ \alpha \otimes ( i \alpha(j) )^{\prime} \otimes P_{ij} 
\eea
in $ \mbC ( S_n ) \otimes \mbC ( S_n ) \otimes  \mbC ( S_n )$, 
which is an analog of the operator $ \bp$ in $  \mbC ( S_n ) \otimes  \mbC ( S_n )$ introduced before. 

Now that we have expressed the  $\cH_2$ action  in terms of 
  $ \mbC ( S_n ) $ quantities, we can transform it to the Fourier basis. Repeating the derivation along the lines of (\ref{eq:cp_derriv1}), (\ref{eq:cp_derriv2}):
\begin{equation}\begin{split}
	\cH_2 & \cO_{ \Lambda , M_{\Lambda} ,  R , \tau  }
=
	{ \sqrt {d_R } \over n! } \sum_{k\neq l} \sum_{\alpha, \vec{a}}
		S^{ R~  R ~ \L ,\;  \tau }_{ ~ i ~ j ~ m } 
		D^{R}_{ij} ( \alpha ) 
		C^{ \vec a }_{ \Lambda , M_{\Lambda } , m } 
		\tr_n  ( \mbX_{ P_{kl}(\vec a) }\, (k \alpha(l))' \alpha\, )
\\	&= 	
{ \sqrt {d_R } \over n! } \sum_{k\neq l} \sum_{\alpha, \vec{a}}
		S^{ R~  R ~ \L ,\;  \tau }_{ ~ i ~ j ~ m } 
		D^{R}_{ij} ( \alpha ) 
		C^{ \vec a }_{ \Lambda , M_{\Lambda } , m } 
\\ & \hskip2.0cm 
	\times \sum_{\L_1, \tilde M_{ \L_1 }, R_1,\tau_1}
	\sqrt { d_{R_1}  }   
	C_{ P_{kl}(\vec a) }^{ \L_1 , \tilde M_{ \L_1 } , m_1  } 
	D^{R_1}_{i_1 j_1  } ( (k \alpha(l))' \alpha  ) 
	S^{ ~ R_1 ~ R_1 ~ \L_1, \tau_1 }_{ ~~i_1   ~ j_1   ~ m_1 }
	\cO_{ \L_1  , \tilde M_{ \L_1 } ,  R_1 , \tau_1  } 
\\ 
& = 
	 \sum_{R_1,\tau_1}
	 	\cO_{ \L, M_{ \L} ,  R_1 , \tau_1  } 
{ \sqrt {d_R d_{R_1} } \over n! } 
\\ & \hskip2.0cm 
	\times \sum_{k\neq l} \sum_{\alpha}	
		S^{ R~  R ~ \L ,\;  \tau }_{ ~ i ~ j ~ m } 
		D^{R}_{ij} ( \alpha ) D^\Lambda_{m m_1 }(P_{kl} )
 D^{R_1 }_{ i_1 j_1   } ( (k \alpha(l))'\,\alpha  )	
 S^{ ~ R_1 ~ R_1 ~ \L , \tau_1 }_{ ~~i_1  ~ j_1  
 ~ m_1 } 
\\
&\equiv 
\sum_{\L_1,M_{\L_1},R_1,\tau_1}
(\cH_2)^{\L_1,\tilde M_{\L_1},R_1,\tau_1}_{\L,M_{\L},R,\tau}
	\cO_{\L_1,\tilde M_{\L_1},R_1,\tau_1}.
\end{split} 
\end{equation}
We have thus calculated the transformation matrix to be:
\begin{equation}\begin{split}\label{mixresult} 
&(\cH_2)^{\L_1,M_{\L_1},R_1,\tau_1}_{\L_2,\tilde M_{\L_2},R_2,\tau_2}
	=
	\delta_{\L_1,\L_2}\delta_{M_{\L_1},\tilde M_{\L_2}}
	{ \sqrt {d_{R_1} d_{R_2} } \over n! } 
\\
 & \hskip2.0cm \times	
	\sum_{k\neq l} \sum_{\alpha \in S_n}	
	 D^{\L_1}_{m_1 m_2 }(P_{kl} )
\left(		
 D^{R_1 }_{ i_1 j_1   } ( (k \alpha(l))'\,\alpha  )\,
 S^{ ~ R_1 ~ R_1 ~ \L_1 , \tau_1 }_{ ~~i_1  ~ j_1  ~ m_1 } 
 \right)
\left(
	D^{R_2}_{i_2j_2} ( \alpha )\,
	S^{ ~ R_2 ~ R_2 ~ \L_1 , \tau_2 }_{ ~~i_2  ~ j_2  ~ m_2 } 		
\right).
\end{split}\end{equation}
This expression should be compared to (\ref{eq:cP_alternative}). One can see the $D^\L(P_{kl})$ appearing in the same place where we had $D^\L(\mP_\alpha)$ before, because that is the permutation acting on $\vec{a}$ in the trace basis. A new ingredient here is the $(k \alpha(l))'$ operator acting on $\alpha$, which was absent in the symmetrization action.

\section{Examples of BPS operators}
\label{sec:app_example}

\subsection{$\L=[2,2]$}

In this section we will provide an example of explicit calculation of BPS operators in the representation $\Lambda = [2, 2]$. Since the operators throughout the section will all have the same $\L$, also $M_\L$ will be the highest weight state and $\tau=1$ because there is no multiplicity, we will abbreviate
\begin{equation}
\begin{split}
	\cO_{R} &\equiv \cO_{\L=[2,2],M_\L={\rm HWS},R,\tau=1}.
\end{split}
\end{equation}

\subsubsection{Large $N$ operators}

 First we will assume $N \ge 3$, but will keep all $1/N$ corrections. Then in the next section we will also explain what happens for lower $N$.

We start with the basis in the free theory. Using (\ref{eq:ob_definition}) the operators are:
\begin{equation}
\label{eq:l22_ob_from_tr}
\begin{split}	
	\cO_{R=[3,1]} &= \frac{1}{4\sqrt{3}N^2} \left(
		- \tr(X_i X^i X_j X^j) + \tr(X_i X_j) \tr(X^i X^j) + \tr(X_i X_j) \tr(X^i) \tr(X^j)
	\right)
\\	
	\cO_{R=[2,2]} &= \frac{1}{2\sqrt{6}N^2} \left(
		\tr(X_i X^i X_j X^j) + \tr(X_i X_j) \tr(X^i) \tr(X^j)
	\right)	
\\	
	\cO_{R=[2,1,1]} &= \frac{1}{4\sqrt{3}N^2} \left(
		- \tr(X_i X^i X_j X^j) - \tr(X_i X_j) \tr(X^i X^j) + \tr(X_i X_j) \tr(X^i) \tr(X^j)
	\right).
\end{split}
\end{equation}
Here for compactness we are using upper-lower index notation meaning $X_i X^i = X_1 X_2 - X_2 X_1$, which is a singlet in the quarter-BPS sector.

The symmetrization matrix $\cP$ can be evaluated using (\ref{eq:cp_definition}) or (\ref{eq:cP_alternative}). In the $\Lambda=[2,2]$ sector we get:
\begin{equation}
\cP = 
\left(
\begin{array}{ccc}
 \frac{3}{4} & \frac{1}{2 \sqrt{2}} & -\frac{1}{4} \\
 \frac{1}{2 \sqrt{2}} & \frac{1}{2} & \frac{1}{2 \sqrt{2}} \\
 -\frac{1}{4} & \frac{1}{2 \sqrt{2}} & \frac{3}{4}
\end{array}
\right).
\end{equation}
It can be verified that transforming the free operator basis with this matrix indeed symmetrizes terms in each trace. In this example that just means that the $\tr(X_i X_j X^i X^j)$ is dropped.
\begin{equation}
\left( \begin{array}{ccc}
	\cO_{[3,1]}^\rmS & \cO_{[2,2]}^\rmS & \cO_{[2,1,1]}^\rmS
\end{array}\right)
=
\left( \begin{array}{ccc}
	\cO_{[3,1]} & \cO_{[2,2]} & \cO_{[2,1,1]}
\end{array}\right)
\left(
\begin{array}{ccc}
 \frac{3}{4} & \frac{1}{2 \sqrt{2}} & -\frac{1}{4} \\
 \frac{1}{2 \sqrt{2}} & \frac{1}{2} & \frac{1}{2 \sqrt{2}} \\
 -\frac{1}{4} & \frac{1}{2 \sqrt{2}} & \frac{3}{4}
\end{array}
\right)
\end{equation}

\begin{equation}
\begin{split}
\cO_{[3,1]}^\rmS 	
	&= \frac{1}{4\sqrt{3}N^2} \left(
		\tr(X_i X_j) \tr(X^i X^j) + \tr(X_i X_j) \tr(X^i) \tr(X^j)
	\right)
\\	
\cO_{[2,2]}^\rmS 	
	&= \frac{1}{2\sqrt{6}N^2} \left(
		\tr(X_i X_j) \tr(X^i) \tr(X^j)
	\right)	
\\	
\cO_{[2,1,1]}^\rmS 	
	&= \frac{1}{4\sqrt{3}N^2} \left(
		- \tr(X_i X_j) \tr(X^i X^j) + \tr(X_i X_j) \tr(X^i) \tr(X^j)
	\right).
\end{split}
\end{equation}
Clearly, there are only 2 independent symmetric operators.
 This corresponds to the fact that at large $N$ there
 are only 2 BPS operators when the coupling is turned on,
 compared to 3 operators in the free theory.
It is also worth noting that the null eigenvector 
\bea 
\begin{pmatrix} { 1\over 2 } \\ { - 1 \over \sqrt{2} } \\ { 1 \over 2 }
\end{pmatrix} 
\eea 
 corresponding to  
$ { 1\over 2 } \cO_{[3,1]} - { 1\over \sqrt{2} } \cO_{ [ 2,2]} +
 { 1\over 2} \cO_{[2,1^2]} $ is the descendant operator. This is a general fact 
that the descendants can be characterized as $ \Ker ( \cP ) $.

We can write down the precise expressions for the BPS operators using (\ref{eq:obps_result}):
\begin{equation}
		\cO^{\rm BPS }_{ \Lambda , M_{\Lambda} ,  R , \tau  }   = \sum_{R_1,\tau_1} ( \cG \cP )^{\L , R_1 , \tau_1}_{\L ,  R , \tau} \cO_{ \L , M_{ \L } , R_1 , \tau_1 }
\end{equation}
with the diagonal matrix
\begin{equation}
\begin{split}
\cG &= {\rm diag}\left\{ \frac{N^n d_R}{n!\,\Dim R} \right\}
	= \frac{N^3}{N^2 - 1}
\left(
\begin{array}{ccc}
	\frac{1}{N+2} & 0 & 0 \\
	0 & \frac{1}{N} & 0 \\
	0 & 0 & \frac{1}{N-2}
\end{array}
\right)
\end{split}
\end{equation}
Let us arbitrarily pick $\cO_{[3,1]}^{\rm BPS}$ and $\cO_{[2,1,1]}^{\rm BPS}$ as the two linearly independent operators. In terms of the free operators we get
\begin{equation}
\label{eq:l22_bps_from_b}
\begin{split}
\cO_{[3,1]}^{\rm BPS} &= \frac{N^3}{N^2 - 1} \left(
	\frac{3}{4}\frac{\cO_{[3,1]}}{N+2}
	+ \frac{1}{2\sqrt{2}}\frac{\cO_{[2,2]}}{N}
	- \frac{1}{4}\frac{\cO_{[2,1,1]}}{N-2}	
\right),
\\
\cO_{[2,1,1]}^{\rm BPS} &= \frac{N^3}{N^2 - 1} \left(
	-\frac{1}{4}\frac{\cO_{[3,1]}}{N+2}
	+ \frac{1}{2\sqrt{2}}\frac{\cO_{[2,2]}}{N}
	+ \frac{3}{4}\frac{\cO_{[2,1,1]}}{N-2}	
\right).
\end{split}
\end{equation}
Explicitly in terms of products of traces they are
\begin{equation}
\label{eq:l22_bps_from_tr}
\begin{split}
\cO_{[3,1]}^{\rm BPS} &= 
	\frac{1}{4\sqrt{3}(N^2-1)(N^2-4)}
	\left(
		2(N-1) \tr(X_i X^i X_j X^j) 
	\right.
\\	& \quad\quad
	\left.
		+ N(N-1) \tr(X_i X_j) \tr(X^i X^j) 
		+ (N^2 - 2N - 2) \tr(X_i X_j) \tr(X^i) \tr(X^j)
	\right),
\\	
\cO_{[2,1,1]}^{\rm BPS} &= 
	\frac{1}{4\sqrt{3}(N^2-1)(N^2-4)}
	\left(
		-2(N+1) \tr(X_i X^i X_j X^j) 
	\right.
\\	& \quad\quad
	\left.
		- N(N+1) \tr(X_i X_j) \tr(X^i X^j) 
		+ (N^2 + 2N - 2) \tr(X_i X_j) \tr(X^i) \tr(X^j)
	\right).
\end{split}	
\end{equation}

To the leading order in $N$ the operators $\cO_R^{\rm BPS}$ are equal to the symmetric combinations $\cO_R^\rmS$, but there are further $1/N$ corrections which make the state exactly annihilated by the one-loop dilatation operator. Note how the antisymmetric combinations $\tr(X_i X_j X^i X^j)$ appear in the exact operator, although only in the subleading-$N$ order.

\subsubsection{Finite $N$ cutoff}

The expressions for BPS operators (\ref{eq:l22_bps_from_b}), (\ref{eq:l22_bps_from_tr}) are valid whenever $N \ge 3$.
 Clearly something goes wrong when $N=2$ as indicated by $N-2$ in the denominator -- this is where the finite-$N$ cutoff takes effect. 
 We will demonstrate here how it is dealt 
with according to the general procedure 
described in Section~\ref{sec:finiteN}. It turns out that at $N=2$ 
there is only one BPS operator remaining.

The cutoff is best seen in the free operator basis: 
$\cO_{\Lambda,M_{\Lambda },R,\tau}$ becomes 0 whenever 
Young diagram $R$ has more than $N$ rows. 
In our case one of the three free operators drops out:
\begin{equation}
	\cO_{[2,1,1]}= 0.
\end{equation}
Note that implies a relationship between products of traces, using (\ref{eq:l22_ob_from_tr}):
\begin{equation}
\label{eq:l22_tr_relationship}
- \tr(X_i X^i X_j X^j) - \tr(X_i X_j) \tr(X^i X^j) + \tr(X_i X_j) \tr(X^i) \tr(X^j) = 0.
\end{equation}
In the expressions for BPS operators exactly these free operators which vanish are accompanied by 0 in denominator because of $\Dim R = 0$. 

It is not immediately obvious how then to find correct BPS combinations. One could try to just naively cut off the terms in (\ref{eq:l22_bps_from_b}) and write
\begin{equation}
\nonumber
\begin{split}
\cO_{[3,1]}^{\rm BPS} &\sim 
	\frac{3}{4}\frac{\cO_{[3,1]}}{N+2}
	+ \frac{1}{2\sqrt{2}}\frac{\cO_{[2,2]}}{N},
\\
\cO_{[2,1,1]}^{\rm BPS} &\sim
	-\frac{1}{4}\frac{\cO_{[3,1]}}{N+2}
	+ \frac{1}{2\sqrt{2}}\frac{\cO_{[2,2]}}{N},
\end{split}	
\end{equation}
however, this is \emph{not} correct. If it were true, it would seem that we are still left with two BPS operators, however, it turns out that even at $N=2$ neither of them is annihilated by one-loop dilatation operator!

What we have to do instead is to pick linear combinations of $\cO_R^{\rm BPS}$ which do not depend on these $\cO_R$ that are set to zero. These linear combinations will be ``protected" from what happens to the terms with $\Dim R =0$ in the denominator. This procedure corresponds exactly to picking vectors in the intersection space $\Im(\cP_I)$ described in Section~\ref{sec:finiteN}, where it is shown to give all the BPS operators. In our example we have to take the combination:
\begin{equation}\begin{split}
\label{eq:n2_bps_operator}
	\cO^{\BPS; (N=2)} = \cO_{[3,1]}^\BPS + \frac{1}{3} \cO_{[2,1,1]}^\BPS &= \frac{N^3}{N^2 - 1} \left(
	\frac{2}{3} \frac{\cO_{[3,1]}}{N+2} + \frac{\sqrt{2}}{3} \frac{\cO_{[2,2]}}{N} \right)
\\
	&= \frac{4}{9} (\cO_{[3,1]} + \sqrt{2} \cO_{[2,2]})
\end{split}\end{equation}
This is the single remaining BPS operator at $N=2$. To be more precise, in the language of Section~\ref{sec:finiteN}, we find the projector $\cP_I$ for the intersection of $\Im(\cI_{(N)})$ and $\Im(\cP)$ to be
\begin{equation}
\cP_I^{(N=2) } = 
\left(
\begin{array}{ccc}
 \frac{2}{3} & \frac{\sqrt{2}}{3} & 0 \\
 \frac{\sqrt{2}}{3} & \frac{1}{3} & 0 \\
 0 & 0 & 0
\end{array}
\right).
\end{equation}
Then we get the BPS operators by acting with $(\cG \cP_I)$ on the full Hilbert space, which gives precisely the one-dimensional space spanned by the operator (\ref{eq:n2_bps_operator}).

There is a related subtlety worth pointing out. It is true that counting BPS operators is the same as counting states in the chiral ring, where the matrices $X_i$ can be taken to be diagonal. One may wonder whether this is the same as counting independent products of traces, where the terms in each trace are symmetrized, that is, the $\cO_{\Lambda;R,\tau}^{\rmS}$ operators. It turns out that this is \emph{not} the case when the finite-$N$ cutoff applies. In our present example at $N=2$ there are still two independent symmetric operators: $\tr(X_i X_j) \tr(X^i X^j)$ and $\tr(X_i X_j) \tr(X^i) \tr(X^j)$. However, one has to eliminate one linear combination - the one that is equal to $\tr(X_i X^i X_j X^j)$ by the relationship (\ref{eq:l22_tr_relationship}). This leaves us with one operator.

\subsubsection{Orthogonal BPS basis and $\cP \cG \cP$ eigenvalues}
\label{sec:app_l22_orth}

In order to find an orthogonal basis for the BPS operators we have to look for the eigenvectors of the two-point function matrix on the BPS operator space $\cP \cG \cP$. In the case of our present example $\L=[2,2]$ we can calculate the matrix to be:
\begin{equation}
\cP \cG \cP =
\frac{N^2}{(N^2-1)(N^2-4)} 
	\left(
		\begin{array}{ccc}
		 \frac{3N^2-4N-2}{4}  & \frac{N^2-2N-2}{2 \sqrt{2}} & -\frac{N^2+2}{4} \\
		 \frac{N^2-2N-2}{2 \sqrt{2}} & \frac{N^2-2}{2} & \frac{N^2-2N-2}{2 \sqrt{2}} \\
		 -\frac{N^2+2}{4} & \frac{N^2-2N-2}{2 \sqrt{2}} & \frac{3N^2-4N-2}{4}
		\end{array}	
	\right).
\end{equation}
Its characteristic polynomial is
\begin{equation}
	{\rm det}(\cP \cG \cP - \lambda \mbI) = 
	\frac{\lambda}{(N^2-4)(N^2-1)^2} \left(
		- (N^2-4)(N^2-1)^2 \lambda^2 
		+ 2N^2(N^2-1)^2 \lambda 
		- N^6
	\right) 
\end{equation}
and we find eigenvalues
\begin{equation}\begin{split}
\label{eq:l22_pgp_evalues}
	\lambda_{1,2} &= \frac{N^2}{(N^2-4)(N^2-1)}\left( N^2 - 1 \mp \sqrt{2N^2 + 1} \right), \\
	\lambda_3 &= 0.
\end{split}\end{equation}
The eigenvectors with zero eigenvalues are the same as those of $\cP$ and span the non-BPS combinations. We are interested in the other two eigenvectors
\begin{equation}\begin{split}
\label{eq:l22_pgp_evectors}
\cO_{(1,2)}^\BPS &= 
		\frac{-3N\mp2\sqrt{2N^2+1}}{N} \cO_{[3,1]}^\BPS
		+\frac{-\sqrt{2}(N+1 \pm \sqrt{2N^2+1})}{N} \cO_{[2,2]}^\BPS
		+\frac{N-2}{N} \cO_{[2,1,1]}^\BPS 
\\	
&= 	\frac{N^2}{N^2 - 1} \left(
		(-3N \mp 2\sqrt{2N^2+1}) \frac{\cO_{[3,1]}}{N+2}
		-\sqrt{2}(N+1 \pm \sqrt{2N^2+1}) \frac{\cO_{[2,2]}}{N}
		+ \cO_{[2,1,1]}
	\right).
\end{split}\end{equation}
These two operators provide us with an \emph{orthogonal basis} for the BPS sector. The normalization that we picked so far is arbitrary, except that we adjusted it to allow for a nice limit to $N=2$ as we will see shortly.

\subsubsection{$N = 2$ limit revisited}

As discussed already, at $N=2$ there is only one BPS operator in the $\L=[2,2]$ representation given by the combination (\ref{eq:n2_bps_operator}). On the other hand, the operators $\cO_{(1)}^\BPS$, $\cO_{(2)}^\BPS$ provide an $N$-dependent basis of orthogonal states from a calculation done assuming $N \ge n$. According to the discussion in Section~\ref{PIandP} we can in fact take the limit of the orthogonal basis to $N\rightarrow 2$ and find the one remaining BPS state as the surviving eigenvector with finite eigenvalue. We will see an example of that here.

Let us simply take the expressions for eigenvalues and eigenvectors (\ref{eq:l22_pgp_evalues}), (\ref{eq:l22_pgp_evectors}) and calculate the limit $N \rightarrow 2$. We find
\begin{equation}
	\lambda_1^{(N=2)} = \frac{8}{9}, \quad
	\lambda_2^{(N=2)} \sim \frac{2}{N-2} \rightarrow \infty
\end{equation}
and
\begin{equation}\begin{split}
	\cO_{(1)}^{\BPS;(N=2)} &= - 4 \cO_{[3,1]} - 4 \sqrt{2} \cO_{[2,2]} + \frac{4}{3} \cO_{[2,1,1]},
	\\
	\cO_{(2)}^{(N=2)} &= \frac{4}{3} \cO_{[2,1,1]}.
\end{split}\end{equation}
The operator $\cO_{[2,1,1]}$ is identically zero  in this case and so can be dropped from the operators. The first eigenvector $\cO_{(1)}^{\BPS;(N=2)}$, which has a finite eigenvalue, is indeed the right BPS state (\ref{eq:n2_bps_operator}) up to an overall normalization. The second eigenvector, with divergent 
eigenvalue, disappears from the Hilbert space. This reflects a pattern 
that we expect to be general. 
\subsubsection{$N \rightarrow \infty$ limit revisited}

Another interesting limit of the orthogonal BPS operators that we can take is $N\rightarrow\infty$. We know already that at large $N$ the leading term of the BPS operators is always a symmetrized trace. So a priori we can take any orthogonal combination of $\tr(X_i X_j) \tr(X^i X^j)$ and $\tr(X_i X_j) \tr(X^i) \tr(X^j)$ to be the large-$N$ basis. What is interesting about taking the limit of (\ref{eq:l22_pgp_evectors}) is that it provides us with a unique \emph{preferred} basis. We get
\begin{equation}\begin{split}
\cO_{(1,2)}^{\BPS;(N=\infty)} &= 
		(-3 \mp 2\sqrt{2}) \cO_{[3,1]} + (-\sqrt{2} \mp 2) \cO_{[2,2]} + \cO_{[2,1,1]}		
\\ 
	&= - \frac{2 \pm \sqrt{2}}{2\sqrt{3}N^2} \tr(X_i X_j) \tr(X^i X^j) - \frac{1 \pm \sqrt{2}}{\sqrt{3}N^2} \tr(X_i X_j) \tr(X^i) \tr(X^j) .
\end{split}\end{equation}
Both eigenvalues approach $1$, because $\cP\cG\cP \rightarrow \cP$. These states can also be found as the eigenstates of $\cP\cG_1\cP$, as discussed in Section~\ref{sec:pg1p}. The $\cG_1$ matrix is in this case:
\begin{equation}
	\cG_1 = \left(	
\begin{array}{ccc}
	-2 & 0 & 0 \\
	0 & 0 & 0 \\
	0 & 0 & 2
\end{array}
\right).
\end{equation}

\subsection{$\L=[3,2]$}

Here we work out the calculations for the $\L=[3,2]$ sector. In this section we again abbreviate the operators:
\begin{equation}
\begin{split}
	\cO_{R,\tau} &\equiv \cO_{\L=[3,2],M_\L={\rm HWS},R,\tau}  \\
	\cO_{R} &\equiv \cO_{\L=[3,2],M_\L={\rm HWS},R,\tau=1}
\end{split}
\end{equation}
with the $U(3)$ label $M_\L$ in the highest weight state.

The free operators:
\begin{equation}\begin{split}
\nonumber
\cO_{R=[4,1]} &= \frac{1}{2\sqrt{10}N^{5/2}} \left(
	-\tr(X_1 X_i X^i X_j X^j)
	-\frac{2}{3} \tr(X_i X^i X_j X^j)\tr(X_1)
	\right . \\ & \left. \quad
	+\tr(X_1 X_i X_j)\tr(X^i X^j)
	+\frac{1}{3} \tr(X_1 X_i X_j)\tr(X^i)\tr(X^j)
	\right . \\ & \left. \quad
	+\frac{2}{3} \tr(X_i X_j) \tr(X^i X^j) \tr(X_1)
	+\frac{1}{3} \tr(X_i X_j)\tr(X^i)\tr(X^j)\tr(X_1)
\right),
\\
\cO_{R=[3,2]} &= \frac{1}{4\sqrt{2}N^{5/2}} \left(
	-\tr(X_1 X_i X^i X_j X^j)
	-\frac{2}{3} \tr(X_i X^i X_j X^j)\tr(X_1)
	\right . \\ & \left. \quad
	-\frac{2}{3} \tr(X_1 X_i X_j)\tr(X^i)\tr(X^j)
	\right . \\ & \left. \quad
	-\frac{1}{3} \tr(X_i X_j) \tr(X^i X^j) \tr(X_1)	
	-\frac{2}{3} \tr(X_i X_j)\tr(X^i)\tr(X^j)\tr(X_1)
\right),
\end{split}\end{equation}
\begin{equation}\begin{split}
\cO_{R=[3,1^2],\tau=1} &= \frac{1}{4\sqrt{5}N^{5/2}} \left(
	\tr(X_1 X_i X^i X_j X^j)
	-2 \tr(X_1 X_i X_j)\tr(X^i)\tr(X^j)
	\right . \\ & \left. \quad
	+ \tr(X_i X_j) \tr(X^i X^j) \tr(X_1)
\right)
\\
\cO_{R=[3,1^2],\tau=2} &= \frac{1}{\sqrt{30}N^{5/2}} \left(
	\tr(X_i X^i X_j X^j)\tr(X_1)
	+\tr(X_1 X_i X_j)\tr(X^i X^j)	
	\right . \\ & \left. \quad
	-\frac{1}{2} \tr(X_i X_j)\tr(X^i)\tr(X^j)\tr(X_1)
\right),
\end{split}\end{equation}
\begin{equation}\begin{split}
\nonumber
\cO_{R=[2^2,1]} &= \frac{1}{4\sqrt{2}N^{5/2}} \left(
	-\tr(X_1 X_i X^i X_j X^j)
	+\frac{2}{3} \tr(X_i X^i X_j X^j)\tr(X_1)
	\right . \\ & \left. \quad
	-\frac{2}{3} \tr(X_1 X_i X_j)\tr(X^i)\tr(X^j)
	\right . \\ & \left. \quad
	-\frac{1}{3} \tr(X_i X_j) \tr(X^i X^j) \tr(X_1)	
	+\frac{2}{3} \tr(X_i X_j)\tr(X^i)\tr(X^j)\tr(X_1)
\right),
\\
\cO_{R=[2,1^3]} &= \frac{1}{2\sqrt{10}N^{5/2}} \left(
	\tr(X_1 X_i X^i X_j X^j)
	-\frac{2}{3} \tr(X_i X^i X_j X^j)\tr(X_1)
	\right . \\ & \left. \quad
	+\tr(X_1 X_i X_j)\tr(X^i X^j)
	-\frac{1}{3} \tr(X_1 X_i X_j)\tr(X^i)\tr(X^j)
	\right . \\ & \left. \quad
	-\frac{2}{3} \tr(X_i X_j) \tr(X^i X^j) \tr(X_1)
	+\frac{1}{3} \tr(X_i X_j)\tr(X^i)\tr(X^j)\tr(X_1)
\right).
\end{split}\end{equation}

The symmetrization matrix:
\begin{equation}
\cP = 
\left(
\begin{array}{cccccc}
 \frac{2}{3} & -\frac{\sqrt{5}}{6} & \frac{1}{5 \sqrt{2}} & \frac{2}{5 \sqrt{3}} & -\frac{1}{6 \sqrt{5}} & \frac{1}{15} \\
 -\frac{\sqrt{5}}{6} & \frac{7}{12} & \frac{1}{2 \sqrt{10}} & \frac{1}{\sqrt{15}} & -\frac{1}{12} & \frac{1}{6 \sqrt{5}} \\
 \frac{1}{5 \sqrt{2}} & \frac{1}{2 \sqrt{10}} & \frac{9}{10} & 0 & \frac{1}{2 \sqrt{10}} & -\frac{1}{5 \sqrt{2}} \\
 \frac{2}{5 \sqrt{3}} & \frac{1}{\sqrt{15}} & 0 & \frac{3}{5} & -\frac{1}{\sqrt{15}} & \frac{2}{5 \sqrt{3}} \\
 -\frac{1}{6 \sqrt{5}} & -\frac{1}{12} & \frac{1}{2 \sqrt{10}} & -\frac{1}{\sqrt{15}} & \frac{7}{12} & \frac{\sqrt{5}}{6} \\
 \frac{1}{15} & \frac{1}{6 \sqrt{5}} & -\frac{1}{5 \sqrt{2}} & \frac{2}{5 \sqrt{3}} & \frac{\sqrt{5}}{6} & \frac{2}{3}
\end{array}
\right),
\end{equation}
which has rank 4, corresponding to four BPS operators in this sector. The $\cG$ matrix:
\begin{equation}
\cG = \frac{N^4}{(N^2 - 1)(N^2 - 4)}
	\left(\begin{array}{cccccc}
			\frac{N-2}{N+3} & 0 & 0 & 0 & 0 & 0 \\
			0 & \frac{N-2}{N} & 0 & 0 & 0 & 0 \\
			0 & 0 & 1 & 0 & 0 & 0 \\
			0 & 0 & 0 & 1 & 0 & 0 \\
			0 & 0 & 0 & 0 & \frac{N+2}{N} & 0 \\
			0 & 0 & 0 & 0 & 0 & \frac{N+2}{N-3}
	\end{array}\right)	
\end{equation}
The BPS operators are again given by
\begin{equation}
		\cO^{\rm BPS}_{ R , \tau  }   = \sum_{R_1,\tau_1} ( \cG \cP )^{ R_1 , \tau_1}_{R , \tau} \cO_{ R_1 , \tau_1 }
\end{equation}
The resulting combinations are best seen by inspecting matrices $\cP$ and $\cG$ and not worth writing out explicitly.

The two-point function matrix on the BPS states will be
\begin{equation}
\hspace*{-3.0cm}\begin{split}
	&\cP\cG\cP =
	\frac{N^3}{(N^2-1)(N^2-4)(N^2-9)} \times
\\ &	
	\left(
\begin{array}{cccccc}
 \frac{10 N^3-37 N^2+11 N+36}{15} & -\frac{5 N^3-23 N^2+4 N+54}{6 \sqrt{5}} & \frac{N^3-3 N^2-6}{5 \sqrt{2}} & \frac{2N^3-4 N^2+4 N-18}{5 \sqrt{3}} & -\frac{N^3-3 N^2-4 N-18}{6 \sqrt{5}} & \frac{N^3+11N}{15} \\
  & \frac{7 N^3-16 N^2-37 N+72}{12} & \frac{N^3-15 N+12}{2 \sqrt{10}} & \frac{N^3+N^2-13 N+9}{\sqrt{15}} & -\frac{N^3-19N}{12} & \frac{N^3+3 N^2-4 N+18}{6 \sqrt{5}} \\
  & & \frac{9 N^3-75N}{10} & -\frac{ 6N^2-18}{5\sqrt{6}} & \frac{N^3-15 N-12}{2 \sqrt{10}} & -\frac{N^3+3 N^2+6}{5 \sqrt{2}} \\
 & & & \frac{3 N^3-19N}{5} & -\frac{N^3-N^2-13 N-9}{\sqrt{15}} & \frac{2N^3+4 N^2+4 N+18}{5 \sqrt{3}} \\
 & \ldots  & & &  \frac{7 N^3+16 N^2-37 N-72}{12} & \frac{5 N^3+23 N^2+4 N-54}{6 \sqrt{5}} \\
 & & & & & \frac{10 N^3+37 N^2+11 N-36}{15}
\end{array}
\right)
\end{split}\end{equation}
The characteristic polynomial of this matrix is:
%
%
\begin{equation}\begin{split}
\label{eq:l32_pgp_poly}
&\det(\cP\cG\cP - \lambda \mbI) =  
\frac{\lambda^2}{(N^2-9)(N^2-4)^3(N^2-1)^4}
\left(
	(N^2-9)(N^2-4)^3(N^2-1)^4\lambda^4
\right. \\ & \left. \quad\quad
	- 4N^4(N^2-4)^3(N^2-1)^3 \,\lambda^3
	+ 3N^6(2N^4-3N^2+4)(N^2-4)(N^2-1)^2 \,\lambda^2
\right. \\ & \left. \quad\quad	
	- 2N^{10}(2N^4-3N^2+4)(N^2-1) \,\lambda
	+ N^{16}
\right),
\end{split}\end{equation}

The $\lambda^2$ prefactor indicates two zero-eigenvalue states corresponding to the kernel of $\cP$. The remaining fourth-degree characteristic polynomial equation can not be solved explicitly, but it is useful for inspecting what happens at finite values of $N$. When $N$ approaches critical values, some eigenvalues of $\cP\cG\cP$ run off to infinity while the number of eigenvalues that remain finite is equal to the number of surviving BPS operators. One can see that (\ref{eq:l32_pgp_poly}) reduces to a third degree polynomial when $N\rightarrow 3$ and to a first degree polynomial when $N\rightarrow 2$. We can conclude the following multiplicities of BPS operators depending on $N$:
\begin{center}
	\begin{tabular}{c|l}
		$N$ & \# of BPS states \\
		\hline
		$> 3$ & 4 \\
		3 & 3 \\
		2 & 1 \\
		1 & 0
	\end{tabular}
\end{center}

Finally, we find the orthogonal BPS states in the limit of $N\rightarrow\infty$. That is, the preferred basis of symmetrized traces. We calculate the $O(1/N)$ term of $\cG$:
\begin{equation}
	\cG = \mbI + \frac{1}{N}\cG_1 + O(1/N^2),
\end{equation}
\begin{equation}
	\cG_1 = \left(\begin{array}{cccccc}
			-5 & 0 & 0 & 0 & 0 & 0 \\
			0 & -2 & 0 & 0 & 0 & 0 \\
			0 & 0 & 0 & 0 & 0 & 0 \\
			0 & 0 & 0 & 0 & 0 & 0 \\
			0 & 0 & 0 & 0 & 2 & 0 \\
			0 & 0 & 0 & 0 & 0 & 5
	\end{array}\right).
\end{equation}
Then the eigenvectors of
\begin{equation}
	\cP \cG_1 \cP = \cP + \frac{1}{N}\cP \cG_1 \cP + O(1/N^2)
\end{equation}
approach the eigenvectors of $\cP\cG_1\cP$ as $N\rightarrow\infty$. Since $\cP$ and $\cP\cG_1\cP$ commute, they can be simultaneously diagonalized. That means that we can resolve the degeneracy of the symmetrized traces which all have $\cP$ eigenvalue $+1$, by finding combinations of them which are at the same time eigenstates of $\cP\cG_1\cP$, but with different eigenvalues. We find these eigenstates reexpressed in terms of multitraces to be:
\begin{equation}\begin{split}
&\cO^{\BPS;(N=\infty)}_{(1)} = 
	10 \tr(X_1 X_i X_j) \tr(X^i X^j)
	+ (-\sqrt{19}+3\sqrt{11})\tr(X_1 X_i X_j) \tr(X^i) \tr(X^j)	
\\ & \quad
	+ \frac{1}{2} (3\sqrt{19}+\sqrt{11}) \tr(X_i X_j)\tr(X^i X^j)\tr(X_1)
	+ \frac{1}{2} (-3 + \sqrt{209})  \tr(X_i X_j)\tr(X^i) \tr(X^j)\tr(X_1),
\\
&\cO^{\BPS;(N=\infty)}_{(2)} = 
	10 \tr(X_1 X_i X_j) \tr(X^i X^j)
	+ (-\sqrt{19}-3\sqrt{11})\tr(X_1 X_i X_j) \tr(X^i) \tr(X^j)	
\\ & \quad
	+ \frac{1}{2} (3\sqrt{19}-\sqrt{11}) \tr(X_i X_j)\tr(X^i X^j)\tr(X_1)
	+ \frac{1}{2} (-3 - \sqrt{209})  \tr(X_i X_j)\tr(X^i) \tr(X^j)\tr(X_1),
\\
&\cO^{\BPS;(N=\infty)}_{(3)} = 
	10 \tr(X_1 X_i X_j) \tr(X^i X^j)
	+ (\sqrt{19}+3\sqrt{11})\tr(X_1 X_i X_j) \tr(X^i) \tr(X^j)	
\\ & \quad
	+ \frac{1}{2} (-3\sqrt{19}+\sqrt{11}) \tr(X_i X_j)\tr(X^i X^j)\tr(X_1)
	+ \frac{1}{2} (-3 - \sqrt{209})  \tr(X_i X_j)\tr(X^i) \tr(X^j)\tr(X_1),
\\
&\cO^{\BPS;(N=\infty)}_{(4)} = 
	10 \tr(X_1 X_i X_j) \tr(X^i X^j)
	+ (\sqrt{19}-3\sqrt{11})\tr(X_1 X_i X_j) \tr(X^i) \tr(X^j)	
\\ & \quad
	+ \frac{1}{2} (-3\sqrt{19}-\sqrt{11}) \tr(X_i X_j)\tr(X^i X^j)\tr(X_1)
	+ \frac{1}{2} (-3 + \sqrt{209})  \tr(X_i X_j)\tr(X^i) \tr(X^j)\tr(X_1).	
\end{split}\end{equation}

\subsection{Computational remarks}

The main challenge in the calculations was getting the $\cP$ matrix. In our implementation we used \emph{Mathematica} to evaluate the expression (\ref{eq:cP_alternative}). It involves three ingredients: calculation of $\mP_\alpha$, the representation matrices $D^R_{ij}(\alpha)$ and the Clebsch-Gordan (CG) coefficients $S^{ R ~ R ~ \L, ~ \tau }_{~  i ~ j ~ k }$. We evaluated the algebra element $\mP_\alpha$ directly using the definition (\ref{eq:mpa_defined}). The representation matrices were calculated using the method described in Section~7-7 of \cite{hammermesh}.

Finally, the CG coefficients $S^{ R ~ R ~ \L, ~ \tau }_{~  i ~ j ~ k }$ were calculated using the recursion relationships according to the method described in Section~7-14 of \cite{hammermesh}. This is the most computationally intensive procedure. We also found that the relevant equations (7-226), (7-233) needed a small clarification for dealing with multiplicities, which is worth pointing out here. The coefficients $K^{ \gamma ~ \alpha ~ \beta}_{\,i ~ a ~ b }$ used in the recursion relationship have, in fact, \emph{two} extra multiplicity indexes. The equation (7-226) should read as:
\begin{equation}
	S^{ \,\gamma ~\; \alpha ~\; \beta, ~ \tau }_{ij ~ ab ~ ef } =
	\sum_{\tau'} 
	K^{ \gamma ~ \alpha ~ \beta, ~\tau'}_{\,i ~ a ~ e }(\tau) \;
	S^{ \gamma_i ~ \alpha_a ~ \beta_e, ~ \tau' }_{\,j ~\; b ~~ f },
\end{equation}
where $\tau'$ runs over the multiplicity associated with $(\gamma_i,\alpha_a,\beta_e)$. Then $K^{ \gamma ~ \alpha ~ \beta, ~\tau'}_{\,i ~ a ~ e }(\tau)$ is the $\tau$'th solution of the equation (7-233) which should read as:
\begin{equation}
\begin{split}
	&K^{ \gamma ~ \alpha ~ \beta, \;\tau'}_{\,i ~ c ~ g }
	S^{ \gamma_i ~ \alpha_c ~ \beta_g, \;\tau' }_{jt ~ dr ~ hs }
	(f^\alpha_{cd}f^\beta_{gh} - f^\gamma_{ij})
+	
	K^{ \gamma ~ \alpha ~ \beta, \;\tau'}_{\,i ~ c ~ h }
	S^{ \gamma_i ~ \alpha_c ~ \beta_h, \;\tau' }_{jt ~ dr ~ gs }
	f^\alpha_{cd}g^\beta_{hg}
\\
&+
	K^{ \gamma ~ \alpha ~ \beta, \;\tau'}_{\,i ~ d ~ g }
	S^{ \gamma_i ~ \alpha_d ~ \beta_g, \;\tau' }_{jt ~ cr ~ hs }
	g^\alpha_{dc}f^\beta_{gh}
+
	K^{ \gamma ~ \alpha ~ \beta, \;\tau'}_{\,i ~ d ~ h }
	S^{ \gamma_i ~ \alpha_d ~ \beta_h, \;\tau' }_{jt ~ cr ~ gs }
	g^\alpha_{dc}g^\beta_{hg}
=
	K^{ \gamma ~ \alpha ~ \beta, \;\tau'}_{\,j ~ c ~ g }
	S^{ \gamma_j ~ \alpha_c ~ \beta_g, \;\tau' }_{it ~ dr ~ hs }	
	g^{\gamma}_{ij},
\end{split}	
\end{equation}
with sums over $i, j, c, d, g, h, t, r, s$ and the extra multiplicity $\tau'$.

The \emph{Mathematica} code is available from the authors upon request.

\section{Multiplicity tables}
\label{sec:app_multiplicities}

In this appendix we give some counting information.

First we write out the evaluated universal element as defined in (\ref{eq:p_definition}). We organize the terms in increasing $n$.
\begin{equation}\nonumber
\begin{split}	
\mP(S_2) &= \frac{1}{2!} \left( 2\Sigma_{[2]} + 2\Sigma_{[1,1]} \right), \\
\mP(S_3) &= \frac{1}{3!} \left( 
		2\Sigma_{[3]} + 3\Sigma_{[2,1]} + 5\Sigma_{[1,1,1]}
	\right), \\
\mP(S_4) &= \frac{1}{4!} \left(
		3 \Sigma _{[4]}+3 \Sigma _{[3,1]}+7 \Sigma _{[2,2]}+7 \Sigma _{[2,1,1]}+15 \Sigma _{[1,1,1,1]}
	\right), \\
\mP(S_5) &= \frac{1}{5!} \left(
		2 \Sigma _{[5]}+4 \Sigma _{[4,1]}+5 \Sigma _{[3,2]}+7 \Sigma _{[3,1,1]}+12 \Sigma _{[2,2,1]}+20 \Sigma _{[2,1,1,1]}+52 \Sigma _{[1,1,1,1,1]}
	\right),
\end{split}\end{equation}	
\begin{equation}\begin{split}\nonumber
\mP(S_6) &= \frac{1}{6!} \left(
		4 \Sigma _{[6]}+3 \Sigma _{[5,1]}+9 \Sigma _{[4,2]}+9 \Sigma _{[4,1,1]}+8 \Sigma _{[3,3]}+10 \Sigma _{[3,2,1]}+20 \Sigma _{[3,1,1,1]} 
		\right. \\ & \quad\quad \left.
		+31 \Sigma _{[2,2,2]}+31 \Sigma _{[2,2,1,1]}+67 \Sigma _{[2,1,1,1,1]}+203 \Sigma _{[1,1,1,1,1,1]}
	\right), \\	
\mP(S_7) &= \frac{1}{7!} \left(
		2 \Sigma _{[7]}+5 \Sigma _{[6,1]}+5 \Sigma _{[5,2]}+7 \Sigma _{[5,1,1]}+7 \Sigma _{[4,3]}+15 \Sigma _{[4,2,1]}+25 \Sigma _{[4,1,1,1]}
		\right. \\ & \quad\quad \left.
		+13 \Sigma _{[3,3,1]}+19 \Sigma _{[3,2,2]}+27 \Sigma _{[3,2,1,1]}+67 \Sigma _{[3,1,1,1,1]}+59 \Sigma _{[2,2,2,1]}+97 \Sigma _{[2,2,1,1,1]}
		\right. \\ & \quad\quad \left.
		+255 \Sigma _{[2,1,1,1,1,1]}+877 \Sigma _{[1,1,1,1,1,1,1]}
	\right),
\end{split}\end{equation}	
\begin{equation}\label{eq:p_samples}\begin{split}	
\mP(S_8) &= \frac{1}{8!} \left( 
		4 \Sigma _{[8]}+3 \Sigma _{[7,1]}+11 \Sigma _{[6,2]}+11 \Sigma _{[6,1,1]}+5 \Sigma _{[5,3]}+10 \Sigma _{[5,2,1]}+20 \Sigma _{[5,1,1,1]}
		\right. \\ & \quad\quad \left.
		+16 \Sigma _{[4,4]}+13 \Sigma _{[4,3,1]}+38 \Sigma _{[4,2,2]}+38 \Sigma _{[4,2,1,1]}+82 \Sigma _{[4,1,1,1,1]}+21 \Sigma _{[3,3,2]}
		\right. \\ & \quad\quad \left.
		+33 \Sigma _{[3,3,1,1]}+43 \Sigma _{[3,2,2,1]}+87 \Sigma _{[3,2,1,1,1]}+255 \Sigma _{[3,1,1,1,1,1]}+164 \Sigma _{[2,2,2,2]}
		\right. \\ & \quad\quad \left.
		+164 \Sigma _{[2,2,2,1,1]}+352 \Sigma _{[2,2,1,1,1,1]}+1080 \Sigma _{[2,1,1,1,1,1,1]}+4140 \Sigma _{[1,1,1,1,1,1,1,1]}
	\right). \\	
\end{split}
\end{equation}

In (\ref{tab:multiplicities}) we give the multiplicities $\cM_\Lambda$ for various $U(3)$ representations of the BPS states organized by total charge $n$. According to (\ref{eq:M_lambda_from_P}) they are equal to $\chi_\Lambda(\mP)$, using the elements in (\ref{eq:p_samples}). 
\begin{equation}
\label{tab:multiplicities}
\begin{tabular}{l|l}
	$n$ & $U(3)$ representations  \\
	\hline
	1 & $1\,[1]$ \\
	\hline
	2 & $2\,[2]$ \\
	\hline
	3 & $3\,[3] \oplus 1\,[2,1]$ \\
	\hline
	4 & $5\,[4] \oplus 2\,[3,1] \oplus 2\,[2,2]$ \\
	\hline
	5 & $7\,[5] \oplus 5\,[4,1] \oplus 4\,[3,2]$ \\
	  & $\; \oplus 1\,[2,2,1]$ \\
	\hline
	6 & $11\,[6] \oplus 8\,[5,1] \oplus 10\,[4,2] \oplus 2\,[3,3]$ \\
	  & $\; \oplus 1\,[4,1,1] \oplus 2\,[3,2,1] \oplus 2\,[2,2,2]$ \\	  
	\hline
	7 & $15\,[7] \oplus 15\,[6,1] \oplus 17\,[5,2] \oplus 10\,[4,3]$ \\
	  & $\; \oplus 2\,[5,1,1] \oplus 7\,[4,2,1] \oplus 1\,[3,3,1] \oplus 4\,[3,2,2]$ \\	  
	\hline
	8 & $22\,[8] \oplus 23\,[7,1] \oplus 32\,[6,2] \oplus 20\,[5,3] \oplus 12\,[4,4]$ \\
	  & $\; \oplus 5\,[6,1,1] \oplus 14\,[5,2,1] \oplus 9\,[4,3,1] \oplus 11\,[4,2,2] \oplus 2\,[3,3,2]$ \\	  
\end{tabular}
\end{equation}

In (\ref{tab:op_labels}) we list combinations of $R, \tau$ which are allowed at various values of $\Lambda$ to write the free operators $\cO_{\Lambda,M_\Lambda;R,\tau}$. The $\tau$ multiplicity for any $\Lambda,R$ is given by $c(R,R,\Lambda)$. We skip $\Lambda=[n]$ because in that case $R$ runs over all partitions of $n$ with multiplicity 1, and we will constrain ourselves to the $U(2)$ sector $\Lambda=[\lambda_1,\lambda_2]$. Note that the number of free operators is larger than $\cM_\Lambda$.
\begin{equation}
\begin{tabular}{l|l}
	$\Lambda$ & Operators \\
	\hline
	$\Lambda=[2, 1]$ & $\cO_{R=[2,1],\tau=1}$ \\
	\hline
	$\Lambda=[3, 1]$ & $\cO_{[3,1]}, \cO_{[2,1,1]}$ \\
	\hline
	$\Lambda=[2, 2]$ & $\cO_{[3,1]}, \cO_{[2,2]}, \cO_{[2,1,1]}$ \\
	\hline
	$\Lambda=[4, 1]$ & $\cO_{[4,1]}, \cO_{[3,2]}, \cO_{[3,1,1]}, \cO_{[2,2,1]}, \cO_{[2,1,1,1]}$ \\	
	\hline
	$\Lambda=[3, 2]$ & $\cO_{[4,1]}, \cO_{[3,2]}, \cO_{[3,1,1],\tau=1}, \cO_{[3,1,1],\tau=2}, \cO_{[2,2,1]}, \cO_{[2,1,1,1]}$ \\
	\hline
	$\Lambda=[5, 1]$ & $\cO_{[5,1]}, \cO_{[4,2]}, \cO_{[4,1,1]}, \cO_{[3,2,1],\tau=1}, \cO_{[3,2,1],\tau=2}, \cO_{[3,1,1,1]}, \cO_{[2,2,1,1]}, \cO_{[2,1,1,1,1]}$ \\
	\hline
	$\Lambda=[4, 2]$ & $\cO_{[5,1]}, \cO_{[4,2],\tau=1}, \cO_{[4,2],\tau=2}, \cO_{[4,1,1],\tau=1}, \cO_{[4,1,1],\tau=2}, \cO_{[3,3]}, \cO_{[3,2,1],\tau=1}, \cO_{[3,2,1],\tau=2},$ \\
		& $\cO_{[3,2,1],\tau=3}, \cO_{[3,1,1,1],\tau=1}, \cO_{[3,1,1,1],\tau=2}, \cO_{[2,2,2]}, \cO_{[2,2,1,1],\tau=1}, \cO_{[2,2,1,1],\tau=2}, \cO_{[2,1,1,1,1]}$ \\	
	\hline
	$\Lambda=[3, 3]$ & $\cO_{[4,1,1]}, \cO_{[3,2,1],\tau=1}, \cO_{[3,2,1],\tau=2}, \cO_{[3,1,1,1]}$		
\end{tabular}
\label{tab:op_labels}
\end{equation}

\section{Integer sequences in coefficients of $\mP$ } 
\label{sec:integerseq}  

The integer coefficients of $\mP$ which we called $t_p$ contain a wealth of combinatoric data. The first few examples of $t_p$ can be seen in (\ref{eq:p_samples}). Here we explore a few known integer sequences found among them. For more information about the sequences we will refer to \cite{OEIS}.

An easily identifiable two-dimensional class of coefficients is found by considering $p=[j^n]$. For example:
\begin{align}
	t_{[1^n]} &= \{ 1, 2, 5, 15, 52, 203, 877, 4140, \ldots \} \\
	t_{[2^n]} &= \{ 2, 7, 31, 164, 999, 6841, \ldots \} \\
	t_{[3^n]} &= \{ 2, 8, 42, 268, 1994, 16852, \ldots \} \\
	t_{[4^n]} &= \{ 3, 16, 111, 931, 9066, 99925, \ldots \}.
\end{align}
According to the formula (\ref{eq:TPcompact}) we can express the generating functions for this class of sequences as
\begin{equation}
	t(y_j) = \exp\left( \sum_{d|j} \frac{1}{d}\left(e^{d\,y_j} -1\right)\right),
\end{equation}
where $d$ are divisors of $j$. For example:
\begin{align}
	t(y_1) &= \exp\left( e^{y_1} - 1 \right) \\
	t(y_2) &= \exp\left( (e^{y_2} - 1) + (e^{2 y_2} - 1)/2 \right) \\
	t(y_3) &= \exp\left( (e^{y_3} - 1) + (e^{3 y_3} - 1)/3 \right) \\
	t(y_4) &= \exp\left( (e^{y_4} - 1) + (e^{2 y_4} - 1)/2 + (e^{4 y_4} - 1)/4 \right).
\end{align}

We find the sequences $t_{[1^n]}, t_{[2^n]}, t_{[3^n]}, t_{[4^n]}$ are identified in the \cite{OEIS} as \emph{A000110}, \emph{A002872}, \emph{A002874}, \emph{A141003} respectively. The whole class of such coefficients $t_{[j^n]}$ are in fact called ``Sorting numbers'' \cite{sorting}.

A couple of interesting special cases are worth mentioning. For $j=1$ the sequence is the Bell or exponential numbers. It counts the total number of ways that $n$ distinguishable elements can be divided into any number of indistinguishable subsets.

For $j=2$ we have a description as ``number of partitions of $2n$ objects invariant under a permutation consisting of $n$ $2$-cycles''.


It is also noteworthy that the 
 individual factors in (\ref{eq:TPcompact})  have a combinatoric 
and group theoretic meaning. 
\begin{equation} 
\exp {  { 1 \over d } \left ( x^d -1 \right ) } 
= \sum_{ k=0 }^{\infty}      B ( d , k ) {  x^k \over k ! }  
\end{equation}  
The numbers $ B ( d , i )$ are discussed in \cite{Kerber1978} 
which defines them 
as 
\begin{equation} 
B ( i , k ) = { i^k \over (ik)! } \sum_{ \sigma \in S_{ik} } a_i ( \sigma )^ k 
\end{equation} 
where $a_i ( \sigma ) $  is the number of $i$-cycles in the symmetric group element $ \sigma $. 

They are  related  to Stirling numbers of the second kind $ S ( k , j ) $. 
\begin{equation} 
B ( i , k ) = i^k \sum_{ j =1 }^{ k } {  S ( k , j ) \over i^j }  
\end{equation}  
The Stirling numbers of the second kind $ S ( k , j ) $
\cite{Wiki-stirling} count the number 
of ways to partition a set of $k$ elements into $j$ 
non-empty subsets. An explicit 
formula is 
\begin{equation} 
    S(k,j) = \frac{1}{j!}\sum_{i=0}^{j}(-1)^{j-i}{j \choose i} i^k. 
  \end{equation}

\section{Fourier basis two-point function}
\label{sec:free2pt}

Here we present a derivation of (\ref{diag2pt}) in the conventions of this paper. The Fourier basis operators are defined as (\ref{eq:ob_definition}).
\begin{equation}
\label{eq:appE_ob_definition}
\cO_{ \Lambda , M_{\Lambda} ,  R , \tau  } =  
\frac{ \sqrt {d_R } }{ n! }
\sum_{\alpha, \vec{a}}
S^{ ~ R~  R ~ \L ,~  \tau }_{ ~~  i ~ j ~ m } D^{R}_{ij} ( \alpha )
C^{ \vec a }_{ \Lambda , M_{\Lambda } ,
 m }  \cO_{ \vec a , \alpha }
\end{equation} 
The two-point function in the trace basis can be evaluated from Wick contractions to be:
\begin{equation}\begin{split}
\label{eq:appE_tr_2pt}
	\langle \cO_{ \vec b , \beta } | \cO_{ \vec a , \alpha } \rangle &= 
	\frac{1}{N^n} \langle \tr_n ( \mbX_{ \vec{b} } \beta ) | \tr_n ( \mbX_{ \vec{a} } \alpha ) \rangle
\\
	&= \sum_{ \gamma \in S_n  } \delta_{ \gamma ( \vec a )  , \vec b  } 
N^{C(\beta^{-1} \gamma^{-1} \alpha \gamma ) - n}		
\\
	&= \sum_{ \gamma, \sigma \in S_n  } \delta_{ \gamma ( \vec a )  , \vec b  } 
N^{C(\sigma) - n} \delta( \beta^{-1} \gamma^{-1} \alpha\gamma \sigma)
\\
	&= \sum_{ \gamma \in S_n  } \delta_{ \gamma ( \vec a )  , \vec b  } 
\delta(\beta^{-1} \gamma^{-1} \alpha\gamma  \Omega).
\end{split}\end{equation}
Here $C(\sigma)$ is the number of cycles in permutation $\sigma$ and we used the definition (\ref{omegadef})
\begin{equation}
\Omega = \sum_{ \sigma } N^{ C(\sigma) - n } \sigma.
\end{equation} 

We can evaluate now the two-point function of operators (\ref{eq:appE_ob_definition}) using (\ref{eq:appE_tr_2pt}):
\begin{equation}\begin{split}
	& \langle \cO_{ \Lambda_2 , M_{\Lambda_2} ,  R_2 , \tau_2  } | \cO_{ \Lambda_1 , M_{\Lambda_1} ,  R_1 , \tau_1  } \rangle 
\\
& = \frac{\sqrt{d_{R_1}d_{R_2}}}{(n!)^2} 
	\sum_{\vec{a}, \alpha, \vec{b}, \beta, \gamma}
	S^{ ~ R_2 ~  R_2 ~ \L_2 ,~  \tau_2 }_{ ~~  i_2 ~ j_2 ~ m_2 } 
	S^{ ~ R_1 ~  R_1 ~ \L_1 ,~  \tau_1 }_{ ~~  i_1 ~ j_1 ~ m_1 } 
	D^{R_2}_{i_2j_2} ( \beta )
	D^{R_1}_{i_1j_1} ( \alpha )
	\left(C^{ \vec b }_{ \Lambda_2 , M_{\Lambda_2} , m_2 }\right)^*
	C^{ \vec a }_{ \Lambda_1 , M_{\Lambda_1} , m_1 }
	\\ & \quad\quad\quad\quad
	\times \delta_{ \gamma ( \vec a )  , \vec b  } \delta( \beta^{-1} \gamma^{-1} \alpha\gamma \Omega)
\\
& = \frac{\sqrt{d_{R_1}d_{R_2}}}{(n!)^2} 
	\sum_{\vec{a}, \alpha, \gamma}
	S^{ ~ R_2 ~  R_2 ~ \L_2 ,~  \tau_2 }_{ ~~  i_2 ~ j_2 ~ m_2 } 
	S^{ ~ R_1 ~  R_1 ~ \L_1 ,~  \tau_1 }_{ ~~  i_1 ~ j_1 ~ m_1 } 
	D^{R_2}_{i_2j_2} ( \gamma^{-1} \alpha \gamma \Omega )
	D^{R_1}_{i_1j_1} ( \alpha )
	\left(C^{ \gamma(\vec a) }_{ \Lambda_2 , M_{\Lambda_2} , m_2 }\right)^*
	C^{ \vec a }_{ \Lambda_1 , M_{\Lambda_1} , m_1 }	
\\
& = \frac{\sqrt{d_{R_1}d_{R_2}}}{(n!)^2} 
	\sum_{\vec{a}, \alpha, \gamma}
	S^{ ~ R_2 ~  R_2 ~ \L_2 ,~  \tau_2 }_{ ~~  i_2 ~ j_2 ~ m_2 } 
	S^{ ~ R_1 ~  R_1 ~ \L_1 ,~  \tau_1 }_{ ~~  i_1 ~ j_1 ~ m_1 } 
	D^{R_2}_{ki_2}(\gamma) D^{R_2}_{kl}(\alpha) D^{R_2}_{lj_2}(\gamma)
	\frac{\chi_{R_2}(\Omega)}{d_{R_2}}
	D^{R_1}_{i_1j_1} ( \alpha )
	\\ & \quad\quad\quad\quad \times 
	D^{\L_2}_{m'_2m_2}(\gamma) 
	\left(C^{ \vec a }_{ \Lambda_2 , M_{\Lambda_2} , m'_2 }\right)^*
	C^{ \vec a }_{ \Lambda_1 , M_{\Lambda_1} , m_1 }	
\\
& = \frac{\chi_{R_2}(\Omega)}{d_{R_2}}	
	\delta_{\L_1,\L_2} \delta_{M_{\Lambda_1},M_{\Lambda_2}} \delta_{R_1,R_2} 
	S^{ ~ R_2 ~  R_2 ~ \L_2 ,~  \tau_2 }_{ ~~  i_2 ~ j_2 ~ m_2 } 
	S^{ ~ R_2 ~  R_2 ~ \L_2 ,~  \tau_1 }_{ ~~  i_1 ~ j_1 ~ m_1 } 
	\sum_{\tau'} 
	S^{ ~ R_2 ~  R_2 ~ \L_2 ,~  \tau' }_{ ~~  i_1 ~ j_1 ~ m_1 } 
	S^{ ~ R_2 ~  R_2 ~ \L_2 ,~  \tau' }_{ ~~  i_2 ~ j_2 ~ m_2 } 
\\
& = \frac{\chi_{R_2}(\Omega)}{d_{R_2}}	
	\delta_{\L_1,\L_2} \delta_{M_{\Lambda_1},M_{\Lambda_2}} \delta_{R_1,R_2} 
	\delta_{\tau_1,\tau_2}
\end{split}\end{equation}
in agreement with (\ref{diag2pt}). The following steps were performed:
\begin{itemize}
	\item In the second line the $\beta$ and $\vec{b}$ sums were performed using delta functions.
	\item In the third line we expanded the product $D^{R_2}_{i_2j_2}(\gamma^{-1}\alpha\gamma\Omega)$ and used:
\begin{equation}
	D^R_{ij}(\gamma^{-1}) = D^R_{ji}(\gamma),
\end{equation}
\begin{equation}
	D^R_{ij}(\Omega) = \delta_{ij} \frac{\chi_R(\Omega)}{d_R},
\end{equation}
because $\Omega$ is a central element. We also use the following transformation property of the CG coefficient:
\begin{equation}
	C^{ \gamma(\vec a) }_{ \Lambda_2 , M_{\Lambda_2} , m_2 } = 	 C^{ \vec a }_{ \Lambda_2 , M_{\Lambda_2} , m'_2 } D^{\L_2}_{m'_2m_2}(\gamma).
\end{equation}
This is easily understood using the bracket notation, where $C^{ \vec a }_{ \Lambda , M_{\Lambda} , m }$ is the overlap between $V^{\otimes n}$ basis $|\vec{a}\rangle$ and the $V^{U(M)}_\Lambda \otimes V^{S_n}_\Lambda$ basis $|\Lambda , M_{\Lambda} , m\rangle$:
\begin{equation}\begin{split}
C^{ \gamma(\vec a) }_{ \Lambda , M_{\Lambda} , m } 
&\equiv \langle \gamma(\vec{a}) | \Lambda , M_{\Lambda} , m \rangle
\\
&= \langle \vec{a} | \gamma | \Lambda , M_{\Lambda} , m \rangle
\\
&= \langle \vec{a} | \Lambda , M_{\Lambda} , m' \rangle \langle \Lambda , M_{\Lambda} , m' |  \gamma | \Lambda , M_{\Lambda} , m \rangle
\\
&\equiv 
C^{ \vec a }_{ \Lambda , M_{\Lambda} , m' } 
D^\L_{m'm}(\gamma).
\end{split}\end{equation}
Here it is important to note that according to our definition of $\gamma(\vec{a})$ the permutation $\gamma$ acts by the \emph{right action} on the vector space $|\vec{a}\rangle$, that is:
\begin{equation}
	\langle\gamma(\vec{a})| = \langle\vec{a}|\gamma, 
	~~~~~
	|\gamma(\vec{a})\rangle = \gamma^{-1}|\vec{a}\rangle.
\end{equation}

	\item In the fourth line we have performed the sums over $\vec{a},\alpha,\gamma$ using the symmetric group identities:
\begin{align}
	\sum_{\alpha\in S_n} D^{R_1}_{i_1j_1}(\alpha) D^{R_2}_{i_2j_2}(\alpha) &=
	\frac{n!}{d_{R_1}} \delta_{R_1,R_2} \delta_{i_1,i_2} \delta_{j_1,j_2}
\\	
	\sum_{\alpha\in S_n} D^{R_1}_{i_1j_1}(\alpha) D^{R_2}_{i_2j_2}(\alpha) D^{R_3}_{i_3j_3} &= n!
	\sum_{\tau} 
	S^{ ~ R_1 ~  R_2 ~ R_3 ,~  \tau }_{ ~~  i_1 ~ i_2 ~ i_3 } 
	S^{ ~ R_1 ~  R_2 ~ R_3 ,~  \tau }_{ ~~  j_1 ~ j_2 ~ j_3 }. 
\end{align}
The Clebsch-Gordan coefficients $S^{ ~ R_1 ~  R_2 ~ R_3 ,~  \tau }_{ ~~  i_1 ~ i_2 ~ i_3 }$ are again for coupling representations $R_1 \otimes R_2 \otimes R_3 \rightarrow 1$ and \emph{not} as more usual $R_1 \otimes R_2 \rightarrow R_3$. They are related by:
\begin{equation}
	S^{ ~ R_1 ~  R_2 ~ R_3 ,~  \tau }_{ ~~  i_1 ~ i_2 ~ i_3 } =
	\frac{1}{\sqrt{d_{R_3}}} S^{ ~ R_1 ~  R_2; ~ R_3  \tau }_{ ~~  i_1 ~ i_2; ~ i_3 }.
\end{equation}
Also we have used that our $C_{\L,M_\L,m}^{\vec{a}}$ are defined in an orthonormal basis and thus
\begin{equation}
	\sum_{\vec{a}} \left( C^{ \vec a }_{ \Lambda_2 , M_{\Lambda_2} , m_2 } \right)^*
	C^{ \vec a }_{ \Lambda_1 , M_{\Lambda_1} , m_1 } = 
	\delta_{\L_1,\L_2} \delta_{M_{\L_1},M_{\L_2}} \delta_{m_1,m_2}.
\end{equation}

	\item Finally, in the last line we used the orthogonality of the Clebsch-Gordan coefficients:
\begin{equation}
	\sum_{i,j,k} 
	S^{ ~ R_1 ~  R_2 ~ R_3 ,~  \tau_1 }_{ ~~  i ~~ j ~~ k } 	
	S^{ ~ R_1 ~  R_2 ~ R_3 ,~  \tau_2 }_{ ~~  i ~~ j ~~ k } 
	= \delta_{\tau_1,\tau_2}.
\end{equation}
	
\end{itemize}

A useful reference for properties of CG coefficients is \cite{hammermesh}.

\end{appendix}

\end{document}